\documentclass[aps,prd,reprint,nofootinbib,superscriptaddress,preprintnumbers]{revtex4-2} %
\usepackage{graphicx,url,amssymb,amsmath,rotating,color,units,wasysym,epsfig,multirow,epstopdf, stackengine}
\usepackage{amsmath}
\usepackage{cancel}
\usepackage{physics}
\usepackage{appendix}
\usepackage{amssymb, mathtools}
\usepackage[colorlinks,urlcolor=blue,citecolor=blue,linkcolor=blue]{hyperref}
\usepackage[dvipsnames,svgnames]{xcolor}

\usepackage[normalem]{ulem}
\usepackage{tabularx}
\usepackage{enumitem}
\usepackage{tensor}
\usepackage{subcaption}
\usepackage{booktabs,makecell,float}

\usepackage[utf8]{inputenc}
\usepackage{uniinput}
\usepackage{microtype}

\usepackage{tikz}
\usetikzlibrary{decorations,decorations.pathmorphing}
\tikzset{snake it/.style={decorate, decoration=snake}}

\newcommand{\nn}{\nonumber}
\newcommand{\ph}[1]{\phantom{#1}}
\newcommand{\ud}{\mathrm{d}}
\newcommand{\be}{\begin{equation}}
\newcommand{\ee}{\end{equation}}
\newcommand{\ba}{\begin{array}}
\newcommand{\ea}{\end{array}}
\newcommand{\bea}{\begin{eqnarray}}
\newcommand{\eea}{\end{eqnarray}}
\newcommand{\ds}{\displaystyle}
\newcommand{\pa}[1]{\left(#1\right)}
\newcommand{\paq}[1]{\left[#1\right]}
\newcommand{\mysc}[1]{{\textrm{\scshape #1}}}
\newcommand{\intk}{\int_{\boldsymbol{k}}}
\newcommand{\eikx}{\exp(i\boldsymbol{k}\!\cdot\!\boldsymbol{x})}
\newcommand{\sumk}{\boldsymbol{δ}\!\left(\boldsymbol{k}_1 \!+\! \boldsymbol{k}_2 \!+\! \boldsymbol{k}_3\right)}

\makeatletter
\renewcommand{\dddot}[1]{%
  {\mathop{#1}\limits^{\vbox to-1.4\ex@{\kern-\tw@\ex@
   \hbox {\normalfont .\kern-.05em.\kern-.05em.}\vss}}}}
\makeatother

\usepackage{cleveref}

\newcommand{\Meudon}{\affiliation{Laboratoire d'étude de l'Univers et des phénomènes eXtrêmes (LUX), Observatoire de Paris, Université PSL, Sorbonne Université, CNRS, 92190 Meudon, France}}
\newcommand{\PI}{\affiliation{Perimeter Institute for Theoretical Physics, 31 Caroline St., Waterloo, ON, N2L 2Y5, Canada}}
\newcommand{\Uppsala}{\affiliation{Institutionen för fysik och astronomi, Uppsala Universitet, Box 803, SE-751 08 Uppsala, Sweden}}

\newcommand{\UNESP}{\affiliation{Instituto de Fisica Teorica, UNESP - Universidade Estadual Paulista, Sao Paulo 01140-070, SP, Brazil}}

\newcommand{\ICTPSAIFR}{\affiliation{ICTP South American Institute for Fundamental Research, Sao Paulo 01140-070, SP, Brazil}}

\begin{document}

\title{Generic EFT-motivated beyond General Relativity gravitational wave tests and their curvature dependence: from observation to interpretation}

\author{Laura Bernard}
\email{laura.bernard@obspm.fr }
\Meudon

\author{Suvendu Giri}
\email{sgiri@perimeterinstitute.ca}
\PI
\Uppsala

\author{Luis Lehner}
\email{llehner@perimeterinstitute.ca}
\PI

\author{Riccardo Sturani}
\email{riccardo.sturani@unesp.br}
\UNESP
\ICTPSAIFR

\begin{abstract}
    \noindent
	We present a ``dictionary'' to expedite the identification of potential deviations in gravitational waveforms from those predicted by General Relativity (GR) during the inspiral phase of black hole binaries. Assuming deviations from GR can be described by a local Effective Field Theory (EFT) formulated in terms of curvature operators (and possibly additional scalar fields), this dictionary characterizes how deviations scale with the masses of the binary components and identifies the leading order Post-Newtonian corrections in generic theories constructed within the EFT framework. By establishing a direct connection between observations and candidate theories beyond GR, this dictionary also aids in distinguishing genuine physical effects from systematic errors. These results can be readily incorporated into essentially all existing tests for the inspiral regime and, in particular, facilitate a more efficient combination of data from multiple events.
\end{abstract}
\preprint{UUITP-20/25}
\maketitle

\section{Introduction}

The search for deviations from General Relativity (GR) with gravitational waves (GWs) faces many challenges (see e.g.~\cite{Berti:2015itd}). Among these, is a vast landscape of potential alternative theories~\cite{Yunes:2013dva,Shankaranarayanan:2022wbx}. 
There is a scarcity of detailed predictions within each potential theory that can be readily utilized in data analysis and there are, in turn, poorly understood degeneracies with astrophysical effects.

A large fraction of existing tests rely on theory-agnostic searches for general anomalies with respect to predictions within GR~\cite{LIGOScientific:2016lio,LIGOScientific:2018dkp,LIGOScientific:2019fpa,LIGOScientific:2020tif,LIGOScientific:2021sio}. These tests have paved the way for investigating deviations, but typically overlook physical expectations regarding the possible behavior of realistic deviations. Consequently, derived constraints are generally less than optimal and also difficult to interpret with respect to theories of interest~\cite{Perkins:2022fhr,Mehta:2022pcn}. Complementary tests exploit available (limited) knowledge  from a small set of theories where special solutions limited to specific regimes (inspiral, ringdown), motivate specific targets for analysis. In many cases, these solutions are not necessarily robust, as mathematical obstacles intrinsic to most theories often get in the way to even check their physical soundness. Even with recent advances aiming to control mathematical pathologies which plague most potential theories~\cite{Kovacs:2020ywu,East:2020hgw,Cayuso:2023xbc,Corman:2024cdr,Figueras:2024bba}, creating faithful template banks would remain an exceedingly expensive endeavor. It is relevant to recall that most theories being considered are typically derived within the Effective Field Theory (EFT) framework, considering higher order curvature corrections (with the potential addition of other fields) to the Einstein-Hilbert action to capture physics beyond GR. For instance by defining such a beyond GR theory by a Lagrangian density given by ${\cal L }= {\cal L}_{\textrm{EH}} + F({\Psi,{\cal R}})$ with ${\cal L}_{\textrm{EH}}$ the Einstein-Hilbert Lagrangian and $F$ denoting a scalar density with derivatives of order $2p$ of the metric field $g_{ab}$ and possibly additional degrees of freedom denoted by $\Psi$. The former are constructed from curvature contributions given by independent combinations of Riemann tensors; though, potentially, derivatives of both of these can also be considered.

Within this framework, all independent corrections at a given order, in principle, should be included (e.g.~\cite{Donoghue:1995cz,CarrilloGonzalez:2022fwg}). However, in practice—both on theoretical and computational fronts—only specific terms are typically considered, as a complete treatment would entail a prohibitively complex effort.\footnote{Some reduction may be possible under specific symmetry assumptions or fine tuning, though the naturalness of such assumptions would require justification.} For instance, at quadratic order in the curvature and with the addition of a scalar degree of freedom, Einstein Scalar Gauss Bonnet and dynamical Chern Simons are studied independently, even through from an EFT point of view, they should be pieces of a larger theory that contains both.

Faced with the challenge of offering theoretical guidance that minimizes the aforementioned shortcomings, while still enhancing both the detectability and interpretability of possible signatures, a complementary approach can be pursued. Here, the main target is to determine the ``curvature/gradient order" of corrections in the Lagrangian—within some perturbative scheme—at an agnostic level. Thus, regardless of: (i) the particular order of operators considered for going beyond GR, and (ii) the mathematical pathologies triggered by the behavior of a particular theory at short scales. This approach aims to unearth leading effects of beyond GR theories with corrections at a defined order of curvature. Our strategy exploits the fact that observable effects depend on the mass of the binary with a specific scale  determined by the order of operators ($p$) and the scale ($\ell$) of new physics they account for. Such behavior can then help discern beyond GR features from astrophysical and systematic effects as multiple events are analyzed. Specific investigations, tied to a subset of theories and regimes of the dynamics, have a connection with this idea~\cite{Stein:2013wza,Carullo:2021dui,Maselli:2023khq} but we here argue for its general applicability, stress its value and discuss a general way to tie potential observations to theory. Initial steps in this direction have been taken in \cite{Dideron:2022tap,Payne:2024yhk}.

Recall that current observations of gravitational wave mergers, predominantly from binary black holes, are well captured by GR, thus the scale of new physics—where corrections to GR are order ${\cal O}(1)$ is not encountered in the inspiral regime. This suggests that the strain can be expressed to leading order as $h = h_{\mysc{gr}} + ({\ell}/m)^q \, \delta$. Here, $m$ represents a characteristic mass of the system, $\delta$ a quantity depending on the physical parameters of the system (but independent of the new-physics scale) and $q(p)$ an {\em integer number} which is determined by $p$ and whether extra degrees of freedom are assumed. This will be a generic feature of any observable of interest, and crucially will be less pronounced for more massive systems as heavier objects have weaker curvature  at horizon scales. Further, since the merger of two black holes produce a heavier one, the inspiraling phase is particularly important for estimating the value of $q$. Such a stage is particularly amenable to be studied through a perturbative treatment. Specifically, one can use the tools of Non-Relativistic General Relativity (NRGR)~\cite{Goldberger:2004jt,Porto:2016pyg}, also known as PN-EFT, to efficiently compute how the quantity $q$ depends on the order of curvature corrections $p$, as well as to determine the leading Post-Newtonian order at which deviations from GR first appear.

The goal of this work is to provide a self-contained set of  broad rules for interpreting the underlying causes of potential deviations from GR, and connect them to specific theories. To this end, we derive the leading order corrections induced from generic EFT-inspired extensions of GR, focusing on their imprint on the strain in particular. This also helps distinguish genuine deviations due to modified gravity from systematic effects arising from, for example, astrophysical parameter uncertainties or waveform modeling inaccuracies (for a comprehensive enumeration see~\cite{Gupta:2024gun}).

This work is organized as follows. We first include an executive summary, giving the main results in succinct form in \cref{sec:execsumm}. Next, in \cref{sec:PNEFT} we review the main ingredients of the Post Newtonian Effective Field Theory (PN-EFT)  formalism, illustrating with two examples the basic rules that can be applied generically to corrections of any order. In the process, we also highlight additional considerations necessary for applying the PN-EFT framework to extensions of GR. Before discussing the general case, we review the construction of Lagrangian densities to capture general corrections at a given order in \cref{sec:lagrangian}. We then provide in \cref{sec:rulesgeneral} the general rules, apply them to some additional cases \cref{sec:examples} and conclude in \cref{sec:final}.
In the appendices, we present a few complementary explicit derivations to show:
(i) that the Gauss-Bonnet invariant $G_3$ is not an independent combination of Riemann invariants
distinct from $I_1\coloneqq R^{ab}_{~~cd} R^{cd}_{~~ef} R^{ef}_{~~ab}$ (\cref{app:cubicapp});
(ii) how a contribution that can be ``field-redefined" away
indeed gives no effect (clarifying confusing results recently presented) (\cref{app:nullRK});
(iii) how each term in the Gauss-Bonnet invariant independently evaluates to zero (\cref{app:quadratic}.)

Throughout this work, lowercase Latin indices from the beginning of the alphabet denote spacetime indices, while indices from the middle of the alphabet are used for spatial components only. Where needed, a lower case Latin index will be used to label objects (in the binary) and distinguishing it from a spacetime index is natural in context. {Last, we adopt $c=1$ throughout this work except when communicating final results related to observational consequences, where we make $c$ explicit in relevant factors of $(v/c)$.}

\section{Executive summary}\label{sec:execsumm}
The rest of this work details how to obtain the expected observable scaling with $(\ell/m)$, as well as the leading Post-Newtonian (PN) corrections induced by EFT-motivated corrections to GR.  For the reader interested in quickly grasping the main takeaways, they are summarized below (the first two cover the most common classes of EFT-motivated theories). 

Consider the Lagrangian density ${\cal L }= {\cal L}_\mysc{eh} + F({\Psi,{\cal R}})$ where $F$ is a correction to the Einstein-Hilbert Lagrangian ($\mathcal{L}_\mysc{eh}$) that involves $2p$ total derivatives—which includes derivatives of the metric field $g_{ab}$, and possibly additional degrees of freedom denoted by $\Psi$. Below, we denote such a correction with $2p$ derivatives to be order $p$ (as it elicits the order at which the curvature contributions would appear in the first two cases).
\begin{enumerate}[leftmargin=*]
	\item Given a correction of order $p\ge 3$ in the curvature\footnote{A correction consistent with $p=1$ corresponds to a deviation from GR that is independent of the mass scale, potentially indicating an exotic compact object. As we will discuss, $p=2$ does not induce any changes in the vacuum case.} and no extra degrees of freedom (and without derivatives of curvature contributions), departures from GR in the strain scale as $δh/h_\mysc{n} \sim (\ell/m)^{2(p-1)}$ and the leading order PN correction is always 5PN. For $p=3$ this correction comes from both potential effects as well as tidal ones. For $p>3$ it stems from purely tidal effect.
    
	\item Given a correction of order $p \ge 2$ in the curvature and the addition of a  scalar degree of freedom linearly coupled to the curvature contributions\footnote{{Higher order couplings modify this dependence, which can be estimated  by examining solutions of the scalar field's equation of motion.}} (not considering derivatives of the curvature or scalar field), departures from GR—directly arising from the higher-curvature interaction term in the action—in the plus and cross polarization modes of the strain scale as $(\ell^2/(m\,m_\mysc{l}))^{2(p-1)}$, where $m_\mysc{l}$ is the lightest object in the binary. The potential breathing mode\footnote{Potentially detectable when matter is assumed to couple to the Jordan frame and the scalar field decays appropriately at far distances by e.g. multiple detectors, or in a single one when the gravitational wavelength is not small with respect to a single detector's arm~\cite{Essick:2017wyl}.} of the strain scales as $(\ell/m)^{2(p-1)}$, and the leading affected PN order is 3PN for non-spinning binaries in the inspiral regime.\footnote{In addition, we also expect the well-known 1PN correction present in any scalar-tensor theory, proportional to the scalar charge.}

	\item Given a correction of ``net order'' $p$ constructed out of curvature contributions to order $(p-n)$ and $2n$ derivatives—with no extra degrees of freedom—the strain also scales as $(\ell/m_\mysc{l})^{2(p-1)}$. However, the leading PN correction appears at order $\min(3p-n-4,5)$

	\item Given a correction of ``net order'' $p$ constructed out of $a$ derivatives of $b$ scalar fields, and $c$ derivatives of curvature terms of order $d$, the strain scales with a PN correction of order $\min(2p+d+b-4,3)$. However, the scaling with mass requires an understanding of the scalar field equation of motion and their dependence on the coupling parameter, necessitating a case-by-case analysis.

\end{enumerate}
The above ``dictionary'' provides a direct way to constrain a potential set of sub-theories as consistent with observations (as well as distinguish inconsistent ones). For instance, deviations indicating a dependence on mass as $m^{-4}$ is consistent with, e.g. Einstein Scalar Gauss Bonnet gravity~\cite{Kanti:1995vq}, dynamical Chern-Simons gravity~\cite{Alexander:2009tp}—both involving quadratic curvature corrections with an extra scalar degree of freedom\footnote{And as such, from an EFT point of view, not distinct theories in their own right (unless a particular symmetry or extreme fine tuning singles them out) but as being parts of a general theory consistent with that order and the existence of scalar degrees of freedom.}—and cubic gravity~\cite{Bueno:2016xff}. However, if such deviations show departures from GR beginning at 5PN, it  singles cubic gravity as the consistent option. Further, this dictionary can help motivate improved strategies to search for potential deviations~\cite{Silva:2022srr,Dideron:2022tap,Payne:2024yhk,Maenaut:2024oci} and distinguish them from systematic effects. Notice also that the case $p=1$ would indicate a correction independent of mass, which would indicate beyond GR effects are not linked to curvature but other effects, for instance exotic compact binaries instead of black holes.

\section{The PN-EFT formalism}\label{sec:PNEFT}

As already mentioned in the introduction, our focus on the inspiral regime of a binary system aims to capture the leading order impact of deviations on the gravitational waveforms, as induced by theories beyond GR motivated from EFTs. In particular, we aim to answer two key questions: (i) how do deviations scale with the order of modifying operators? (ii) at which post-Newtonian order do these deviations first appear? To this end, a particularly convenient approach is to employ what is known as NRGR, also known as the PN-EFT formalism. This approach exploits EFT techniques that enable an efficient computation of the binary dynamics within the PN approximation scheme, under the assumption of a separation of scales. Namely, the essential contributions to the Lagrangian—and therefore to the equations of motion of the system, can be cleanly targeted in three regions:
(i) an \emph{``internal'' zone} (with scale $r_s$)—of the size of the source;
(ii) a \emph{``near'' or ``potential'' zone} at orbital scales  (of size $r$) and
(iii) a \emph{``radiation'' zone} where one can identify gravitational waves produced by objects moving with relative speed $v$, with radiation wavelength $\lambda_{\mysc{gw}} \simeq r/v$.

In the inspiral regime (where $v \ll 1$), $r_s\sim v^2 r \ll r \ll \lambda_{\mysc{gw}}$ the EFT machinery can be exploited to handle multiple scales in a ``tower-like'' fashion. In particular, we will separate the long wavelength (radiation) modes propagating degrees of freedom, from the short wavelength (potential) modes which contribute to the binding energy. The former will be on-shell and their typical four-momentum will scale as $\left(ω,|\boldsymbol{k}|\right)\sim\left(v/r,v/r\right)$, while the latter ones are off-shell with a four-momentum scaling as $\left(ω,|\boldsymbol{k}|\right)\sim\left(v/r,1/r\right)$. The EFT formalism makes a convenient use of these different scalings by going to Fourier space. For a detailed presentation of the approach, see~\cite{Goldberger:2004jt,Foffa:2013qca,Porto:2016pyg,Levi:2018nxp}. Below, we review the main building blocks of the formalism in GR to later be able to compute both potential and radiative contributions beyond GR. For simplicity, we restrict to non-spinning objects in this work; and indeed the overall discussion as well as a majority of the results carry over to the general case.

First, focusing on the aforementioned regime, we assume that we have already solved the internal problem and we model the individual objects by point-particles, eventually augmented by finite size effects. The corresponding worldline action is then,
\begin{align}\label{eq:Spp}
	S_\mysc{pp} = -\sum_{a=1,2}m_a\int\ud\tau_a\,.
\end{align}
We augment the bulk action (the usual Einstein-Hilbert action) with a gauge-fixing term (we choose the harmonic gauge $0 = \Gamma^{a} \coloneqq g^{cd}\Gamma^{a}_{~cd}$):
\begin{subequations}
	\begin{align}\label{eq:EH}
		S_{\rm bulk} & = \frac{1}{16πG_\mysc{n}}\int\ud^4 x\sqrt{-g}  R\,,                                    \\ \label{eq:GF}
		S_\mysc{gf}  & = \frac{1}{16πG_\mysc{n}}\int\ud^4 x \sqrt{-g} \left( - \frac{1}{2}Γ^a Γ_a \right) \,,
	\end{align}
\end{subequations}
where $R$ is the Ricci scalar. Defining $Λ^{-2} \coloneqq 32\pi G_\mysc{n}$, the normalization constant can be written more conveniently as $2Λ²$, which gives the gauge-fixed EH action
\begin{equation}\label{eq:EH+GF}
	S_{\rm bulk} + S_\mysc{gf} = 2Λ²\int\ud^4 x\sqrt{-g} \left[ R - \frac{1}{2}Γ^a Γ_a \right]\,.
\end{equation}
Throughout this article, we will work in the harmonic gauge, and all computations are performed with this gauge fixing term implicitly added to the action.

Next, we decompose the metric using the Kol-Smolkin (KS) parametrization~\cite{Kol:2007bc},
\begin{align}
	g_{\mu\nu} = e^{2ϕ/Λ}
	\begin{pmatrix}
		-1                   &  & \dfrac{A_j}{\Lambda}    \\[8pt]
		\dfrac{A_i}{\Lambda} &  & \left(\delta_{ij}+\dfrac{\sigma_{ij}}{\Lambda}\right)e^{-4ϕ/Λ} +\dfrac{A_i A_j}{\Lambda^2}
	\end{pmatrix}\,,
\end{align}
where $i,j$ run over spatial indices. In passing this also allows  defining $h_{\mu \nu} = g_{\mu \nu} - \eta_{\mu \nu}$. We insert the above metric decomposition in the bulk and point-particle actions. Then, we expand the Lagrangians up to a certain order in the fields $\{\phi, A_i, \sigma_{ij}\}$, and we use the result to derive a set of Feynman rules for the fields by applying the Fourier transform. These rules constitute the basic ingredients to compute the binding potential, that is obtained by integrating out the gravitational fields.

As an example, expanding the GR Lagrangian in \cref{eq:EH+GF} gives, at quadratic order in fields
\begin{align}\label{eq:Sbquad}\nn
	S^{(\rm quad)}_{\rm bulk} \supset & \int\ud^4x \biggl\{ -4\bigl(\partial_{i}\phi\partial^{i}\phi -\partial_t\phi\partial_t\phi\bigr)+\partial_{i}A_{j}\partial^{i}A^{j}    \\\nn
	& -\partial_t A_{i}\partial_t A^{i} +\frac{1}{4}\bigl(\partial_{i}\sigma_{~j}^{j}\partial^{i}\sigma_{~k}^{k}-2\partial_{i}\sigma_{jk}\partial^{i}\sigma^{jk} \\
	  & -\partial_t\sigma_{~j}^{j}\partial_t\sigma_{~k}^{k} -2\partial_t\sigma_{jk}\partial_t\sigma^{jk}\bigr) \biggr\} \,.
\end{align}

An advantage of using the KS parametrization is that in GR, propagators are diagonal in the fields—meaning that at quadratic order the fields $ϕ, A_i, σ_{ij}$ don't mix.\footnote{Adding higher derivative terms to the EH action can introduce mixing at quadratic order. See \cref{app:quadratic} for an explicit example in Gauss-Bonnet gravity.}
We can now read off the Feynman propagator for the scalar field from the above Lagrangian
\begin{align} \label{eq:prop}\nn
	\langle ϕ(x) ϕ(y) \rangle & = -\frac{i}{8} \int \frac{\dd^4 k}{\left(2π\right)^4} \frac{\exp(ik(x-y))}{k² + iϵ}    \\ \nn
	  & \hspace{-45pt} = -\frac{i}{8} \int \frac{\dd ω}{2π} e^{-iω(t_1-t_2)} \int_{\boldsymbol{k}} \frac{\exp(i\boldsymbol{k}\!\cdot\!(\boldsymbol{x}-\boldsymbol{y}))}{\boldsymbol{k}² - ω² + iϵ} \\ \nn
	  & \hspace{-45pt} \approx -\frac{i}{8}\int_{\boldsymbol{k}} \frac{\exp(i\boldsymbol{k}\!\cdot\!(\boldsymbol{x}-\boldsymbol{y}))}{\boldsymbol{k}²} \\ \nn
	  & \times \int\frac{\dd\omega}{2\pi} \left[ 1 +\frac{ω²}{\boldsymbol{k}²} +\cdots \right]e^{-i\omega(t_1-t_2)}    \\
	  & \hspace{-45pt} \approx -\frac{i}{8}\cdot δ(t_1-t_2) \int_{\boldsymbol{k}} \frac{\exp(i\boldsymbol{k}\!\cdot\!(\boldsymbol{x}-\boldsymbol{y}))}{\boldsymbol{k}²}    \\ \nn
	  & \hspace{-45pt} = -\frac {iδ(t_1-t_2)}{32\pi r}\,,
\end{align}
where $\{x,k\}$ are four vectors, the corresponding  bold letters $\{\boldsymbol{x},\boldsymbol{k}\}$ represent their spatial parts ($\pa{x,y}_0=t_{1,2}$), $r\equiv |\boldsymbol{x}-\boldsymbol{y}|$, and we have defined $\int_{\boldsymbol{k}} \coloneqq \dd^3 \boldsymbol{k}/\left(2π\right)^3$.
Since these modes are off-shell, the $iϵ$ prescription is irrelevant and trading $\omega \to \mp i\dd/\dd t_{1,2}$ shows that $k=\left(ω,{\boldsymbol{k}}\right)\sim\left(v/r,1/r\right)$. We will work in the limit $v \ll 1$, thus justifying the propagator expansion using $ω \ll |{\boldsymbol{k}}|$. The last line in \cref{eq:prop} gives the leading order scaling of the propagator—this will be useful later. Propagators for $A_i$, and $σ_{ij}$ differ from that of $ϕ$ only in their tensor structure:
\begin{subequations}
	\begin{align}
		\langle A_i(x) A_j(y) \rangle       & \approx \frac{δ_{ij}}{2}\cdot δ(t_1-t_2) \int_{\boldsymbol{k}} \frac{i \exp(i\boldsymbol{k}\cdot(\boldsymbol{x}-\boldsymbol{y}))}{\boldsymbol{k}²}\,,     \\
		\langle σ_{ij}(x) σ_{kl}(y) \rangle & \approx \mathcal{D}_{ij;kl}\cdot δ(t_1-t_2) \int_{\boldsymbol{k}} \frac{i \exp(i\boldsymbol{k}\cdot(\boldsymbol{x}-\boldsymbol{y}))}{\boldsymbol{k}²} \,,
	\end{align}
\end{subequations}
where $\mathcal{D}_{ij;kl} \coloneqq -\left( δ_{ik}δ_{jl} + δ_{il}δ_{jk} - 2 δ_{ij}δ_{kl} \right)/2$.

This gives Feynman rules for the propagators (in the static limit)
\begin{equation}\label{eq:phi-prop}
	\approx i δ^{(3)} \left(\boldsymbol{k}+\boldsymbol{q}\right) δ(t_1-t_2) \frac{\mathcal{A}}{\boldsymbol{k}²}
	\sim \frac{r^5}{t}\,,
\end{equation}
where the right hand side shows the scaling with $r$ and $\dd t$, using $δ^{(3)}(\boldsymbol{k}) \sim 1/\boldsymbol{k}^3 \sim r^3$, and $δ(t) \sim 1/t$, and $\mathcal{A}$ depends on the field:
\begin{table}[H]
	\centering
	\begin{tabularx}{0.6\linewidth}{@{}
		>{\centering\arraybackslash}m{0.1\linewidth}
		>{\centering\arraybackslash}m{0.2\linewidth}
		>{\centering\arraybackslash}X@{}}
		\toprule
		        & $\mathcal{A}$         
                & diagram \\ \midrule
		$ϕ$     & $-1/8$                
                & $\begin{tikzpicture}[thick]
				    \draw [black, dashed] (0,0) -- (2,0);
			    \end{tikzpicture}$   \\ \addlinespace[5pt]
		$A_i$   & $δ_{ij}/2$
                & $\begin{tikzpicture}[thick]
                    \draw [black, double, dashed] (0,0) -- (2,0);
			    \end{tikzpicture}$   \\ \addlinespace[5pt]
		$σ_{ij}$ & $\mathcal{D}_{ij;kl}$ 
                & $\begin{tikzpicture}[thick]
                    \draw [black, decorate, decoration={snake, amplitude=0.5mm, segment length=2mm}, shorten >=1pt] (0,0) -- (2,0);
			    \end{tikzpicture}$ \\ \bottomrule
	\end{tabularx}%
\end{table}
\begin{figure*}
	\centering
	\begin{subfigure}{0.19\linewidth}
		{
			\begin{tikzpicture}[thick,]
				\draw [black] (0,2) -- (2,2);
				\draw [black, dashed] (1,2) -- (1,0.7);
				\node[align=left] at (0.7,1.3) {$ϕ$};
			\end{tikzpicture}
			\vspace{-10pt}
		}
		\caption{
			\label{fig:scalar-matter-coupling}}
	\end{subfigure}
    \hfill
	\begin{subfigure}{0.19\linewidth}
		{
			\begin{tikzpicture}[thick]
				\draw [black] (0,2) -- (2,2);
				\draw [black, dashed] (1,2) -- (.5,0.7);
				\draw [black, dashed] (1,2) -- (1.5,0.7);
				\node[align=left] at (0.5,1.3) {$ϕ$};
				\node[align=left] at (1.5,1.3) {$ϕ$};
			\end{tikzpicture}
			\vspace{-10pt}
		}
		\caption{
			\label{fig:scalar-matter-quadratic-coupling}}
	\end{subfigure}
	\hfill
	\begin{subfigure}{0.19\linewidth}
		{
			\begin{tikzpicture}[thick]
				\draw [black] (0,2) -- (2,2);
				\filldraw (1,2) circle (2pt);
				\draw [black, dashed] (1,2) -- (.5,0.7);
				\draw [black, dashed] (1,2) -- (1.5,0.7);
				\node[align=left] at (0.5,1.3) {$ϕ$};
				\node[align=left] at (1.5,1.3) {$ϕ$};
			\end{tikzpicture}
			\vspace{-10pt}
		}
		\caption{
			\label{fig:scalar-matter-tidal-coupling}}
	\end{subfigure}
	\hfill
	\begin{subfigure}{0.19\linewidth}
		{
			\begin{tikzpicture}[thick]
				\draw [black, dashed] (1,  1.3) -- (1, 2);
				\draw [black, dashed] (0.5,0.8) -- (1,1.3);
				\draw [black, dashed] (1.5,0.8) -- (1,1.3);
				\node[align=left] at (0.4,1) {$ϕ$};
				\node[align=left] at (1.6,1) {$ϕ$};
				\node[align=left] at (0.8,1.9) {$ϕ$};
			\end{tikzpicture}
		}
		\caption{
			\label{fig:3-point-vertex-phi3}}
	\end{subfigure}
	\hfill
	\begin{subfigure}{0.19\linewidth}
		{
			\begin{tikzpicture}[thick]
				\draw [black, dashed] (0.5,0.8) -- (1,1.3);
				\draw [black, dashed] (1.5,0.8) -- (1,1.3);
				\draw [black, decorate, decoration={snake, amplitude=0.5mm, segment length=2mm}, shorten >=1pt] (1,  1.3) -- (1, 2);
				\node[align=left] at (0.4,1) {$ϕ$};
				\node[align=left] at (1.6,1) {$ϕ$};
				\node[align=left] at (1.35,1.9) {$σ_{ij}$};
			\end{tikzpicture}
		}
		\caption{
			\label{fig:3-point-vertex-phi2-sigma}}
	\end{subfigure}
	\caption{Interaction vertices at lowest orders: (a) linear scalar-matter coupling, (b) quadratic scalar-matter coupling, both coming from \cref{eq:KS} (c) tidal coupling coming from \cref{eq:Stidal}, (d) bulk $\phi^3$ interaction, and (e) bulk $\phi^2\sigma$ interaction both coming from \cref{eq:GR-cubic}.
		\label{fig:all-vertices}}
\end{figure*}
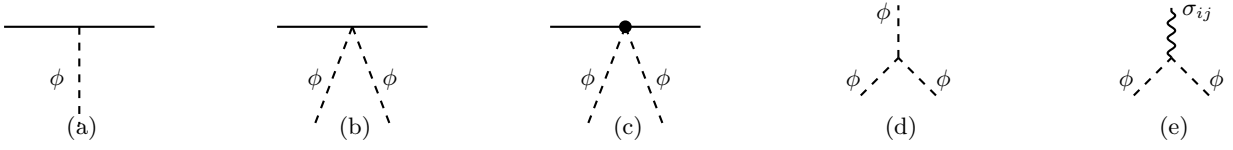
Continuing to expand \cref{eq:EH+GF} to next order (cubic) in field power gives three-point interactions of the fields. Here is a subset of the terms that will be relevant for the following discussion
\begin{align}\label{eq:GR-cubic}\nn
	S^{(\rm cub)}_{\rm bulk} \supset \frac{1}{\Lambda} \int\ud^4x\biggl\{ & -16\phi\partial_t\phi\partial_t\phi +4\sigma^{ij}\partial_{i}\phi\partial_{j}\phi                         \\
	   & + 2\sigma^{i}_{~i}\left(\partial_t\phi\partial_t\phi -\partial_{j}\phi\partial^{j}\phi\right)\biggr\} \,,
\end{align}
These give the $ϕ³$ and $ϕ² σ$ interactions shown in \cref{fig:3-point-vertex-phi3,fig:3-point-vertex-phi2-sigma}:

\begin{itemize}[leftmargin=*]
	\item \textbf{\boldmath$\phi^3$--cubic interaction:}
	      \begin{align} \nn \label{GRcubvertf3}
		      \textrm{\cref{fig:3-point-vertex-phi3}}
		      \propto & \frac{ω_1 ω_2 + ω_2 ω_3 + ω_3 ω_1}{\Lambda \boldsymbol{k}_1^2\boldsymbol{k}_2^2\boldsymbol{k}_3^2} \\ \nn
		              & \times \sumk δ(t_1-t_2) δ(t_1-t_3)                                                                 \\
		              & \sim \frac{v² r^4}{Λ} \frac{r^3}{t²}\,,
	      \end{align}
	\item \textbf{\boldmath$\sigma\phi^2$--cubic interaction:} %
	      \begin{align}\label{GRcubvertf2s}\nn
		      \textrm{\cref{fig:3-point-vertex-phi2-sigma}}
		      \propto & \frac{2\boldsymbol{k}_1^{(i}\boldsymbol{k}_2^{j)}-\left( \boldsymbol{k}_1 \cdot \boldsymbol{k}_2 - ω_1 ω_2\right)\delta^{ij}}{\Lambda \boldsymbol{k}_1^2\boldsymbol{k}_2^2\boldsymbol{k}_3^2} \\ \nn
		              & \times \sumk δ(t_1-t_2) δ(t_1-t_3)                                                                                                                                                            \\
		              & \sim \frac{r^4}{Λ} \frac{r^3}{t²}\,,
	      \end{align}
\end{itemize}
where $\boldsymbol{δ}\coloneqq δ^{(3)}$ represents the three dimensional Dirac delta distribution. Note that these interaction vertices include the propagators of the respective fields.

Notice that the $\phi^3$--vertex is subdominant in GR because the potential off-shell modes have $ω\ll |\boldsymbol{k}|$.
Finally, the point-particle action of \cref{eq:Spp} expanded to second order in the fields is
\begin{align}\label{eq:KS}\nn
	S^{(\rm quad)}_{\rm pp} = & -\sum_{a}m_a\int\ud t\biggl\{ 1 -\frac{1}{2}v_a^2 -\frac{1}{8}v_a^4                                                                           \\\nn
	                          & +\frac{1}{\Lambda}\biggl[\bigl(1+\frac{3}{2}v_a^2\bigr)\phi - \bigl(1+\frac{1}{2}v_a^2\bigr)A_iv_a^i -\frac{1}{2}v_a^iv_a^j\sigma_{ij}\biggr] \\
	                          & +\frac{1}{\Lambda^2}\biggl[\frac{1}{2}\bigl(1-\frac{9}{2}v_a^2\bigr)\phi^2 - \phi A_iv_a^i \biggr]
	\biggr\} \,.
\end{align}
This gives the coupling (matter vertices) between the worldline of the point particle and the gravitational fields, generically represented by \cref{fig:scalar-matter-coupling,fig:scalar-matter-quadratic-coupling}:
\begin{itemize}[leftmargin=*]
	\item \textbf{\boldmath$\phi$–matter coupling:}
	      \begin{equation}\label{eq:scalar-matter-coupling}
		      \propto -i \frac{m_a}{\Lambda}\int\ud t \intk \eikx
		      \sim \frac{t}{r³} \frac{m}{\Lambda}\,,
	      \end{equation}
	\item \textbf{\boldmath$A_i$–matter coupling:}
	      \begin{equation}
		      \propto i \frac{m_a}{\Lambda}\int\ud t \intk \eikx v_a^i
		      \sim \frac{t}{r³} \frac{m}{\Lambda}v\,,
	      \end{equation}
	\item \textbf{\boldmath$\sigma_{ij}$–matter coupling:}
	      \begin{equation}
		      \propto i \frac{m_a}{2\Lambda}\int\ud t \intk \eikx v_a^i\,v_a^j
		      \sim \frac{t}{r³} \frac{m}{\Lambda}v^2\,.
	      \end{equation}
\end{itemize}
Recall that the factor of $1/r³$ comes from the integral $\int_{\boldsymbol{k}} \coloneqq \dd^3 \boldsymbol{k}/\left(2π\right)^3$. The $A_i$ –matter and $σ_{ij}$ –matter vertices contain factors of $v$, and are therefore subleading compared to the $ϕ$–matter vertex.

To consider finite-size effects, on should complement the point-particle action in \cref{eq:Spp}, by the following tidal (quadrupolar) action,
\begin{equation}\label{eq:internaleffects}
	{S}_{\rm tid } =\sum_{a=1,2}\int\ud\tau_a \left[c^{(a)}_E E_{ab} E^{ab} +c^{(a)}_B B_{ab} B^{ab} \right]\,,
\end{equation}
where $c^{(a)}_E$ and $c^{(a)}_B$ are the coupling constant related to the electric and magnetic tidal Love numbers of body $a$, $E_{ab} \coloneqq R_{acbd}u^{c} u^{d}$ and $2 B_{ab} \coloneqq \epsilon_{acef} R^{ce}_{~~~bg}u^{g} u^{f}$. We recall that for non-spinning black holes in 4D GR, $c_E=c_B=0$~\cite{Damour:2009vw,Binnington:2009bb,Kol:2011vg}. Using the KS parameterization, we get that in the static limit
\begin{align} \label{eq:Stidal}
	S^{(\rm static)}_{\rm tid} = \sum_{a=1,2} \frac{c_E^{(a)}}{Λ²} \int\ud t\;
	\pa{\partial_a\partial_b\phi}\pa{\partial^a\partial^b\phi}\,.
\end{align}
This gives the matter tidal vertex shown in \cref{fig:scalar-matter-tidal-coupling}
\begin{align}\label{eq:ce-vertex}
	 & =
	2i\frac{c^{(a)}_{E}}{Λ²}\int\ud t \int_{\boldsymbol{k}_1,\boldsymbol{k}_2}\hspace{-10pt} \exp[i(\boldsymbol{k}_1+\boldsymbol{k}_2)\!\cdot\!\boldsymbol{x}] \left( \boldsymbol{k}_1 \cdot \boldsymbol{k}_2 \right)² \\ \nn
	 & \sim \frac{t}{r^6}\frac{c^{(a)}_{E}}{Λ² r^4}\,,
\end{align}
where, again, the integral $\int_{\boldsymbol{k}} \coloneqq \dd^3 \boldsymbol{k}/\left(2π\right)^3$ gives a factor of $\boldsymbol{k}^6 \sim 1/r^6$.

We are also interested in determining the radiative dynamics—the gravitational flux and the waveform. To do so, we split the metric perturbation into potential and radiative modes, $h^{\mu\nu} =  H^{\mu\nu}+ \overline{h}^{\mu\nu}$. Differently from potential modes ($H^{μν}$), radiative ones, $\overline{h}^{\mu\nu}$, evolve on scales set by the gravitational waves, whose wavelength $\lambda_{\textrm{\scshape gw}}\sim r/v$, and carry on-shell four-momenta, $k = \left(ω,\,\boldsymbol{k}\right) \sim \left( v/r, v/r \right)$.

The radiation effective action is obtained by integrating out the potential modes while keeping one external radiative mode, propagating on-shell, in order to compute the one-graviton emission amplitude. The resulting effective action couples linearly source to radiation
\begin{equation}\label{eq:Seffrad}
	S_{\rm eff} = \frac{1}{2\Lambda}\int\ud^4x T^{ab}\overline{h}_{ab} \,,
\end{equation}
where $T^{ab} \coloneqq \left( 2/\sqrt{-g} \right) \left(\delta S_{\rm tot}/\delta g_{ab}\right)$ is the total stress-energy pseudo-tensor, which includes contribution both from matter and non-linear gravity terms. The latter can come from GR but could also have their origin in the higher curvature terms that we will introduce in the next section. For long wavelength radiation $\lambda_\mysc{gw}\gg r$, it is useful to Taylor
expand the radiative field (taking the center of mass of the binary to be at rest at the origin $x^i_{\textrm{cm}} = \dot{x}^i_{\textrm{cm}} = 0$), whose derivatives $\overline h_{\mu\nu,i_1\dots i_l}$ couple to source moments via $\sum_{l=0}^{+\infty}\frac{1}{l!} T^{ab,i_1\dots i_l}\overline h_{ab,i_1\ldots i_l}$, with
\begin{align}
	T^{ab,i_1\dots i_l}(t) \coloneqq \int\ud^3x T^{ab}(t,\boldsymbol{x})x^{i_1}\cdots x^{i_l}\,,
\end{align}
the small parameter of this expansion being the source internal velocity $v$, with a factor $v^2$ needed to hop from one PN order to the next.

\begin{figure*}
	\centering
	\begin{subfigure}{0.1\linewidth}
		{
			\begin{tikzpicture}[thick,baseline=(current bounding box.center)]
				\draw [black] (0,0.05) -- (2,0.05);
				\draw [black] (0,0) -- (2,0);
				\filldraw (0.95,-0.025) rectangle ++(3pt,3pt);
				\draw [black, decorate, decoration={snake, amplitude=0.5mm, segment length=2mm}, shorten >=1pt] (1,0.025) -- (1,1);
				\node[align=left] at (1.35,0.5) {$σ_{ij}$};
			\end{tikzpicture}
		}
	\end{subfigure}
	\begin{subfigure}{0.07\linewidth}{$=$} \end{subfigure}
	\begin{subfigure}{0.1\linewidth}
		{
			\begin{tikzpicture}[thick,baseline=(current bounding box.center)]
				\draw [fill=lightgray,lightgray] (0.5,0) rectangle (1.5,1);
				\draw [black] (0,1) -- (2,1);
				\draw [black] (0,0) -- (2,0);
				\draw [black, decorate, decoration={snake, amplitude=0.5mm, segment length=2mm}, shorten >=1pt] (1,1) -- (1.7,1.7);
				\node[align=left] at (1.1,1.5) {$σ_{ij}$};
			\end{tikzpicture}
		}
	\end{subfigure}
	\begin{subfigure}{0.05\linewidth}{$=$} \end{subfigure}
	\begin{subfigure}{0.1\linewidth}
		{
			\begin{tikzpicture}[thick,baseline=(current bounding box.center)]
				\draw [black] (0,1) -- (2,1);
				\draw [black] (0,0) -- (2,0);
				\draw [black, decorate, decoration={snake, amplitude=0.5mm, segment length=2mm}, shorten >=1pt] (1,1) -- (1.7,1.7);
				\node[align=left] at (1.1,1.5) {$σ_{ij}$};
			\end{tikzpicture}
		}
	\end{subfigure}
	\begin{subfigure}{0.05\linewidth}{$+$} \end{subfigure}
	\begin{subfigure}{0.1\linewidth}
		{
			\begin{tikzpicture}[thick,baseline=(current bounding box.center)]
				\draw [black] (0,1) -- (2,1);
				\draw [black] (0,0) -- (2,0);
				\draw [black, dashed] (1,0) -- (1,1);
				\draw [black, decorate, decoration={snake, amplitude=0.5mm, segment length=2mm}, shorten >=1pt] (1,0.5) -- (2,0.5);
				\node[align=left] at (0.8,0.25) {$ϕ$};
				\node[align=left] at (0.8,0.75) {$ϕ$};
				\node[align=left] at (1.4,0.7) {$σ_{ij}$};
			\end{tikzpicture}
		}
	\end{subfigure}
	\begin{subfigure}{0.1\linewidth}{$+\quad \cdots$} \end{subfigure}
	\caption{Leading order coupling of radiative (wavy line) mode to a binary system of massive particles (solid straight lines) mutually interacting via  potential modes (dashed lines).
		\label{fig:Tmunu}}
\end{figure*}
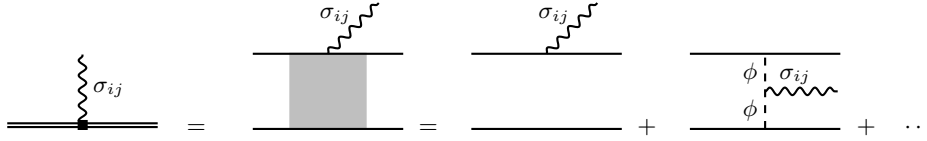

To determine the multipoles that directly enter the computation of the flux and gravitational wave modes, one needs to match it to the source quantities. One can straightforwardly determine the matching at lowest order, for instance, by computing the diagrams in Fig.~\ref{fig:Tmunu}. Working in the transverse-traceless gauge (TT), only space-space components of the energy momentum are radiative sources, and using energy momentum tensor conservation in the form
\begin{align}
	T^{\mu\nu}_{\ \ ;\nu}=0\,,
\end{align}
one can trade $T^{ij}$ and its momenta for momenta of (derivatives of) $T^{00}$. For instance in linearised GR, integrating by parts, it is straightforward to show that
\begin{eqnarray}
	\ba{lcl}
	\ds\int{\rm d}^3x\paq{\dot T^{00} x^i-T^{0i}}&=&0\,,\\
	\ds\int{\rm d}^3x\paq{\dot T^{00} x^ix^j-2T^{0(i}x^{j)}}&=&0\,,\\
	\ds\int{\rm d}^3x\paq{\dot T^{0(i}x^{j)}-T^{ij}}&=&0\,,
	\ea
\end{eqnarray}
at the lowest orders in multipole moment expansion (an overdot stands for a time derivative and an index pair in round brackets stands for symmetrization).

At higher orders one must decompose the multipole moments in irreducible representation of rotations, labeled by the integer $l$. The leading order quadrupole source $Q^{ij}\coloneqq \int {\rm d}^3xT^{00}x^ix^j$, $\ddot Q^{ij}=2\int {\rm d}^3x T^{ij}$, ($l=2$) receives the lowest order (1PN) corrections from higher multipole from the $l=2$ part of $T^{ij}x^kx^l$ and $T^{0i}x^jx^k$, which give the standard PN expanded expression \cite{Thorne:1980ru}
\begin{align}\label{eq:quad1PNGR}\nn
	I^{ij}_{|1\textrm{PN}} = & \int\ud^3\boldsymbol{x}\left(T^{00}_{|1PN}+T^{ll}_{|0PN}\right)x^{\langle ij \rangle}          \\\nn
	                         & -\frac{4}{3}\int\ud^3\boldsymbol{x}\left(\dot{T}^{0l}_{|0PN} x^l \right)x^{\langle ij \rangle} \\
	                         & +\frac{11}{42}\int\ud^3\boldsymbol{x}\ddot{T}^{00}_{|0PN}\boldsymbol{x}^2x^{\langle ij \rangle}
\end{align}
where $\langle ij \rangle $ refers to the symmetric trace-free (STF) combination,  $x_a^{\langle ij \rangle}= x_a^{(i}x_a^{j)}-(1/3)\delta^{ij}\boldsymbol{x}_a^2$.

Finally, the quadrupole moment is the key ingredient to compute the gravitational flux and waveform and to determine corrections from higher curvature terms in the emitted radiation. At leading order we have for the gravitational flux,
\begin{align}\label{eq:gwstrain}
	{\cal F_{\rm grav}} =\frac{G}{5c^5} \dddot{I}{\!}_{ij} \dddot{I}^{ij}\,,
\end{align}
and for the gravitational waveform,
\begin{align}\label{eq:waveform}
	\overline{h}_{ij}^{\mysc{tt}} = \frac{2 G}{c^4 r} {\cal P}_{ij}^{\phantom{ij}kl} \ddot{I}_{kl}\,,
\end{align}
where ${\cal P}_{ij}^{\phantom{ij}kl}=\left(P_i^{~k}P_{j}^{~l}+P_i^{~l}P_{j}{^k}-P_{ij}P^{kl}\right)/2$ is the tensor transverse-traceless projector, with $P_{ij} = \delta_{ij}-\boldsymbol{n}_i \boldsymbol{n}_j$ being the vector transverse projector, with $\boldsymbol{n}$ the unit vector of the radiation propagation direction.
In the rest of this work, we apply the techniques of PN-EFT to obtain the leading PN effects and scaling of observables in terms of the order of operators considered to account for potential deviations. This allows one to lay down simple rules that can be readily exploited in data analysis to not only connect observations to potential theories but also to distinguish them from systematic effects. To fix ideas, we begin discussing two particular examples that have already been explored in the literature but here we do so with the aim of addressing general theories. (For other examples of PN-EFT applied to specific beyond GR-theories see, e.g.~\cite{Endlich:2017tqa,Diedrichs:2023foj,Almeida:2024cqz,Lins:2020omt}.)

The first theory contemplates potential deviations through {\em cubic curvature corrections}—and no extra degree of freedom, and the second a {\em scalar-tensor theory} with an extra (scalar) degree of freedom coupled to a quadratic curvature correction. After discussing the relevant Feynman rules, we obtain corrections to the gravitational potential, source quadrupole and radiation from the system.

\section{Cubic gravity}\label{sec:cubicgravity}

To start we consider corrections that are of order cubic in curvature without the addition of extra degrees of freedom. To make contact with different literature, e.g.~\cite{Bueno:2016xff,Cano:2019ore,Emond:2019crr,Brandhuber:2019qpg}, we consider three terms that are: the only two invariants constructed in terms of Riemmann tensors (and its dual) $\{ {\cal R}^{(3)},{\cal {}^*R}^{(3)}\}$, together with a (redundant in vacuum\footnote{This is however expressible in terms of the first one together with further contributions at least linear in Ricci tensors (see \cref{app:cubicapp}).}) third one considered in the so-called  Einstein cubic gravity $\{{\cal \widetilde{R}}^{(3)}\}$,
\begin{align}\label{eq:cubic-action}\nn
	S_\textrm{bulk} & = 2Λ²\int\ud^{4}x\sqrt{-g}   \\
	                & \qquad \times \left[R + \alpha_{(3)}{\cal R}^{(3)} + \beta_{(3)}{\cal \widetilde{R}}^{(3)} + \gamma_{(3)}{\cal {}^*R}^{(3)}\right] \,,
\end{align}
where $Λ$ is the reduced Planck mass and
\begin{align}\nn
	{\cal R}^{(3)}             & \coloneqq
	R_{abcd}\,R^{cdef}\,R_{ef}^{~~~ab}\,,                                                                  \\ \nn
	{\cal \widetilde{R}}^{(3)} & \coloneqq
	R_{abcd}\,R^{bgde}\,R_{g~e}^{~a~c},                                                                    \\ \nn
	{\cal {}^*R}^{(3)}         & \coloneqq \epsilon_{abcd}\,R^{cd}_{\ph{ef}ef}\,R^{efgh}\,R_{gh}^{~~ab}\,.
\end{align}
Notice that
$\alpha_{(3)}$, $\beta_{(3)}$ and $\gamma_{(3)}$ all carry dimensions of $[{\rm length}]^4$; we thus define associated dimensionless coupling constants as,
\begin{equation}\label{eq:dimless-coupling-cubic}
	\overline{X}_{(3)} \coloneqq \left(\frac{c^2}{G{M_Λ}}\right)^{4}\,X_{(3)} \coloneqq \frac{X_{(3)}}{\ell^4}\,,
\end{equation}
where $X_{(3)} \in \{α_{(3)}, β_{(3)},γ_{(3)}\}$, and $M_Λ$ is the mass scale at which this new physics appears.

\subsection{Feynman rules}

As the EH and matter parts of the action are not modified by the introduction of cubic gravity corrections, the propagators and matter coupling presented in \cref{sec:PNEFT} are still valid.
The first correction will come from the bulk vertices, that are obtained by expanding the bulk action to cubic or higher orders in the gravitational fields. In particular, we have at leading order for the first two terms:
\begin{align}\label{eq:S-cubic}\nn
	S^{(3)}_{\rm cub} & \supset \frac{1}{\Lambda} \int\ud^4x \left[ 12\left(4\alpha_{(3)}+\beta_{(3)}\right)∇²ϕ\,\partial_{cd}\phi\partial^{cd}\phi \right.           \\
	                  & \left. + 4\beta_{(3)}(∇²ϕ)^3  -16\left(2\alpha_{(3)}+\beta_{(3)}\right)\partial_{b}\partial^{c}\phi\partial_{ac}\phi\partial^{ab}\phi \right]
	\,,
\end{align}
Hence, in addition to the GR cubic vertices in \cref{GRcubvertf3,GRcubvertf2s}, the cubic gravity terms give additional vertices that scale at least proportionally to $\alpha_{(3)}$, $\beta_{(3)}$ or $\gamma_{(3)}$:\footnote{the vertex carries a factor of $-i$, and the three propagators each bring $-i$ to give an overall factor of $(-i)^4 = 1$.}
\begin{itemize}[leftmargin=*]
	\item \textbf{\boldmath$\alpha_{(3)}$--cubic vertex:} the leading order (LO) contribution comes from $\left(\partial^2\phi\right)^3$, which gives the following Feynman rule (including the three propagators)
	      \begin{align}\nn
		       & =\frac{3}{16 Λ\, \boldsymbol{k}_1^2\boldsymbol{k}_2^2\boldsymbol{k}_3^2} \left[ \boldsymbol{k}_1² (\boldsymbol{k}_2 \!\cdot\! \boldsymbol{k}_3 )^2
		      \!+\! \boldsymbol{k}_3² (\boldsymbol{k}_1 \!\cdot\! \boldsymbol{k}_2 )^2 \!+\! \boldsymbol{k}_2² (\boldsymbol{k}_3 \!\cdot\! \boldsymbol{k}_1 )^2  \right.                 \\ \nn
		       & \quad \left.  - 2 (\boldsymbol{k}_1 \!\cdot\! \boldsymbol{k}_2)(\boldsymbol{k}_1 \!\cdot\! \boldsymbol{k}_3)(\boldsymbol{k}_2 \!\cdot\! \boldsymbol{k}_3) \right] \sumk \\ \nn
		       & \quad \times δ(t_1-t_2) δ(t_1-t_3) \, α_{(3)}                                                                                                                           \\ \label{eq:cubic-alpha-rule}
		       & \sim \frac{α_{(3)}}{\Lambda}\frac{r^{3}}{t²}\,,
	      \end{align}
	\item \textbf{\boldmath$\beta_{(3)}$--cubic vertex:} the LO contribution comes from $\left(\partial^2\phi\right)^3$, and gives the following Feynman rule
	      \begin{align}\nn
		       & =\frac{3}{64 Λ\, \boldsymbol{k}_1^2\boldsymbol{k}_2^2\boldsymbol{k}_3^2} \left[ \boldsymbol{k}_1² (\boldsymbol{k}_2 \!\cdot\! \boldsymbol{k}_3 )^2
		      \!+\! \boldsymbol{k}_3² (\boldsymbol{k}_1 \!\cdot\! \boldsymbol{k}_2 )^2 \!+\! \boldsymbol{k}_2² (\boldsymbol{k}_3 \!\cdot\! \boldsymbol{k}_1 )^2 \right.   \\ \nn
		       & \quad \left.   - 4 (\boldsymbol{k}_1 \!\cdot\! \boldsymbol{k}_2)(\boldsymbol{k}_1 \!\cdot\! \boldsymbol{k}_3)(\boldsymbol{k}_2 \!\cdot\! \boldsymbol{k}_3) + \boldsymbol{k}_1²\boldsymbol{k}_2²\boldsymbol{k}_3² \right] \\ \nn
		       & \quad \times \sumk δ(t_1-t_2) δ(t_1-t_3) \, \beta_{(3)}  \\ \label{eq:cubic-beta-rule}
		       & \sim \frac{β_{(3)}}{\Lambda}\frac{r^{3}}{t²}\,,
	      \end{align}
	\item \textbf{\boldmath$\gamma_{(3)}$--cubic vertex:} due to the presence of the fully antisymmetric tensor $\epsilon^{\mu\nu\lambda\delta}$, the LO contribution has to include a contribution from $h^{0i}$ that will couple to a matter vertex containing a $v^i$ factor. Such a contribution corresponds to terms of the form $-4\epsilon_{0i}^{~\ph{i}jk}\pa{\partial_{km}A^{l}\,\partial^i\partial_l\phi\,\partial_{jm}\phi +4\partial^{il}\phi\,\partial_j\dot\phi\,\partial_{kl}\phi}$, which gives the following Feynman rule
	      \begin{align}\nn
		       & \propto \frac{\epsilon_{0ijk}}{8 \Lambda\, \boldsymbol{k}_1^2\boldsymbol{k}_2^2\boldsymbol{k}_3^2} \biggl(
		      2k_1^i k_2^jk_3^k(\boldsymbol{k}_1\cdot\boldsymbol{k}_3)(\boldsymbol{k}_2\cdot\boldsymbol{v}) +{\rm  perms.}\biggr) \\ \nn
		       & \qquad \times \sumk δ(t_1-t_2) δ(t_1-t_3)\,\gamma_{(3)}  \\
		       & \sim v\frac{γ_{(3)}}{\Lambda}\frac{r^{3}}{t²}\,.
		      \label{eq:par_break}
	      \end{align}
\end{itemize}
As before, all of the above vertices contain the propagators of the associated scalar legs.

\subsection{Correction to the gravitational binding potential}\label{sec:cubic-potential}

The binding potential is obtained by integrating out the potential modes, which are off-shell. Recall for these modes, we have the scaling rules, $ k = \left(k_0,\,\boldsymbol{k}\right) \sim \left(v/r,1/r\right) $.

\begin{figure}
	\centering
	\begin{subfigure}{0.45\linewidth}
		{
			\begin{tikzpicture}[thick]
				\draw [black] (0,2) -- (2,2);
				\draw [black, dashed] (1,2) -- (1,1.45);
				\draw [black, dashed] (1,1.45) -- (.5,0.7);
				\draw [black, dashed] (1,1.45) -- (1.5,0.7);
				\draw [black] (0,0.7) -- (2,0.7);
				\node[align=left] at (0.4,1) {$ϕ$};
				\node[align=left] at (1.6,1) {$ϕ$};
				\node[align=left] at (0.8,1.7) {$ϕ$};
			\end{tikzpicture}
		}
		\caption{
			\label{fig:binding-potential-3-vertex}}
	\end{subfigure}
	\begin{subfigure}{0.45\linewidth}
		{
			\begin{tikzpicture}[thick]
				\draw [black] (0,2) -- (2,2);
				\filldraw (1,2) circle (2pt);
				\draw [black, dashed] (1,2) -- (.5,0.7);
				\draw [black, dashed] (1,2) -- (1.5,0.7);
				\draw [black] (0,0.7) -- (2,0.7);
				\node[align=left] at (0.5,1.3) {$ϕ$};
				\node[align=left] at (1.5,1.3) {$ϕ$};
			\end{tikzpicture}
		}
		\caption{
			\label{fig:binding-potential-3-vertex-tidal}}
	\end{subfigure}
	\caption{Leading correction to the potential energy in (a) a cubic theory of gravity, and (b) from finite size effects.
		\label{fig:potential-cubic}
	}
\end{figure}
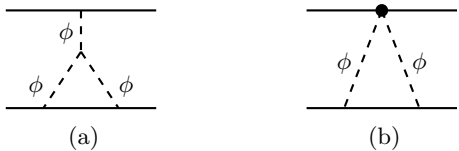

From this and using the Feynman rules, we can compute the leading-order correction due to the cubic gravity term in the action. The leading contribution comes from graviton exchange between the binaries via a cubic interaction in the bulk shown in \cref{fig:binding-potential-3-vertex}. We can now compute it using the building blocks of this diagram i.e. \cref{fig:3-point-vertex-phi3,fig:scalar-matter-coupling}. Using the Feynman rules derived in \cref{eq:cubic-alpha-rule,eq:cubic-beta-rule,eq:scalar-matter-coupling} we obtain the leading order correction shown in \cref{fig:binding-potential-3-vertex}:\footnote{Using the Feynman rules gives $i\, S_\textrm{eff} = i \int \dd t \mathcal{L}_\textrm{eff}$, from which we extract $V = - \mathcal{L}_\textrm{eff}$ presented above. We will do the same for all subsequent computations.\label{footnote:seff}}
\begin{align} \label{eq:V-beta} \nn
	V_{(∂²ϕ)³} & = -\frac{3}{32}\frac{m_1m_2(m_1+m_2)}{Λ^4} \int_{\boldsymbol{k,q}} \frac{\exp[i\boldsymbol{k}\!\cdot\!\boldsymbol{r}]}
	{{\boldsymbol{k}²\boldsymbol{q}²\left(\boldsymbol{k}\!+\!\boldsymbol{q}\right)²}} \\ \nn
	           & \times \left[\pa{α_{(3)}+\frac{\beta_{(3)}}4}\boldsymbol{k}^2\pa{\pa{\boldsymbol{k} \!+\! \boldsymbol{q}} \!\cdot\! \boldsymbol{q}}^2\right. \\ \nn
	           & \quad\left.-\pa{2\alpha_{(3)}+\beta_{(3)}}(\boldsymbol{k} \!\cdot\! \boldsymbol{q})(\boldsymbol{k} \!\cdot\! (\boldsymbol{k} \!+\! \boldsymbol{q}))(\boldsymbol{q} \!\cdot\! ({\boldsymbol{k} \!+\! \boldsymbol{q}}))\right] \\
	           & = \frac{9}{1024}\frac{m_1m_2(m_1+m_2)}{\pi^2\Lambda^4r^6} \beta_{(3)}\,,
\end{align}

where $\boldsymbol{r} \equiv (\boldsymbol{x}-\boldsymbol{y})$. Again, remember that the interaction vertices in \cref{eq:cubic-alpha-rule,eq:cubic-beta-rule} already include the propagators.
The integrals are evaluated using standard master integrals, and dimensional regularization has been used to treat the infinities (see e.g. appendix B of \cite{Porto:2016pyg}).
The $α_{(3)}$ integral vanishes, indicating that it contributes at higher PN order. This is due to the specific tensor structure of this term. Generically, if we consider every possible contraction of Riemann tensors to get the most general cubic gravity theory, \cref{fig:binding-potential-3-vertex} will give the leading order contribution to the potential from one of the terms (which happens to be $β_{(3)}$ in this case).
For a general theory, it is natural to expect that a simple scaling argument suffices to extract the leading order contribution. This is where the power of the scalings shown in the Feynman rules comes into play. Using the scalings derived in \cref{eq:cubic-beta-rule,eq:scalar-matter-coupling}, we show this for the cubic gravity case:
\begin{align}\label{eq:cubic-scaling}\nn
	V_{(\partial^2 \phi)^3}
	 & \sim \frac{1}{t}\left(\frac{t}{r³}\right)³\left(\frac{m_1}{\Lambda}\right)\left(\frac{m_2}{\Lambda}\right)^{2}  \!\cdot\! \frac{\beta_{(3)}}{\Lambda} \frac{r³}{t²}
	+ \left(1 \! \leftrightarrow \! 2\right) \\
	 & \sim \frac{m_1 m_2 (m_1 + m_2)}{\Lambda^4\,r^{6}}\,\beta_{(3)}\,.
\end{align}
The factor of $1/t$ removes the overall $\int \dd t$ in the effective action to give the potential as explained in \cref{footnote:seff}.
Writing this in terms of the dimensionless coupling $\overline{β}_{(3)}$ defined in \cref{eq:dimless-coupling-cubic}, it is now straightforward to see that the correction to the Newtonian binding potential is of order 5PN, although it formally appears as a 1PN correction in the binding potential.
Explicitly,
\begin{align}\nn
	V_{(\partial^2 \phi)^3} & \sim \frac{Gm_1 m_2 }{r}\,\frac{G\left(m_1 + m_2\right)}{r}\,\left(\frac{G M_{\Lambda}}{r}\right)^4\,\overline{\beta}_{(3)}  \\
	& \sim \frac{Gm_1 m_2 }{r} \cdot \frac{Gm}{r} \cdot \left(\frac{Gm}{r}\right)^4 \left(\frac{M_{\Lambda}}{m}\right)^4\,\overline{\beta}_{(3)}\,,
\end{align}
where we have also reintroduced the gravitational constant $G$, and defined the total mass $m \coloneqq m_1+m_2$. Using the virial theorem $Gm/r \sim v²$, we can now identify the PN order of the various pieces.
The first factor on the second line is the Newtonian potential, the second one $\sim v²$ is the 1PN correction, and the third one is the cubic gravity correction, which $\sim v^8$, and adds a 4PN correction. With this, we finally see that the correction is order 5PN ($\sim v^{10}$)
\begin{equation}
	V_{(\partial^2 \phi)^3} \sim V_\mysc{n} \left(\frac{M_{\Lambda}}{m}\right)^4\cdot \left(\frac{v^2}{c^2}\right)^{5}\cdot\overline{\beta}_{(3)}\,.
\end{equation}

All other diagrams coming from the cubic gravity term will include either higher order vertices or several cubic vertices and thus will be subleading. For example, the LO contribution to the binding potential coming from the $\alpha_{(3)}$ term will be at 6PN order.

However, we must add to this contribution the one coming from finite-size contributions, represented in the diagrams in \cref{fig:binding-potential-3-vertex-tidal}, where the black dot corresponds to tidal coupling.\footnote{This contribution is often omitted in explicit computations of the potential or radiative effects, as it depends on the (often unspecified) Wilson coefficients $c_E, c_B$.  However, such terms must eventually be included for completeness, which is seldom emphasized in practice, sometimes leading to partial results being mistakenly treated as final. See, for example, the discussion in \cref{app:nullRK}.}
The tidal contribution to the potential can be computed using \cref{eq:scalar-matter-coupling,eq:phi-prop,eq:ce-vertex}
\be\label{eq:V-cubic-tidal}
V_{\rm tidal} = - c_E^{(1)}\frac{3m_2^2}{512\pi^2\Lambda^4 r^6} + \left(1 \leftrightarrow 2\right)\,,
\ee
where $c_E^{(1)}$ is the tidal deformability parameter entering the action is \cref{eq:internaleffects}. We see that such a contribution scales exactly as the previous one in \cref{eq:V-beta} and cannot be distinguished from it. More precisely, there will be cancellation between the two terms, for $c_E^{(i)}=(3/2)m_i\beta_{(3)}$,
which corresponds exactly to the value computed in \cref{app:nullRK}.\footnote{There we compute the contribution of the $R R^{abcd}R_{abcd} \equiv RK$ term, where $K \coloneqq R^{abcd}R_{abcd}$ is the Kretschmann scalar. Given that $\widetilde{R}^{(3)} \supset -(3/8) RK$, we should have the relation $c_E\big|_{\widetilde{R}^{(3)}} = -(3/8) c_E\big|_{RK}$. Using \cref{eq:ce-RK}, this gives $c_E\big|_{\widetilde{R}^{(3)}} = -(3/8) (-4mλ) = (3/2) m λ = (3/2) m_1 β_{(3)}$.}
Incidentally, this cancellation was recently argued from a field-redefinition point of view in~\cite{AccettulliHuber:2020dal,Wilson-Gerow:2025xhr}.

For the term containing the dual Riemann tensor, proportional to $\gamma_{(3)}$, one of the matter coupling will contain a factor of velocity, adding a $0.5$PN order, e.g.:
\begin{align}\nn
	V_{\partial^2 A^{i}(\partial^2 \phi)^2}
	 & \sim  \frac{1}{t}\left(\frac{t}{r³}\right)³\left(\frac{m_1 v_2}{\Lambda}\right)\left(\frac{m_2}{\Lambda}\right)^{2}  \!\cdot\! \frac{γ_{(3)}}{\Lambda} \frac{r³}{t²} \\ \nn
	 & \quad + \left(1 \leftrightarrow 2\right)    \\\nn
	 & \sim \frac{m_1 m_2 (m_1 v_2 + m_2 v_1)}{\Lambda^4\, r^6}\,γ_{(3)}   \\
	 & \sim \frac{G\,m_1 m_2 }{r}\,\frac{G\,\left(m_1 v_2 +m_2 v_1 \right) }{r^{5}}\, γ_{(3)}\,.
\end{align}

\subsection{Correction to the gravitational radiative modes}\label{sec:correction-rad}

\begin{figure}
	\begin{tikzpicture}[thick,baseline=(current bounding box.center)]
		\draw [black] (0,1) -- (2,1);
		\draw [black] (0,0) -- (2,0);
		\draw [black, decorate, decoration={snake, amplitude=0.5mm, segment length=2mm}, shorten >=1pt] (1,1) -- (1.7,1.7);
		\filldraw (1,1) circle (2pt);
		\draw [black, dashed] (1,0) -- (1,1);
		\node[align=left] at (0.8,0.5) {$ϕ$};
		\node[align=left] at (1.1,1.5) {$σ_{ij}$};
	\end{tikzpicture}
	\caption{Radiation from finite size effects (leading order).
		\label{fig:tidal-rad}
	}
\end{figure}
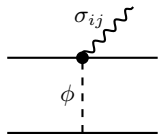

The gravitational radiative modes in Einstein cubic gravity are obtained by computing the correction to the radiative effective action in \cref{eq:Seffrad}. In practice, it consists in computing the last diagram in Fig.~\ref{fig:Tmunu} where a radiative graviton is emitted, with a modified interaction and the new diagram in Fig.~\ref{fig:tidal-rad}. The first diagram originates directly from the new cubic gravity interaction while the second one is a tidal contribution, that has to be included similarly to the binding potential calculation.

Expanding the action in \cref{eq:cubic-action}, the LO trilinear coupling involving one radiative field comes from $\beta_{(3)}{\cal \widetilde{R}}^{(3)}$ (using $\Box\sigma_{ij}=\sigma_{ij,j}=\sigma=0$):\footnote{An aside about notation: as mentioned below \cref{eq:ce-vertex}, we denote the radiative modes with an overline: $\overline{h}_{ij}$. In the KS parameterization, this corresponds to $\left( \overline{ϕ}, \overline{A}_i, \overline{σ}_{ij} \right)$. For brevity, however, we will omit the overlines when referring to radiation modes in the KS parameterization. It will be clear from the context when such modes are being discussed.}
\begin{equation}\label{eq:trilinear-coupling}
	S^{(3)}_{\phi^2\sigma} = \frac{12\beta_{(3)}}{\Lambda} \int\ud^4x \left[ 2\partial_{ik}ϕ\partial_{jk}\phi-
		\partial_{ij}\phi∇²ϕ\right]\ddot \sigma_{ij}\,,
\end{equation}
which adds to the standard GR vertex $ 4 ∂^i ϕ ∂^j ϕ σ_{ij}/Λ$ %
to give:
\begin{align}\nn \label{eq:Seff-cubic-rad}
	S_{\rm eff,\ rad}^{\rm (cub)}
	 & =\frac{1}{\Lambda}\int\ud t \,\biggl\{\pa{\frac{m_1}{2} v_1^i v_1^j-\frac{G m_1m_2}{4r}n^in^j}\sigma_{ij} \\
	 & \quad -\frac{9m_1m_2\beta_3}{64\pi \Lambda^2r^3}n^in^j\ddot{\sigma}_{ij}\biggr\}+1\leftrightarrow 2\,,    %
\end{align}
where $ \boldsymbol{r} \coloneqq \boldsymbol{x}_1 - \boldsymbol{x}_2, r \coloneqq |\boldsymbol{r}|$, and $\boldsymbol{n} \coloneqq \boldsymbol{r}/r$ is a unit vector. Following the discussion around \cref{eq:Seffrad}, this can be identified with
\begin{equation}
	S_{\rm eff,\ rad}^{\rm (cub)} = \frac{1}{2}\int\ud t\, T^{ij}\,\frac{\sigma_{ij}}{\Lambda}\,.
\end{equation}
This is shown in \cref{fig:Tmunu}.

The cubic gravity contribution modifies Newton's force law, which can be read off from the leading order equations of motion following from \cref{eq:V-beta}. In the center of mass frame ($m_1 \boldsymbol{x}_1 + m_2 \boldsymbol{x}_2 = 0)$, these are:
\begin{align}\label{eq:LO-eom-cubic}
	m_1 \ddot{\boldsymbol{x}}_1 & = -m_2 \ddot{\boldsymbol{x}}_2 = -\frac{∂}{∂ \boldsymbol{x}_1}\left(V_{\mysc{gr}} + V_{\textrm{cub}}\right)  \\ \nn
    & = -\frac{G m_1 m_2 \boldsymbol{n}}{r²} + 54 β_{(3)} \frac{G²m_1 m_2 \left( m_1 + m_2 \right)\boldsymbol{n}}{r^7}\,.
\end{align}
Using this, after some algebra, we can rewrite \cref{eq:Seff-cubic-rad} as:\footnote{See eq. (A2) of \cite{Lins:2020omt} for a derivation of \cref{eq:Seffradcub} in GR. In the center of mass frame, $\boldsymbol{x}_1 = m_2 \boldsymbol{r}/M, \boldsymbol{x}_2 = -m_1 \boldsymbol{r}/M $.}
\begin{align}
	\label{eq:Seffradcub}
	S_{\rm eff,\ rad}^{\rm (cub)} = & \frac{1}{\Lambda}\int\ud t\, \biggl\{\frac{\mu}{4}\frac{\ud^2}{\ud t^2}\paq{r^i r^j
	-36\beta_{(3)}\frac{GM}{r^3}n^in^j}\sigma_{ij} \\
	& \qquad\quad -27\beta_{(3)}\frac{G^2M^2\mu}{r^6}n^in^j\sigma_{ij} \ \biggr\}\nn\,,
\end{align}
where $M \coloneqq (m_1 + m_2)$ is the total mass, and $μ\coloneqq m_1 m_2/M$ is the reduced mass of the binary.
The first line corresponds to $\sim I^{ij} \ddot{σ}_{ij}$, from which we see that the mass quadrupole moment receives a leading order correction to its GR value of the form:
\begin{align}\label{eq:Iijcub}
	I_{\rm cub}^{ij}
	\sim & \sum_{a\neq b} m_a\,x_a^{\langle ij \rangle}\, \frac{G m_b}{r}\, \left(\frac{GM_Λ}{r}\right)^4 \overline{\beta}_{(3)}\,,
\end{align}
where $\langle ij \rangle $ refers to the STF combination defined below \cref{eq:quad1PNGR}.
Using \cref{eq:waveform}, this correction to the mass quadrupole gives a correction to the gravitational waveform which scales as
\begin{equation}\label{eq:δh-cubic}
	\delta h_{+,\times}^{\rm TT} \sim {h_{+,\times}^{\rm TT}}\big|_{\textrm {\scshape LO}} \cdot \left(\frac{v^2}{c^2}\right)^{5}\cdot\left(\frac{{M_{\Lambda}}}{m}\right)^{4}\,\overline{β}_{(3)} \,,
\end{equation}
where we have highlighted that it is a 5PN correction to the leading order GR waveform
The phase $\varphi$ is obtained by integrating the equation $\omega \coloneqq \ud\varphi/\ud t$ and using the quadrupole formula \cref{eq:gwstrain} to get $\ud\omega/\ud t = -{\cal F}/(\ud E/\ud \omega)$ (where $E$ is the total energy of the system). This scales the same way as the gravitational waveform.
\begin{equation}\label{eq:phase-cubic}
	\delta \varphi \sim \varphi |_{\textrm {\scshape n}} \cdot x^{5}\cdot\left(\frac{{M_{\Lambda}}}{m}\right)^{4}\,\overline{β}_{(3)} \,,
\end{equation}
where $x \coloneqq \left(G m \omega/c^3\right)^{2/3}$ is the PN parameter, $\varphi |_{\textrm {\scshape n}}=-x^{-5/2}/(32\, \nu)$ is the phase at Newtonian order, and
$ν \coloneqq μ/M$ is the dimensionless reduced mass.

Additionally, the term in the second line of \cref{eq:Seffradcub} arises because Newton's force law is modified as we saw in \cref{eq:LO-eom-cubic}. This term cannot be expressed in terms of the standard multipole moments, yet it still contributes to the gravitational flux and waveform. This term arises due to the non-conservation of the stress-energy tensor, or equivalently, from inserting the leading-order cubic gravity equations of motion, \cref{eq:LO-eom-cubic}, when deriving \cref{eq:Seffradcub}.

This behavior deviates from standard results in PN-EFT for GR and arises when extending the framework beyond GR (though it is sometimes overlooked). For earlier discussions, see~\cite{Cannella:2009he,Lins:2020omt}.

Although this additional contribution falls outside the traditional multipolar expansion, a simple dimensional analysis shows that it contributes to the flux and waveform at the same order as the quadrupole term. Therefore, the scaling of the waveform and phase in \cref{eq:δh-cubic,eq:phase-cubic} remains valid.

The leading-order contribution again comes from the $\beta^{(3)}$ term in the action, while the correction from the $\alpha^{(3)}$ term enters with an additional factor of $\propto v^2$, corresponding to a 1PN correction. From \cref{eq:Iijcub}, although the correction to the mass quadrupole formally appears at 1PN order, it effectively contributes at 5PN order due to an additional scaling of $\ell^4 / r^4$.

Note however that, similar to the binding potential, we also have to include the correction to the radiation coming from tidal effects (\cref{fig:tidal-rad}). Using \cref{eq:V-cubic-tidal} to write the modified equations of motion, and doing manipulations similar to \cref{eq:Seffradcub}, we get
\begin{align}\label{eq:Seff-rad-tid}
	S_{\rm eff,\ rad}^{\rm (tid)} = & \int\ud t\, \biggl\{\frac{1}{4}\frac{\ud^2}{\ud t^2}\biggl[\frac{12Gm_2\,c_E^{(1)}}{r^3}n^in^j\biggr]\sigma_{ij} \nn \\
	& ~~ +\frac{18G^2m_2^2\,c_E^{(1)}}{r^5}n^in^j\sigma_{ij}\biggr\} +(1\leftrightarrow 2)\,.
\end{align}
This correction contributes with the same scaling as the leading-order cubic term, making it indistinguishable at that level. Moreover, the value of the tidal parameter for which the two contributions cancel is the same as in the conservative dynamics: $c_E^{(1)} = (3/2)m_1 \beta_{(3)}$, as expected.

\section{Scalar-tensor gravity}\label{sec:esgb}

We now add a massless scalar field, $\Psi$, to the action. For simplicity here, we only consider a canonical kinetic term and a coupling to the metric field through the Gauss-Bonnet invariant ${G_2 \coloneqq R^2-4R_{ab}R^{ab} +R_{abcd}\,R^{abcd}}$. This is usually referred to as the Einstein-scalar-Gauss-Bonnet (EsGB) theory
\begin{equation}\label{eq:ESGB-action}
	\!\!\! S = 2Λ² \int\ud^4 x\sqrt{-g}\left[R -\frac{1}{4}\nabla_{a}\Psi\nabla^{a}\Psi + χ_{(2)}\,\Psi\,G_2 \right].
\end{equation}
The quadratic gravity coupling parameter $χ_{(2)}$ has now a dimension of $[{\rm length}]^2$ and we define the dimensionless coupling constant  $\overline{χ}_{(2)} \coloneqq  χ_{(2)}\left(G M_Λ/c²\right)^{-2} \coloneqq  χ_{(2)}/ℓ²\,.$

In the presence of a scalar field, the matter action~\eqref{eq:Spp} is generically modified to allow for a dependence of the mass in the scalar field (which, in particular, would capture potential scalarization of black holes, see e.g. \cite{Sotiriou:2013qea}),
\begin{align}\label{eq:Sppscal}
	S_\mysc{pp,s} = -\sum_{a=1,2}\int\ud\tau_a\,m_a\left(\Psi\right)\,.
\end{align}
Similarly, the action for gravitational finite-size effects in Eq.~\eqref{eq:internaleffects} will also contain a dependence of the deformability parameters with respect to the scalar field, i.e. $c_E^{(a)}(\Psi),c_B^{(a)}(\Psi)$. In addition, there will be a new type of finite-size effects, sourced by the scalar field. They are described by the following scalar tidal action,
\begin{equation}\label{eq:scalarinternaleffects}
	{S}_{\rm tid }^{\rm (scal)} =\sum_{a=1,2}\int\ud\tau_a \left[c^{(a)}_s (\Psi)\,\partial_{b}\Psi\partial^{b}Ψ \right]\,,
\end{equation}
where $c^{(a)}_s$ are the deformability parameters linked to scalar tides.
\subsection{Feynman rules}
Next, we expand the scalar field as $\Psi := 1 + ψ/Λ$, where $\psi$ is the perturbed scalar field. By inserting this decomposition into the Lagrangians, we get the new set of Feynman rules coming from the scalar field, while the ones already present in GR and exhibited above remain unchanged.
The propagator for this perturbed scalar field $ψ$ is (cf. \cref{eq:prop})
\begin{align}\nn
	\langle \psi(x) \psi(y) \rangle & =
	-\delta(t_1-t_2)\int_{\boldsymbol{k}}\frac{i\exp({i\boldsymbol{k}\cdot\left(\boldsymbol{x}_1-\boldsymbol{x}_2\right)})}{\boldsymbol{k}^2}\, \\
	                                & \sim \frac{t}{r} \,.
\end{align}
Next, to determine the matter vertices,
we perturbatively expand the mass with respect to the scalar field,
\begin{equation} \label{eq:mass_expansion}
	m_a\left(\Psi\right) = \overline{m}_a\sum_{n\geq0}\frac{d_a^{(n)}}{n!}\left(\frac{\psi}{\Lambda}\right)^n\,.
\end{equation}
In particular, $d_a^{(0)}=1$ and the coefficient $d_a^{(1)}$ correspond to the scalar charge. In this section, we consider only interactions that are linear in $ψ$, and therefore omit the expansion order for notational simplicity; that is, we take $d_a \coloneqq d_a^{(1)}$.\footnote{Higher order couplings of matter to the scalar field are subdominant compared to the leading linear coupling. Since we are only interested in the leading order correction, we neglect these terms in this work.}
Examining the scalar field equation, $\Box\Psi = -2\chi_{(2)}G_{2}$, we see that the scalar field, sourced by the Gauss-Bonnet term, is proportional to the coupling constant $\chi_{(2)}$. Consequently, the scalar charge inherits the same linear dependence, as discussed in e.g.~\cite{Julie:2019sab,Julie:2022huo,Julie:2023ncq}. We make this dependence explicit by defining $\overline{d}_a$ as,
\be\label{d_bar}
d_a \coloneqq  \left(M_{\Lambda}/\overline{m}_a\right)^{2} \overline{\chi}_{(2)}\overline{d}_a \,,
\ee
where $\overline{\chi}_{(2)}$ is the dimensionless coupling constant.~\footnote{The linear dependence is a consequence of the linear coupling to Gauss-Bonnet curvature scalar. Other couplings can lead to different scalings—for example a quadratic coupling leads to $\propto\sqrt{\chi_{(2)}}$; see e.g.\cite{Julie:2019sab,Julie:2022huo,Julie:2023ncq}.}

Then, by expanding the matter action in \cref{eq:Spp} and using \cref{eq:mass_expansion}, we get the scalar-matter couplings:
\begin{itemize}[leftmargin=*]
	\item \textbf{\boldmath $\psi$--matter coupling:}
	      \begin{equation}\nn
		      \propto -i\frac{\overline{m}_a}{\Lambda} \int\ud t \intk \eikx \left(d_a+\cdots\right) \sim \frac{t}{r³} \frac{m}{\Lambda}
	      \end{equation}
	\item \textbf{\boldmath $ψ$-$ϕ$--matter coupling:}
	      \begin{equation}\nn
		      \propto -i\frac{\overline{m}_a}{\Lambda^2}\int\ud t \intk \eikx \left(d_a+\cdots\right) \sim \frac{t}{r³} \frac{m}{\Lambda^2}
	      \end{equation}
\end{itemize}
We emphasize that each time a matter coupling involves a scalar field, it introduces a power of the coupling constant $ χ_{(2)}\propto\ell^2 $. At lowest order, such a coupling can be absorbed into a redefinition of the gravitational constant. However, this is no longer true at higher orders—for instance, starting at 1PN order there is a contribution proportional to the scalar charge (and thus to the coupling constant $χ_{(2)}$) that cannot be absorbed by redefining $G$~\cite{Bernard:2019yfz}.

The strength of the scalar tidal interaction, described in \cref{eq:scalarinternaleffects} gives the following Feynman rule.
\begin{align}\label{eq:cs-vertex}
	 & =
	-2i \frac{c^{(a)}_{s}}{Λ²} \int\ud t \int_{\boldsymbol{k}_1,\boldsymbol{k}_2}\hspace{-10pt} \exp[i(\boldsymbol{k}_1+\boldsymbol{k}_2)\!\cdot\!\boldsymbol{x}] \left( \boldsymbol{k}_1 \cdot \boldsymbol{k}_2 \right) \\ \nn
	 & \sim \frac{t}{r^6}\frac{c^{(a)}_{s}}{Λ²r²}\,,
\end{align}

Finally, we write the leading order cubic bulk vertex induced by the new cubic gravity term in the action. It originates from a term scaling as $\psi(\partial^2 \phi)^2$, which gives the vertex:
\begin{align}\nn
	 & = i\frac{\left(\boldsymbol{k}_2\cdot\boldsymbol{k}_3\right)^2-\boldsymbol{k}_2^2\,\boldsymbol{k}_3^2}{2Λ\boldsymbol{k}_1² \boldsymbol{k}_2² \boldsymbol{k}_3²} \\ \nn
	 & \quad \times \sumk\, δ(t_1-t_2)δ(t_1-t_3)  χ_{(2)}                                                                                                              \\\nn
	 & \sim \frac{t²}{Λ} r^5 χ_{(2)}\,.
\end{align}

\begin{figure*}
	\centering
	\begin{subfigure}{0.18\linewidth}
		{
			\begin{tikzpicture}[thick]
				\draw [black] (0,2) -- (2,2);
				\draw [black, double, decorate, decoration={snake, amplitude=0.5mm, segment length=2mm}] (1,2) -- (1,0.7);
				\draw [black] (0,0.7) -- (2,0.7);
				\node[align=left] at (0.7,1.3) {$ψ$};
			\end{tikzpicture}
		}
		\caption{
			\label{fig:ESGB-1}}
	\end{subfigure}
	\centering
	\begin{subfigure}{0.18\linewidth}
		{
			\begin{tikzpicture}[thick]
				\draw [black] (0,2) -- (2,2);
				\draw [black, double, decorate, decoration={snake, amplitude=0.5mm, segment length=2mm}] (1,2) -- (1,1.45);
				\draw [black, dashed] (1,1.45) -- (.5,0.7);
				\draw [black, dashed] (1,1.45) -- (1.5,0.7);
				\draw [black] (0,0.7) -- (2,0.7);
				\node[align=left] at (0.7,1.7) {$ψ$};
				\node[align=left] at (0.5,1.15) {$ϕ$};
				\node[align=left] at (1.5,1.15) {$ϕ$};
			\end{tikzpicture}
		}
		\caption{
			\label{fig:ESGB-2a}}
	\end{subfigure}
	\centering
	\begin{subfigure}{0.18\linewidth}
		{
			\begin{tikzpicture}[thick]
				\draw [black] (0,2) -- (2,2);
				\draw [black, dashed] (1,2) -- (1,1.45);
				\draw [black, double, decorate, decoration={snake, amplitude=0.5mm, segment length=2mm}] (1,1.45) -- (.5,0.7);
				\draw [black, dashed] (1,1.45) -- (1.5,0.7);
				\draw [black] (0,0.7) -- (2,0.7);
				\node[align=left] at (0.75,1.7) {$ϕ$};
				\node[align=left] at (0.45,1.15) {$ψ$};
				\node[align=left] at (1.5,1.15) {$ϕ$};
			\end{tikzpicture}
		}
		\caption{
			\label{fig:ESGB-2b}}
	\end{subfigure}
	\begin{subfigure}{0.18\linewidth}
		{
			\begin{tikzpicture}[thick]
				\draw [black] (0,2) -- (2,2);
				\draw [black, double, decorate, decoration={snake, amplitude=0.5mm, segment length=2mm}] (1,2) -- (.5,0.7);
				\draw [black, double, decorate, decoration={snake, amplitude=0.5mm, segment length=2mm}] (1,2) -- (1.5,0.7);
				\draw [black] (0,0.7) -- (2,0.7);
				\filldraw (1,2) circle (2pt);
				\node[align=left] at (0.45,1.3) {$ψ$};
				\node[align=left] at (1.55,1.3) {$ψ$};
			\end{tikzpicture}
		}
		\caption{
			\label{fig:ESGB-tidal}}
	\end{subfigure}
	\begin{subfigure}{0.18\linewidth}
		{
			\begin{tikzpicture}[thick]
				\draw [black] (0,2) -- (2,2);
				\draw [black] (0,0.7) -- (2,0.7);
				\draw [black, black, double, decorate, decoration={snake, amplitude=0.5mm, segment length=2mm}] (1,0.7) -- (1,1.3);
				\draw [black, dashed] (1,1.3) -- (1,2);
				\draw [black, decorate, decoration={snake, amplitude=0.5mm, segment length=2mm}, shorten >=1pt] (1,1.3) -- (2,1.3);
				\node[align=left] at (0.7,1) {$ψ$};
				\node[align=left] at (0.7,1.6) {$ϕ$};
				\node[align=left] at (1.6,1.5) {$σ_{ij}$};
			\end{tikzpicture}
		}
		\caption{
			\label{fig:ESGB-rad}}
	\end{subfigure}
	\caption{(a) First correction to Newtonian potential from EsGB. This simply renormalizes $G_\mysc{n}$,
		Leading order correction to (b,c) the potential energy from EsGB,
		(d) the potential from finite size effects, and
		(e) the mass quadrupole.
		\label{fig:ESGB}
	}
\end{figure*}
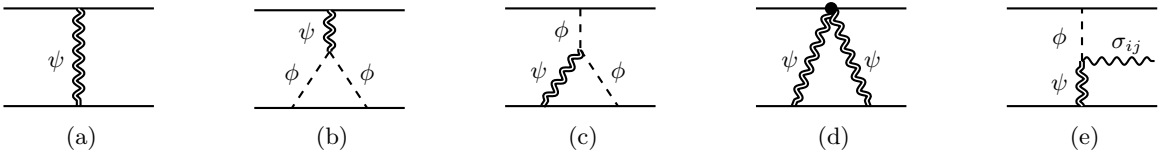

\subsection{Binding potential}

In the presence of a scalar field, the first correction comes in the Newtonian potential (shown in \cref{fig:ESGB-1}), and involves the exchange of a $\psi$--scalar propagator between the two particles (in addition to the $\phi$--scalar exchange):
\begin{align}\label{eq:VscalN}\nn
	V_{\psi, N} = & -\frac{ \overline{m}_1 \overline{m}_2 \,d_1 d_2} {4\pi\Lambda^2}\frac{1}{r}                              \\
	=             & -\frac{8G\,\overline{d}_1\overline{d}_2}{\overline{m}_1\overline{m}_2 r}\,\ell^4 \overline{χ}_{(2)}^2\,, %
\end{align}
where we have taken into account the scaling of the scalar charges with respect to the coupling constant to write the second line.
Such a correction can be absorbed in a renormalization of the Newton gravitational constant, $G_\mysc{n}=G\pa{1+8\,d_1 d_2}$, where as usual $G=(32\pi \Lambda^2)^{-1}$.

Leading order contribution to the binding potential comes from the three vertex involving $ψ(∂²ϕ)$. There are two diagrams depending on which particle the $ψ$ field couples to shown in \cref{fig:ESGB-2a,fig:ESGB-2b}. These contribute to the binding energy between two particles as
\begin{align}\nn
	\label{eq:Vpsiphi2}
	V_{\psi(\partial^2 \phi)^2} & =
	-\frac{\overline{m}_1 \overline{m}_2 χ_{(2)}}{Λ^4}\paq{\overline{M} (d_1+d_2) + \overline{m}_1 d_1 + \overline{m}_2 d_2} \\
	& \quad  \times  \int_{\boldsymbol{k},\boldsymbol{q}}
	\frac{\boldsymbol{k}^2\boldsymbol{q}^2 -(\boldsymbol{k}\cdot\boldsymbol{q})^2}{\boldsymbol{k}^2\boldsymbol{q}^2(\boldsymbol{k}+\boldsymbol{q})^2} e^{i\boldsymbol{k}\cdot(\boldsymbol{x}_1-\boldsymbol{x}_2)}
	\,,
\end{align}
where $\overline{M} \coloneqq \overline{m}_1 + \overline{m}_2$.
The Lagrangian expanded to third order also includes a three-point vertex involving only the $ϕ$ fields, coming from the Gauss-Bonnet term without any scalar field propagators. However, since the Gauss-Bonnet term is a total derivative in four dimensions, it does not contribute to the potential. In fact, the contribution of each term in $G_2$ is individually zero as shown in \cref{app:quadratic}.
Finally, the LO correction to the binding potential is
\begin{align}
	 & V_{\psi(\partial^2 \phi)^2} =
	- \frac{ χ_{(2)}}{r^4}\nn \, \frac{\overline{m}_1 \overline{m}_2}{128\pi^2\Lambda^4}
	\paq{\overline{M} (d_1+d_2) + \overline{m}_1 d_1 + \overline{m}_2 d_2}            \\ \nn
	 & \quad = - 8 \frac{G \overline{m}_1 \overline{m}_2}{r} \frac{G \overline{M}}{r}
	\paq{ d_1 + d_2 + \frac{{\overline m}_1 d_1 + {\overline m}_2 d_2}{\overline{M}}} \frac{ \ell^2  \overline{χ}_{(2)}}{r^2}\,.
\end{align}
Note we have explicitly written it as a correction to the Newtonian potential, and we see that, while formally at 1PN order, it is a 3PN correction, as expected when the dimensionless coupling constant is introduced.
Note, however, as we have already emphasized, the scalar charge is also proportional to the EsGB coupling constant. Hence, this correction will in fact scale as $\ell^4 \overline{χ}_{(2)}^{2}$.

In addition, we also have a contribution coming from scalar tides, i.e. a deformation of one compact object due to the presence of an external scalar dipole moment sourced by the companion as shown in \cref{fig:ESGB-tidal}. It is described by the action given in \cref{eq:scalarinternaleffects} and it gives the contribution
\begin{align}
	V_{\rm scal.\ tid.} = & -\frac{ \left( \overline{m}_2 d_2 \right)^2}{16\pi^2\Lambda^4r^4}c^{(1)}_s +\{1 \leftrightarrow 2\} \,, %
\end{align}
where $c^{(1)}_{s}$ is the scalar tidal deformation parameter, similar to $c_E$ for the electric-type tidal deformability.

\subsection{Radiative modes}

The presence of an additional scalar field not only yields corrections to the gravitational moments but also introduces a new class of moments of scalar origin: the scalar monopole and dipole moments~\cite{Bernard:2019yfz}.
When the scalar propagator is internal (off-shell), it produces a leading-order correction to the mass quadrupole moment, whereas an external (on-shell) scalar contributes to the scalar moments.

Starting with the correction to the mass quadrupole moment, the correction to the radiative effective action from the EsGB term comes from the interaction $-16\psi\partial_{ij}\phi\ddot \sigma_{ij}/Λ$. The corresponding diagram is shown in \cref{fig:ESGB-rad} and evaluates to

\begin{align}\label{Sscaleffrad}\nn
	S_{\rm eff,\ rad}^{\rm (EsGB)} & =\int\ud t\, \biggl\{\frac{1}{2}\frac{\ud^2}{\ud t^2}\biggl[-\frac{32 G {\overline m}_1 {\overline m}_2\left(d_1+d_2\right)}{r^3}\chi_{(2)}x^ix^j\biggr] \\\nn
	& ~~ +\frac{32 G^2 {\overline m}_1 {\overline m}_2({\overline m}_1+{\overline m}_2)}{r^6} \chi_{(2)}\,x^ix^j                                               \\
	  & ~~ \times \left[{\overline m}_1\left(2d_1+d_2\right)
	+ {\overline m}_2 \left( 2d_2+d_1 \right)\right] \biggr\}\,\frac{\sigma_{ij}}{2\Lambda}\,.
\end{align}
and according to \cref{eq:quad1PNGR}, to the mass quadrupole
\begin{align}\label{Iij-scalar-tensor} \nn
	I_{\rm cub}^{ij} \sim & \sum_{a\neq b} G \overline{m}_a \overline{m}_b\,d_a  χ_{(2)} \frac{x_a^{\langle ij \rangle}}{r_{ab}^3} \\
	\sim & \sum_{a\neq b} \overline{m}_a\,x_a^{\langle ij \rangle} \frac{G \overline{m}_b\,d_a}{r} \frac{\ell^2\overline{χ}_{(2)} }{r^2}\,.
\end{align}
This is a 3PN correction to the Newtonian mass quadrupole, coming from a formal 1PN + a 2PN correction linked to the EsGB coupling.

Next, the case when a scalar field is emitted corresponds to a diagram with an external (on-shell) scalar propagator.
It contributes to the scalar moments at leading order as
see e.g. \cite{Kuntz:2019zef}

\begin{align}
	S_{mon}  & =\sum_a\pa{m_ad_a}\,\psi\,, \\
	S_{dip}  & =\sum_a\left(d_am_a \, x^i \right)\,\psi_{,i}\,, \\
	S_{quad} & =\sum_ad_am_a\pa{x^i_ax^j_a-\frac 13x_a^2\delta^{ij}}\psi_{,ij}\,.
\end{align}
Note that the scalar monopole is constant at leading order, contributing to the radiative flux only at the next sub-leading order.

\section{Constructing general theories beyond GR}\label{sec:lagrangian}

A large body of beyond GR theories are naturally understood (and/or explicitly developed) in terms of Effective-field theories where corrections to the Einstein-Hilbert
action are accounted through higher order curvature operators (and also their derivatives), e.g.~\cite{Ruhdorfer:2019qmk,Aguilar-Gutierrez:2023kfn},  with the possible inclusion of extra degrees of freedom, e.g.~\cite{1974IJTP...10..363H}.
In this approach, high energy (i.e., above the cutoff scale) degrees of freedom are integrated out, and their effects are effectively accounted for through such operators
acting on the lower energy degree of freedom represented by the metric tensor.  Further requirements can be introduced to reduce the number of options to consider.
For instance, causality bounds (e.g.~\cite{deRham:2020zyh}); break or preservation of specific symmetries; consistency with the IR limit of GR;
discard terms proportional to the lower order equations of motion through field redefinitions~\cite{Solomon:2017nlh}, etc.
As well, at a given order, only truly independent contributions should be considered in the construction of the Lagrangian of
interest.
In what follows we consider the general case
where any operator, consistent with a given desired option, can contribute and simply make the distinction of whether only curvature terms are present with and without an
extra scalar degree of freedom as well as also the case with curvature operators/scalar are differentiated.

\section{`Rules' for general case}\label{sec:rulesgeneral}

\subsection{No additional scalars or gradients of Riemann}\label{sec:rules-Rp}

We consider the generic higher derivative theory
\begin{equation}
	S = 2Λ²\int\ud^{4}x\sqrt{-g}\left[R + \alpha_{(p)}{\cal R}^{(p)}\right] \,,
\end{equation}
where ${\cal R}^{(p)} \sim \left(R_{abcd}\right)^p$ represents all possible independent scalars constructed from products of $p$ Riemann tensors, and $\alpha_{(p)}$ is some coupling constant coming with a specific scale at which new physics appear and we restrict to $p\geq 3$.\footnote{The case $p=2$ in vacuum spacetimes in $d=4$ does not introduce departures from GR.}
More precisely, by dimensional analysis we get the scaling $[\alpha_{(p)}]=L^{2(p-1)}$, which leads us to define the dimensionless coupling constant
\begin{equation}\label{eq:alphabar}
	\overline{\alpha}_{(p)} \coloneqq \left(\frac{c^2}{G{M_Λ}}\right)^{2(p-1)}\,\alpha_{(p)}\,,
\end{equation}
where ${M_{\Lambda}}$ is the typical mass at which new physics enters.\footnote{Equivalently, we could have defined $\overline{\alpha}_{(p)} \coloneqq \alpha_{(p)} ℓ^{-2(p-1)}$ with $\ell$ the typical length scale at which new physics would appear. To date, gravitational wave observations being consistent with predictions within GR~\cite{LIGOScientific:2021sio} points to this scale being $\ell\lesssim$ km.}
Our goal is to determine the leading contribution coming from the higher derivative terms in the action.
{\em We note that all presently discussed beyond GR theories motivated from an EFT perspective—without extra degrees of freedom—are included in this discussion (e.g.~\cite{Endlich:2017tqa,Bueno:2016xff,Held:2021pht}).}

The LO contribution from ${\cal R}^{(p)}$ will involve a diagram with a $p$--point vertex
\begin{figure}[H]
	\centering
	\begin{minipage}{0.2\linewidth}
		\hspace*{10pt}
		\begin{tikzpicture}[thick]
			\draw [black, dashed] (0.5,0.8) -- (1,1.3);
			\draw [black, dashed] (1.5,0.8) -- (1,1.3);
			\draw [black, dashed] (1,  1.3) -- (1, 2);
			\node (down) at (1,0.6) {$\underbrace{\quad \cdots \quad }$};
			\node (down2) at (1,0.2) {{\footnotesize $(p-1)$ legs}};
		\end{tikzpicture}
		\label{fig:p-pointvertex}
	\end{minipage}
	\hfill
	\begin{minipage}{0.75\linewidth}
		\centering
		\begin{align}\nn
			 & \sim \frac{(∂²h)^p}{Λ^{p-2}} \frac{\boldsymbol{δ}\!\left(\boldsymbol{k}_1\!+\!\cdots\!+\!\boldsymbol{k}_p\right)}{\boldsymbol{k}_1² \ldots \boldsymbol{k}_p²} \\ \nn
			 & \quad\times δ(t_1\!-\!t_2)\!\cdots\!δ(t_1-t_p) α_{(p)}                                                                                                        \\
			 & \sim \frac{r³}{t^{p-1}Λ^{p-2}} \overline{α}_{(p)} \left(\frac{G {M_{\Lambda}}}{c^2}\right)^{2(p-1)}\,,
		\end{align}
	\end{minipage}
\end{figure}
which, along with \cref{eq:scalar-matter-coupling} gives the following leading order contribution to the binding energy:\footnote{here we have assumed comparable masses for the binary components $m_1 \sim m_2 \sim m$. See \cref{app:whichmass} for the general case.}
\begin{align}\label{eq:Rp}\nn
	\delta V_{(\partial^2h)^p}
	 & \sim\frac{m^p}{Λ^{2p-2}c^{2(p-2)}} \frac{1}{r^{3p-3}}\cdot\left(\frac{G {M_{\Lambda}}}{c^2}\right)^{2p-2}  \overline{\alpha}_{(p)}\,                              \\ \nn
	 & \sim \frac{m²}{rΛ²}\cdot \frac{m^{p-2}}{r^{3p-4}Λ^{2p-4}c^{2p-4}}\cdot\left(\frac{G {M_{\Lambda}}}{c^2}\right)^{2p-2}  \overline{\alpha}_{(p)}\,                  \\ \nn
	 & \sim V_{\textrm {\scshape n}}\cdot\left(\frac{G m}{r c^2}\right)^{p-2}\cdot\left(\frac{G {M_{\Lambda}}}{r c^2}\right)^{2p-2}  \overline{\alpha}_{(p)}\, \\
	 & \sim V_{\textrm {\scshape n}}\cdot\left(\frac{v^2}{c^2}\right)^{3p-4}\cdot\left(\frac{{M_{\Lambda}}}{m}\right)^{2p-2} \overline{\alpha}_{(p)}\,,
\end{align}
where $m\sim m_1+m_2$ is the typical mass of the compact object and we have used the virial theorem
$Gm/r \sim v²$ to make the PN order apparent in the second line. $V_{\textrm {\scshape n}} \sim m²/(rΛ²) \sim Gm²/r$ is the Newtonian potential.
As before, we have removed a factor of $t$ corresponding to the final $\int \dd t$ in the effective action, to obtain the potential (cf. \cref{footnote:seff}). Hence, we directly see that the higher derivative term gives a ${(3p-4)}$--PN contribution, while it scales only as $\left(M_{\Lambda}/{m}\right)^{2p-2}$.

As we have seen in the cubic gravity case, the effective radiative Lagrangian will have two new contributions from the higher curvature coupling. The first one directly comes from the diagram with a p-point vertex (with p-1 internal legs and one external radiative graviton). The second term originates from the modified equations of motion that have to be inserted in the radiative action.
However, both corrections give the same scaling in the radiation part, hence the correction to the gravitational waveform will scale as,
\begin{equation}\label{eq:δh-Rp}
	\delta h_{ij}^{\rm TT} \sim {h_{ij}^{\rm TT}}\big|_{\textrm {\scshape n}} \cdot \left(\frac{v^2}{c^2}\right)^{3p-4}\cdot\left(\frac{{M_{\Lambda}}}{m}\right)^{2p-2}\,\overline{\alpha}_{(p)} \,,
\end{equation}
The phase $\varphi$ is obtained from the relation $\omega \coloneqq \ud\varphi/\ud t$ and $\ud\omega/\ud t = -{\cal F}/(\ud E/\ud \omega)$ and we similarly get
\begin{equation}
	\delta \varphi \sim \varphi |_{\textrm {\scshape n}} \cdot x^{3p-4}\cdot\left(\frac{{M_{\Lambda}}}{m}\right)^{2p-2}\,\overline{\alpha}_{(p)} \,,
\end{equation}
where we have introduced the PN parameter $x\coloneqq \left(Gm\omega/c^3\right)^{2/3}$ that is commonly used in the literature.

Note that although such a correction is the LO in the case of cubic gravity (5PN), it will not be the case for higher order derivative theories, such as for instance quartic gravity. In the latter case such a correction is $8$PN, although there will be tidal effects already showing up at $5$PN. Indeed, while in GR tidal Love numbers (TLNs) are zero, it is usually not the case beyond GR and we should consider those contributions. Then the TLN will scale proportionally to the new coupling (see the detailed discussion on tidal effects). More precisely, the correction to the potential from tidal effects will be
\begin{align}
	\delta V_{\rm tid} \sim & \ V_{\textrm {\scshape n}}\cdot\left(\frac{v^2}{c^2}\right)^{5}\cdot\left(\frac{{M_{\Lambda}}}{m}\right)^{2p-2}\,\overline{c}_E^{(a)}\,.
\end{align}
We see that the correction is always at $5$PN order but will scale as $\left(M_{\Lambda}/m\right)^{2p-2}$ with respect to new physics and dominated by the lightest mass of the binary (see~\cref{app:whichmass}).
Such a scaling of the tidal contribution also holds for the radiation part, for which it will contributes at $5$PN beyond the leading GR contribution both for the waveform and the phase. Hence, we expect this term to be dominant for theories with order above $p=3$.

Finally, let us comment on the generic action we have studied and the possibility that ${\cal R}^{(p)}$ contains also dual of the Riemann tensor $({}^* R)_{abcd}\coloneqq\epsilon_{abef}R^{ef}_{~~cd}$. In that case, due to the presence of the Levi-Civita tensor, an extra factor of $v$ will always show up (said differently, one of the legs of the
$p$--vertex will have to couple with a velocity factor from the matter), hence adding an additional half PN order. To summarize, each dual Riemann tensor present in the action adds a $0.5$PN order to the LO correction. This is in agreement with~\cite{Endlich:2017tqa} for quartic gravity.

It is crucial to highlight that we have calculated the generic leading order contribution expected from a higher derivative theory of gravity, without identifying the specific scalar combination of Riemann tensors it originates from. Nevertheless, not all scalar combinations of $p$ Riemann tensors contribute equally at this order–some may cancel each other out or become zero after dimensional regularization. As illustrated by the example in cubic gravity, the coefficient for $α_3$ turned out to be zero following dimensional regularization. Such zeroes and cancellations are generally expected across all $p$.
The core assertion in this section is that \emph{considering every possible scalar combination of $p$ Riemann tensors, the leading order contribution is as expressed in this section.}

\subsection{With an additional scalar field}\label{sec:rules-Rp-scalar}

Let us now study the case when an additional undifferentiated scalar field coupled to independent curvature operators is present in the Lagrangian. Consider the following action
\begin{equation*}
	S = 2\Lambda^2\int\ud^{4}x\sqrt{-g}\left[R -\frac{1}{4}\nabla^{a}\Psi\nabla_{a}\Psi + \beta_{(p)}\,\Psi\,{\cal R}^{(p)}\right] \,,
\end{equation*}
where ${\cal R}^{(p)}$ is still made of products of $p$ Riemann tensors.
	{\em For $p=2$, we naturally accomodate for
		Einstein-scalar-Gauss-Bonnet gravity~\cite{Kanti:1995vq}, dynamical Chern Simmons~\cite{Alexander:2009tp} and the subset of Hordenski gravity~\cite{1974IJTP...10..363H} without derivatives
		of the scalar field.}

Throughout this work, we will assume that the additional scalar field is massless and
it is linearly coupled to the curvature contribution. The mass of the scalar does not introduce a new scale in the inspiral phase (\emph{cf.} §III of \cite{Diedrichs:2023foj}), and it is sufficient to consider a massless scalar to construct the general rules. Let us see why. If the scalar field were massive, its propagator would be $\sim 1/\left(|\boldsymbol{k}|² + m_s²\right)$. If $m_s \gg 1/r$, the field is too heavy to impact the inspiral phase and its effects can be integrated out to construct an EFT without potential interactions. On the other extreme, if $m_s \ll 1/r$, the field is essentially massless on orbital scales. When $m_s \sim 1/r \sim |\boldsymbol{k}| $, the propagator scales in the same way as the massless propagator $\sim 1/|\boldsymbol{k}|²$. Therefore, it is sufficient to consider a massless scalar field to construct the general rules.
Recall that there is a non-trivial dependence on the coupling between the scalar field and the curvature contribution. For a linear coupling, the scalar field  depends linearly on the coupling strength
$\beta_{(p)}$.  If instead, the field were coupled with a different power $Ψ^n$, this dependence would change accordingly. Such a dependence  can be surmised from the solution to the simpler equation $\Psi_{,xx}=\beta_{(p)} \Psi^{(n-1)}$.  Importantly, for sufficiently high $n$ (i.e. $n\ge 3$), $\Psi$ displays a dependence of the form $1/\beta_{(p)}^q$ with $q>0$, thereby also sourcing in the metric field equation an inverse dependence on the coupling. This, in turn, would signal a break-down of the EFT approach, as there would not be a natural limit to GR as the coupling $β_{(p)} \rightarrow 0$ (see also~\cite{Doneva:2023oww} for differences
arising in exponential forms of the coupling).

At this point, we should remark on the expansion of the scalar field. Writing $\Psi=1+ψ/Λ$, we see that the LO contributions will come from a vertex with $p$ branches from the Riemanns and no scalar field, recovering the same results as in the previous subsection for $p$--order gravity. Thus, in the following we will only consider contributions where at least one scalar-field propagator is involved. In that case, we have for the binding potential
\begin{align}\label{eq:gen_scalar}\nn
	\delta V_{\psi(\partial^2h)^p}
	 & \sim V_{\textrm {\scshape n}}\cdot\left(\frac{v^2}{c^2}\right)^{3p-3}
	\hspace{-5pt}\cdot\left(\frac{{M_{\Lambda}}}{m}\right)^{2p-2} \overline{\beta}_{(p)}\,d_a \\
	 & \sim V_{\textrm {\scshape n}}\cdot\left(\frac{v^2}{c^2}\right)^{3p-3}\hspace{-5pt}\cdot\left(\frac{{M_{\Lambda}}^2}{m \,m_a}\right)^{2p-2} \overline{\beta}_{(p)}^2\,\overline{d}_a,
\end{align}
where in the second line, we have used the fact that $d_a$ is linearly proportional to the coupling (because it is sourced by it) to introduce $\overline{d}_a$, such that $d_a \coloneqq  \left(M_{\Lambda}/m_a\right)^{2p-2} \overline{\beta}_{(p)}\overline{d}_a$.
Hence, we see that such a correction will be further suppressed as proportional to square of the coupling. As well, we have introduced
the dimensionless coupling constant ${\overline \beta}_{(p)}=(c^2/(G M_{\Lambda}))^{2(p-1)} \beta_{(p)}$.
Note however, that there will also be lower order contributions coming from the kinetic part with a scalar field exchange. At Newtonian order, it accounts in a rescaling of the Newtonian gravitational constant, see Eq.~\eqref{eq:VscalN}, but at 1PN, it differentiates it from GR.
This result holds in general, for unequal masses in the binary, in which case the $(p+1)$–point potential contribution in \cref{eq:gen_scalar} is dominated by the lightest mass. The tidal contribution to the potential also goes as the lightest mass in the binary, as discussed earlier. See \cref{app:whichmass} for a detailed discussion.

In addition, we must also include tidal contributions. As we have seen for the EsGB case, the scalar field yields a new type of scalar tidal interactions. Such interactions contribute as early as 3PN order due to the presence of a scalar dipole moment.
The resulting contribution in the binding potential is given by \
\begin{align}\nn
	\delta V_{\rm scal.\,tid.} \sim & \ V_{\textrm {\scshape n}}\cdot\left(\frac{v^2}{c^2}\right)^{3}\frac{m_b\,d_b}{m_a}\,\left(\frac{m_a}{m}\right)^{3}\overline{c}_s^{(a)} \\
	\sim & V_{\textrm {\scshape n}}\cdot\left(\frac{v^2}{c^2}\right)^{3}\!\!
	\left(\frac{m_a}{m}\right)^{3}\overline{c}_s^{(a)} \!\cdot\! \frac{m_b
		\overline{d}_b}{m_a}\overline{\beta}_{(p)}
	\left(\frac{{M_{\Lambda}}}{m}\right)^{2p-2}\,,
\end{align}
where we have defined the dimensionless scalar Love number $\overline{c}_s^{(a)}\coloneqq c_s^{(a)} G/(R_a c^2)$, with $R_a$ being the size of the compact object.

To determine the new associated contributions to the radiation, we distinguish between the emission of a graviton and that of a scalar:\\
In the case of graviton emission, the vertex involves $1$ internal scalar legs, $p$ internal graviton legs and one external (on-shell) graviton leg.
\begin{align}\label{eq:δh-PsiRp}\nn
	\delta h_{ij}^{\rm TT} & \sim {h_{ij}^{\rm TT}}\big|_{\textrm {\scshape n}} \cdot \left(\frac{v^2}{c^2}\right)^{3p-3}\cdot\left(\frac{{M_{\Lambda}}}{m}\right)^{2p-2}\,\overline{\beta}_{(p)}d_a   \\
    & \sim {h_{ij}^{\rm TT}}\big|_{\textrm {\scshape n}} \cdot \left(\frac{v^2}{c^2}\right)^{3p-3}\cdot\left(\frac{{M_{\Lambda}^2}}{m m_a}\right)^{2p-2}\,\overline{\beta}_{(p)}^2\overline{d}_a \,.
\end{align}
For the scalar case, the leading order contribution comes from the emission of a scalar field, corresponding to scalar monopole and dipolar radiation. However, the LO scalar monopole scales as
\begin{align}\nn
	I_{\rm (s,\textrm{\scshape n})} & = \sum_a m_a\,d_a \\
	& \sim \sum_a m_a\,\overline{d}_a\left(\frac{M_{\Lambda}}{m}\right)^{2(p-1)}\,\overline{\beta}_{(p)}\,,
\end{align}
Notice that although it depends on the coupling constant, it is constant and hence it does not contribute to the scalar radiation. As a consequence the leading order scalar radiation comes from the dipole, given at leading order by,
\begin{align}\nn
	I^i_{\rm (s,\textrm{\scshape n})} & = \sum_a m_a\,d_a \,x_a^i \\
	  & \sim \sum_a m_a\,\overline{d}_a \,x_a^i\left(\frac{M_{\Lambda}}{m}\right)^{2(p-1)}\,\overline{\beta}_{(p)}\,,
\end{align}
Because of its dipolar structure, such a term will contribute to the (scalar) flux at $-1$PN order with respect to the gravitational flux. Thus the additional scalar radiative (breathing) mode scales as
\begin{align}\label{eq:δpsi-PsiRp}\nn
	\psi & \sim \frac{Gm\nu}{R c^2}\cdot\left(d_a-d_b\right) \cdot \left(\frac{v}{c}\right)    \\
	     & \sim \frac{Gm\nu}{R c^2}\cdot\left(\overline{d}_a -\overline{d}_b\right) \cdot \left(\frac{v}{c}\right) \cdot\left(\frac{{M_{\Lambda}}}{m_a}\right)^{2p-2}\,\overline{\beta}_{(p)}\,.
\end{align}
{We see, as it is already known, the breathing mode starts at $-0.5$PN order compared to the $+,\times$ polarizations.}

\subsection{Accounting for Derivatives of Riemann and scalar}\label{sec:rules-der}

\subsubsection{No additional scalar field}
Let us consider higher derivative theories which contain derivatives of curvature terms in the action
(see ~\cite{Ruhdorfer:2019qmk}). For the case without additional scalar fields,
the most general term contains scalar combinations of the Riemann tensor and its derivatives.\footnote{Of course, care must be exercised to ensure such differentiated terms are truly independent. That is, that not be written
	in terms of undifferentiated terms through Bianchi identities and commutations, and others that could be removed
	via field-redefinitions or integration by parts. For checking this, symbolic packages
	like {\em Invar} are quite useful~\cite{Martin-Garcia:2008yei}.
	See for instance~\cite{Endlich:2017tqa,Solomon:2017nlh,deRham:2019ctd} for related discussions.}
We write this as $α_{(p)}f\left[(∇)^{2n},\textrm{Riem}^{p-n}\right]$, where the derivatives can act on any number of curvature tensors in any combination.
\begin{equation}\label{eq:deriv_action}
	S = 2Λ²\int\ud^{4}x\sqrt{-g}\left[R + α_{(p)}f\left[(∇)^{2n},\textrm{Riem}^{p-n}\right]\right] \,,
\end{equation}
From an EFT perspective, the net gradient of the metric field is $2n+2(p-n) = 2p$, and
the leading order contribution from this term is a $(p-n)$--scalar vertex, which gives the following leading order contribution to the binding energy
\begin{align}\nn
	\delta V \sim & \ V_{\textrm {\scshape n}}\cdot\left(\frac{G m}{r c^2}\right)^{p-n-2}\cdot\left(\frac{G {M_{\Lambda}}}{r c^2}\right)^{2p-2}  \overline{\alpha}_{(p)}\, \\\nn
	\sim          & \ V_{\textrm {\scshape n}}\cdot\left(\frac{v^2}{c^2}\right)^{3p-n-4}\cdot\left(\frac{{M_{\Lambda}}}{m}\right)^{2p-2} \overline{\alpha}_{(p)}           \\
	\sim          & \ V_{\mathcal{R}^{(p)}} \cdot \left(\frac{v^2}{c^2}\right)^{-n}\,,
\end{align}
where, as before, we have introduced the dimensionless coupling $\overline{α}_{(p)}$, the typical mass $m$ of the binary, and the scale of new physics ${M}_{\Lambda}$.
We see that this potentially contribute at $n$ PN order less than the corresponding term without derivatives $\mathcal{R}^{(p)}$ as computed in \cref{eq:Rp}, while the mass scaling remains the same.

Similar to the case without derivatives, the leading order radiation have the same scaling behavior as the potential, giving the following correction to the gravitational waveform
\begin{equation}
	\delta h_{ij}^{\rm TT} \sim {h_{ij}^{\rm TT}}\big|_{\textrm {\scshape n}} \cdot \left(\frac{v^2}{c^2}\right)^{3p-n-4}\cdot\left(\frac{{M_{\Lambda}}}{m}\right)^{2p-2}\,\overline{\alpha}_{(p)} \,,
\end{equation}
and to the phase $\varphi$,
\begin{equation}
	\delta \varphi \sim \varphi |_{\textrm {\scshape n}} \cdot x^{3p-n-4}\cdot\left(\frac{{M_{\Lambda}}}{m}\right)^{2p-2}\,\overline{\alpha}_{(p)} \,.
\end{equation}

\subsubsection{With an additional scalar field}
The most general term in \cref{eq:deriv_action} with additional scalar fields and their derivatives contain terms like $\{ Ψ, ∇^2 Ψ, (∇Ψ)^2, … \},$\footnote{An example being Horndeski gravity~\cite{1974IJTP...10..363H}.} in addition to derivatives of the curvature tensor. As such scalar fields are contemplated to mediate gravitational effects, their derivatives contribute to the net EFT order considered.
On expanding the scalar field as $Ψ \sim 1 + ψ/M_p$, we see that $Ψ$ without derivatives do not contribute at leading order. Leading order corrections come from derivatives of $Ψ$, which enables us to write a general action of the form
\begin{align}\nn\label{eq:general_scalar_tensor_action}
	S & = 2\Lambda^2\int\ud^{4}x\sqrt{-g}\left(R -\frac{1}{4}\nabla_{a}Ψ \nabla^{a}Ψ \right. \\
	  & \qquad \quad \left. + \beta_{(p)} f\left[ \left(∇^{a}, Ψ^b \right); \left(∇^c, \textrm{Riem}^{d} \right)\right]\right) \,,
\end{align}
where $a,b \geq 1$, and every scalar $Ψ$ appears with derivatives (following the argument in \cref{sec:rules-Rp-scalar}). Again, from an EFT point of view,
derivatives\footnote{And care must also be exercised to ensure all terms are truly independent.} of the scalar field or the Riemann tensor contribute to the net order $2p$.
Since the derivatives must appear in pairs, $a+c \coloneqq 2m$. Additionally we define $m+d \coloneqq p$.
The leading order contribution to the binding energy comes from a $(b+d)$--point vertex, which gives
\begin{equation}\nn
	\delta V \sim V_{\textrm {\scshape n}}\cdot\left(\frac{v^2}{c^2}\right)^{2p+d+b-4}
	\cdot\left(\frac{c^2}{G m}\right)^{2(p-1)}
	{\beta}_{(p)} \cdot \left({d}_a\right)^b\,,    %
\end{equation}
in terms of the quantities defined in \cref{eq:gen_scalar}.
Notice, however, that a simple power scaling with $\beta_{(p)}$ is not straightforward to obtain, as it depends on the details of the solution for $\psi$. Here again, an inverse dependence on the coupling might ensue depending on the choice adopted.
For the radiation, there are two possibilities:
\begin{enumerate}[label=(\roman*), leftmargin=15pt]
	\item Emission of a graviton (vertex with $b$ internal scalar leg, $d-1$ internal gravitational legs and one external (on-shell) graviton leg): the scaling is the same as for the binding potential.
	\item
		Emission of a scalar (vertex with $b-1$ internal scalar leg, $d$ internal gravitational legs and one external (on-shell) scalar leg): As for the case with no derivatives, see Sec.~\cref{sec:rules-Rp-scalar}, the leading order correction comes from the emission of a single scalar giving rise to the following additional scalar radiative (breathing) mode
		\begin{align}
			\psi_{\rm breathing} & \sim \frac{Gm\nu}{R c^2}\cdot\left(d_1 - d_2\right) \cdot \left(\frac{v}{c}\right) \,.
		\end{align}
		As mentioned earlier, working out the explicit dependence on the coupling requires further investigation and will depend on the structure of the derivatives acting on the scalar field.
\end{enumerate}

\section{Example illustration}\label{sec:examples}

\subsection{Quartic gravity}
As an example, we now apply the aforementioned general rules to discuss corrections introduced in a EFT-motivated theory with not extra degree of freedom and quartic curvature corrections. We can then contrast our results
with specific calculations carried out in the particular case described in~\cite{Endlich:2017tqa,Lins:2020omt}:
\begin{equation*}\label{quartic}
	S = 2Λ²\int \ud^4x\sqrt{-g} \left( R + \overline{α}_{(4)}\, {M_Λ}^6 \left[ {\cal C}^2 + {\cal C} \widetilde {\cal C} + \widetilde {\cal C}^2 \right ] \right) \,,
\end{equation*}
with ${\cal C}=R_{abcd} R^{abcd}, \widetilde{\cal C} = R_{abcd} (^*R)^{abcd}$, and we have assumed that all the quadratic corrections appear at similar mass scales $\sim M_Λ$. Thus $p=4$ and neither a scalar field, no extra derivatives are included.
\begin{table}
	\begin{tabularx}{\linewidth}{@{}
		>{\centering\arraybackslash}m{0.4\linewidth}
		>{\centering\arraybackslash}m{0.2\linewidth}
		>{\centering\arraybackslash}X@{}}
		\toprule
		$\displaystyle\frac{\textrm{quantity}}{\overline{α}_{(4)}}\left(\frac{m}{M_Λ}\right)^{6}$ &
		\makecell[cc]{quartic \\ (this work)} &
		\makecell[cc]{most relevant \\ order of \cite{Endlich:2017tqa}} \\ \addlinespace[5pt]
		\midrule
		\addlinespace[5pt]
		$\displaystyle δh/h|_{\textrm{GR}} \propto$                                               &
		$\displaystyle \left( \frac{v^2}{c^2} \right)^8$                                          &
		$\displaystyle \left( \frac{v^2}{c^2} \right)^8$
		\\ \addlinespace[8pt]
		$\displaystyle δV/V|_{\textrm{N}} \propto$                                                &
		$\displaystyle \left( \frac{v^2}{c^2} \right)^8$                                          &
		$\displaystyle  \left( \frac{v^2}{c^2} \right)^8$
		\\ \addlinespace[8pt]
		$\displaystyle δV_{\textrm{tid}}/V|_{\textrm{N}} \propto$                                 &
		$\displaystyle \left( \frac{v^2}{c^2} \right)^{5}$                                        &
		$\displaystyle \left( \frac{v^2}{c^2} \right)^{5}$
		\\ \addlinespace[5pt]
		\bottomrule
	\end{tabularx}
	\caption{Predicted leading PN order compared to known results from quartic corrections for non-spinning black holes.
	\label{tab:quartic}}
\end{table}
Using the rules as described in \cref{sec:rules-Rp}, we can state the predicted leading order PN effects for non-spinning black holes as summarized in \cref{tab:quartic}. These results can be compared to those explicitly computed in~\cite{Endlich:2017tqa}, where each contribution is evaluated individually. For each quantity, we indicate the dominant correction found in that work, across all individual terms. Note that for the strain, which scales with the total mass of the system, we present the correction 
\emph{excluding} tidal effects (which would appear at 5PN order); the tidal contribution is shown in the last row of the table and scales with respect to the lightest mass in the binary (see \cref{app:whichmass}).
\begin{table}
	\begin{tabularx}{\linewidth}{@{}
		>{\centering\arraybackslash}>{\hsize=0.45\hsize}X
		>{\centering\arraybackslash}>{\hsize=0.45\hsize}X
		>{\centering\arraybackslash}>{\hsize=0.25\hsize}X
		>{\centering\arraybackslash}>{\hsize=0.2\hsize}X@{}}
		\toprule
		$\displaystyle\frac{\textrm{quantity}}{\overline{α}_{(2)}^2}
        \left(\!\frac{m\,m_\mysc{l}}{M_Λ^2}\!\right)^{\!2}$ &
		\makecell[cc]{scalar + quadratic \\ (this work)} &
        EsGB &
        dCS \\ \addlinespace[3pt]
		\midrule
		\addlinespace[5pt]
		$\displaystyle δh_{\{+,\cross\}}/{h_{\textrm{GR}}} \propto$ &
		$\displaystyle \left( \frac{v^2}{c^2} \right)^3$ &
		$\displaystyle \left( \frac{v^2}{c^2} \right)^{3}$ &
		$ * $
		\\ \addlinespace[8pt]
		$\displaystyle h_{b} \, \overline \alpha_{(2)} \left( \frac{M_{\Lambda}}{m}\right)^2 \propto$ &
		$\displaystyle ~~~\left( \frac{v^2}{c^2} \right)^{-1/2}$ &
		$\displaystyle ~~~\left( \frac{v^2}{c^2} \right)^{-1/2}$ &
		$\displaystyle ~\left( \frac{v^2}{c^2} \right)^{1/2}$
		\\ \addlinespace[8pt]
		$\displaystyle δV/V|_{\textrm{N}} \propto$ &
		$\displaystyle  \left( \frac{v^2}{c^2} \right)^3$ &
		$\displaystyle ~\left( \frac{v^2}{c^2} \right)^{3}$ &
		$ * $
		\\ \addlinespace[8pt]
		$\displaystyle δV_{\textrm{tid}}/V|_{\textrm{N}} \propto$ &
		$\displaystyle \left( \frac{v^2}{c^2} \right)^3$ &
		$\displaystyle ~\left( \frac{v^2}{c^2} \right)^3 $ &
		$\displaystyle \left( \frac{v^2}{c^2} \right)^3 $
		\\ \addlinespace[5pt]
		\bottomrule
	\end{tabularx}
	\caption{Predicted leading PN order compared to known results from quadratic curvature and scalar field corrections for non-spinning black holes. \label{tab:quadratic}}
\end{table}

\subsection{Quadratic gravity with one scalar degree of freedom}
At quadratic order, there are two independent invariant combinations of the curvature tensor—the Euler invariant $G_2$ and the Chern-Pontryagin invariant $P_2$
\begin{align}\nn
	G_2 & = ({}^* R {}^* )_{abcd} R^{abcd} \coloneqq ϵ_{abef}R^{efgh}ϵ_{ghcd}R^{abcd}\,, \\ \nn
	P_2 & = ({}^* R)_{abcd} R^{abcd} \coloneqq ϵ_{abef}R^{ef}_{~~cd}R^{abcd}\,.
\end{align}
Adding $G_2$ to the EH action gives Gauss-Bonnet gravity, while $P_2$ gives Chern-Simons gravity. Being topological invariants, neither of them contribute if added directly to the EH action. However, when coupled to a scalar field, they do contribute—$G_2$ gives the EsGB theory discussed in \cref{sec:esgb}, while $P_2$ gives dynamical Chern Simons (dCS) gravity~\cite{Yagi:2011xp}. Analogous to the EsGB action in \cref{eq:ESGB-action}, the dCS action is
\begin{equation}\nn
	S_{\textrm{dCS}} = 2Λ²\int \ud^4x\sqrt{-g} \left[ R -\frac{1}{2} (\partial Ψ)^2 +  \overline{α}_{(2)}\, {M_Λ}^2 Ψ P_2\right]\,.
\end{equation}
Let us now compare our results for a theory with quadratic curvature corrections and an extra scalar degree of freedom, with the results presented in~\cite{Almeida:2024cqz,Julie:2019sab,Yagi:2011xp,Shiralilou:2021mfl} for ESGB, and in \cite{Yagi:2011xp,Yagi:2012vf}  for dCS gravity (see also~\cite{Cardoso:2017cfl,Tahura:2018zuq,Shiralilou:2021mfl,vanGemeren:2023rhh,Creci:2024wfu,Jain:2024lie}).

Using the rules described in \cref{sec:rules-Rp}, we tabulate the results in \cref{tab:quadratic}. The table shows the predicted leading PN-order corrections (scalar+quadratic column) for non-spinning black holes, compared to available results from the literature for EsGB and dCS theories.

A few clarifications are in order: for dCS gravity, we report literature values assuming spinning black holes. While some references quote a 2PN correction for dCS, this arises from modifications to the gravitational quadrupole induced by a modified scalar profile and resulting scalar charge. To remain consistent with how we contrast other tabulated theories, we omit this specific correction from the table. In addition, as explained in Sec.~\ref{sec:esgb}, a 1PN correction is expected in the binding potential in any scalar-tensor theory, proportional to the scalar charge. Since this is well known and does not come directly from the new higher-curvature coupling, we omit it from the table.

We also display the breathing mode correction relative to the leading GR scaling. In this case, we assume the theory is expressed in the Jordan frame and that the scalar field decays as $1/r$—otherwise, the breathing mode would decay too rapidly to be detectable (see, e.g.~\cite{Barausse:2012da,Wagle:2019mdq}). For strain, which scales with the total mass $m$ of the system, we show the correction \emph{excluding} tidal effects (which would enter at 5PN order); tidal effects, which scale with the lighter mass $m_\mysc{l}$ in the binary, are shown in the last row (see \cref{app:whichmass}).

\section{Final words}\label{sec:final}

Identifying deviations from GR is certainly a tall order. Capturing key features predicted by theories beyond GR helps to find signs of them in the analysis, and tell them apart from potential systematics. In this work we developed a set of rules to estimate the impact that general EFT-inspired beyond GR theories have on waveforms produced in the inspiral phase of compact binaries.\footnote{Clarifying in the process some misconceptions as well as some subtleties of the application of PN-EFT beyond GR.} Importantly, from an EFT perspective, theories often treated as separate are actually part of a larger set, grouped by the order of corrections included. Furthermore, unless there is a special fine-tuning or a protective symmetry, the leading effect in the set is the most important one, as noted here. In particular, we demonstrate that vast sets of theories—those without extra degrees of freedom and no derivatives of the Riemann tensor describing non-spinning or slowly spinning objects—only deviate from GR beginning at a relatively high order (5PN). 

This holds for {\em all theories with corrections at cubic or higher orders},\footnote{Adding to this point, we note that most analyses for deviations consider up to 3.5PN order—as higher order contributions within GR are still incomplete. Nevertheless searches targeting 5PN corrections have already been proposed~\cite{Chia:2023tle}.} in contrast to the conservative assumption that all orders would be affected. On the other hand, deviations showing up at lower PN orders imply that extra degrees of freedom are at play.
These rules can be readily incorporated in refined analysis options currently employed that can target the inspiral regime
(e.g. ~\cite{Agathos:2013upa,Tahura:2018zuq,Dideron:2022tap,Mehta:2022pcn,Chia:2023tle,Payne:2024yhk}).  In this regime, knowing how deviations scale with mass and their leading PN order are important to tell different potential theories apart, and offers key insight when combining data from multiple events. We note however that extracting confidently the leading PN order is more delicate than identifying the overall scale, due to correlations between the PN coefficients. New techniques are being developed to address this challenge, see e.g.~\cite{Saleem:2021nsb}. Additionally, extracting both the scaling and PN order helps distinguish real deviations from systematics—for example, those of astrophysical origin, which are unlikely to depend on an exact integer power of mass~\cite{Saini:2022igm}.

We conjecture that a similar strategy could be applied in the post-merger case by identifying the quasi-normal mode (QNM) frequencies—which also scale with $M_{\Lambda}/m$ as shown in~\cite{Maenaut:2024oci}—alongside ratios of QNM amplitudes, which would contain information of PN effects. However, uncovering these features would require simulations to reveal how the amplitudes change in theories beyond GR, highlighting again the vital role of simulations, even when exploring a small subset of alternative theories.

Although we have not included spinning objects in this work, their effects can be naturally incorporated within the PN-EFT framework, following the same approach used in GR—see~\cite{Porto:2007qi}.

Finally, recall that in the expression for the strain, $h = h_{\mysc{gr}} + ({\ell}/m)^q \, \delta$, the quantity $\delta$ depends on the physical parameters in the system and therefore does not display universal behavior across all events. Its effect can be handled by marginalizing over it or by suitably binning events based on dimensionless combinations of these physical parameters.

\section*{Acknowledgements}

We thank Cliff Burgess, Gregorio Carullo, Will East, Jaume Gomis, Max Isi, R Loganayagam, Nathan Moynihan, Alex Nielsen, Ethan Payne, Rafael Porto, Adam Pound, Ira Rothstein, Thomas Sotiriou,  Louhan Wang and Nico Yunes for discussions.

LB acknowledges financial support from the ANR PRoGRAM project, grant ANR-21-CE31-0003-001 and the EU Horizon 2020 Research
and Innovation Programme under the Marie Sklodowska-Curie Grant Agreement no.
101007855.
The work of SG was conducted with funding awarded by the Swedish Research Council grant VR 2022-06157.
LL acknowledges support from the Natural Sciences and Engineering Research Council of Canada through a Discovery Grant. LL also thanks financial support via the Carlo Fidani Rainer Weiss Chair at Perimeter Institute and CIFAR.
This research was supported in part by Perimeter Institute for Theoretical Physics. Research at Perimeter Institute is supported in part by the Government of Canada through the Department of Innovation, Science and Economic Development and by the Province of Ontario through the Ministry of Colleges and Universities.
LB and RS thank the Perimeter Institute for Theoretical Physics  and LL the ICTP-SAIFR for hospitality where parts of this work were
carried out. 
RS wishes to acknowledge FAPESP grants n. 2022/06350-2 and 2021/14335-0, as well as CNPq grant n.310165/2021-0.

\appendix

\section{Cubic Lagrangian and a single independent combination}\label{app:cubicapp}
In the context of Einstein-cubic gravity, the action is constructed with 2 scalars, $\{I_1, G_3\}$, their expressions are:
\begin{eqnarray}
	I_1 &=& R^{ab}_{~~cd} R^{cd}_{~~lm} R^{lm}_{~~~ab} \\
	G_3 &=& I_1 - 2 R^{\mu \nu \alpha}_{~~~~\beta} R^{\beta \gamma}_{~~~\nu\sigma} R^{\sigma}_{~\mu γ \alpha}
\end{eqnarray}
though the latter expression can be re-expressed as,
\begin{eqnarray}\label{eq:G3-explicit}
	G_3&=& \frac{5}{4} R^3 - 9 R R^{ab} R_{ab} + \frac{3}{4} R R^{abcd} R_{abcd} \nonumber \\
	&& + 8 R^{ab} R_a^{~c} R_{bc} + 6 R^{ab} R_{acbd} R^{cd}.
\end{eqnarray}
Thus, every single term in $G_3$ can be removed from the Lagrangian through a
field redefinition\footnote{Notice ref~\cite{Metsaev:1986yb} misses this fact, taking $I_1$ and $G_3$ as independent of each other even
	under field redefinitions.} (see, e.g.~\cite{Cano:2019ore}).
Of course, if the scenario studied is not in vacuum, field redefinitions would affect the stress energy tensor.

\section{ Null contribution of {\boldmath$R R_{abcd}R^{abcd}$}}\label{app:nullRK}
We explicitly demonstrate that there is no contribution coming from $R R_{abcd}R^{abcd}$ in vacuum.
While the result follows from general arguments, a direct calculation explicitly demonstrates this and clarifies some existing results in the literature. Consider the action, in vacuum:
\begin{equation}
	S = 2Λ²\int d^4x \sqrt{-g} \, (R + λ_{(3)} R K )\,,
\end{equation}
where $K=R_{abcd} R^{abcd}$, and we will relabel the coupling $λ_{(3)} \equiv λ$ for brevity. This action can be derived from the Einstein-Hilbert action by redefining the metric  $g^\prime_{ab} = g_{ab} (1 + λ K(g))$, making $λ$ physically irrelevant
at linear order (in vacuum). Crucially, $g^\prime_{ab} \approx g_{ab}$ at infinity, ensuring that observables remain unchanged.
For a binary system at orbital scales, the PN-EFT action combines gravitational and point-particle terms:
\begin{align}\label{eq:RK+tidal}
	S_{\textrm{eff}} & = 2Λ²\int \dd ^4x \sqrt{-g} \, (R + λ R K ) \nn \\
	   & \quad - \sum_{a=1,2} m_a \int \dd \tau + \int \dd τ(c^{(a)}_E E^2 + c^{(a)}_B B^2)\,.
\end{align}
As outlined in the main text, the system’s dynamics and radiation follow from a systematic matching procedure.
For a concrete physical scenario—like the collision of two non-spinning black holes—the tidal coefficients $\{c_E, c_B\}$ are a key input.

Here, we compute these coefficients in the $RK$ theory directly from the GR effective action.
As here we focus on non-spinning black holes, we set $c_E = c_B = 0$.
Starting with
\begin{equation}
	S_{\textrm{eff}} = 2Λ²\int \dd^4 x \sqrt{-g'} \, R(g')  \nn - \sum_{a=1,2} m_a \int \dd\tau'\,,
\end{equation}
the field redefinition $g^\prime_{ab} = g_{ab} (1+ λ K(g))$ gives
\begin{eqnarray}\label{eq:ce_field_redef} \nn
	S_{\textrm{eff}} &=& 2Λ²\int \dd^4x \sqrt{-g} \, (R + λ R K ) \\ \nn
	&& - \sum_a m_a \int \dd\tau \left( 1+\frac{λK}{2} \right) \\ \nn
	&=& \int \dd^4x \sqrt{-g} \, (R + λ R K ) - \sum_a m_a \int \dd\tau  \\
	&&-  \sum_a  4 m_a λ  \int \dd\tau\left( E^2 - B^2 \right)\,,
\end{eqnarray}
where we have used the identity $K = 8 (E²-B²)$.
Comparing with \eqref{eq:RK+tidal}, we identify the tidal coefficients:\footnote{The explicit computation through standard techniques will appear in~\cite{louhan}.}
\begin{equation}\label{eq:ce-RK}
	c_E = - c_B = -4 m λ.
\end{equation}
With these coefficients in hand, we can now evaluate the corrections to the potential and radiation.
Starting with the action
\begin{equation}
	S=2Λ² \int \sqrt{-g} (R + λ l^4 RK) - m\int \dd τ + c_E \int \dd τ E²\,,
\end{equation}
we can evaluate the potential exactly, which gives,
\begin{align}\label{eq:RK-contribution-cubic}
	V_{\textrm{cubic}} & = - \frac{3}{128}\frac{m_1 m_2²}{r^6} \frac{λl^4}{π²Λ^4}                    \\ \label{eq:RKcontribution-tidal}
	V_{\textrm{tidal}} & = - \frac{3}{512} \frac{m_2²}{r^6} \frac{c_{\textrm{\scshape e}}}{π²Λ^4}\,.
\end{align}
Using $c_E = -4 m_1 λ l^4$ from above,
$$δV = V_{\textrm{cubic}} + V_{\textrm{tidal}} = 0\,.$$
The $c_B B^2$ is velocity suppressed by $\mathcal{O}(v²)$, due to the $\epsilon$ tensor in $B²$, similar to the discussion in \cref{sec:cubic-potential}. This contribution would similarly cancel against higher order contributions from the $RK$ term.

While \cite{Emond:2019crr,Brandhuber:2019qpg} identified the cubic contribution in \cref{eq:RK-contribution-cubic},\footnote{their result is for $S \supset (2/κ²)\int \sqrt{-g} (R+G_3/24)$, which is $1/32 \times \eqref{eq:RK-contribution-cubic}$, since $G_3 \sim (3/4)RK$ from \eqref{eq:G3-explicit}.} and \cite{Liu:2024atc} discussed its effects, they
did not account for the tidal contribution which, as shown above, cancels the cubic one.

The same conclusion was reached previously in~\cite{AccettulliHuber:2020dal,Wilson-Gerow:2025xhr} through a combination of explicit computations of the gravitational potential (using scattering amplitude techniques) and noticing some contributions can be absorbed away through field redefinitions. Of course this observation is not exclusive to the $RK$ contribution, but will be true also for any one proportional to at least the Ricci scalar or tensor.

\section{Quadratic in curvature and the Gauss-Bonnet invariant}\label{app:quadratic}

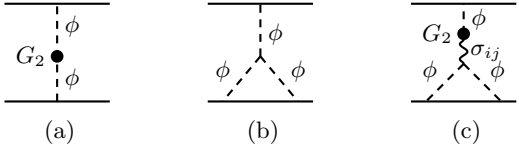
\begin{figure}
	\centering
	\begin{subfigure}{0.3\linewidth}
		{
			\begin{tikzpicture}[thick]
				\draw [black] (0,2) -- (1.4,2);
				\draw [black, dash pattern=on 3pt off 2.5pt] (0.7,0.7) -- (0.7,2);
				\filldraw (0.7,1.3) circle (2pt);
				\draw [black] (0,0.7) -- (1.4,0.7);
				\node[align=left] at (0.9,1.7) {$ϕ$};
				\node[align=left] at (0.35,1.3) {$G_2$};
				\node[align=left] at (0.9,1) {$ϕ$};
			\end{tikzpicture}
		}
		\caption{
			\label{fig:GB-2point}}
	\end{subfigure}
	\begin{subfigure}{0.3\linewidth}
		{
			\begin{tikzpicture}[thick]
				\draw [black] (0,2) -- (1.4,2);
				\draw [black, dash pattern=on 3pt off 2.5pt] (0.7,1.3) -- (0.7,2);
				\draw [black, dash pattern=on 3pt off 2.5pt] (0.7,1.3) -- (0.2,0.7);
				\draw [black, dash pattern=on 3pt off 2.5pt] (0.7,1.3) -- (1.2,0.7);
				\draw [black] (0,0.7) -- (1.4,0.7);
				\node[align=left] at (0.9,1.6) {$ϕ$};
				\node[align=left] at (0.2,1.1) {$ϕ$};
				\node[align=left] at (1.2,1.1) {$ϕ$};
			\end{tikzpicture}
		}
		\caption{
			\label{fig:GB-peace}}
	\end{subfigure}
	\begin{subfigure}{0.3\linewidth}
		{
			\begin{tikzpicture}[thick]
				\draw [black] (0,2) -- (1.4,2);
				\draw [black, dash pattern=on 3pt off 2.5pt] (0.7,1.6) -- (0.7,2);
				\draw [black, decorate, decoration={snake, amplitude=0.5mm, segment length=2mm}] (0.7,1.2) -- (0.7,1.6);
				\filldraw (0.7,1.6) circle (2pt);
				\draw [black, dash pattern=on 3pt off 2.5pt] (0.7,1.2) -- (0.2,0.7);
				\draw [black, dash pattern=on 3pt off 2.5pt] (0.7,1.2) -- (1.2,0.7);
				\draw [black] (0,0.7) -- (1.4,0.7);
				\node[align=left] at (0.9,1.8) {$ϕ$};
				\node[align=left] at (0.35,1.6) {$G_2$};
				\node[align=left] at (1,1.35) {$σ_{ij}$};
				\node[align=left] at (0.25,1.1) {$ϕ$};
				\node[align=left] at (1.15,1.1) {$ϕ$};
			\end{tikzpicture}
		}
		\caption{
			\label{fig:GB-peace+}}
	\end{subfigure}
	\caption{Leading order contribution to the potential in Gauss-Bonnet (GB) gravity. Dotted lines correspond to the KS scalar $ϕ$, the doubled line is the KS tensor $σ_{ij}$, solid lines are the world lines of the particle, and the black dots mark the contribution coming from the quadratic pieces in GB.
	\label{fig:GB-contributions}
	}
\end{figure}

In 4 spacetime dimensions the Gauss-Bonnet invariant
$$G_2=R^2 -4 R_{ab} R^{ab} + R_{abcd} R^{abcd}$$ is a total derivative and thus has no effect.
Moreover, the first two terms can be removed by field redefinitions (in vacuum), suggesting the final term should also vanish. As a consequence, this should hold at all orders in a PN-EFT expansion.
For concreteness, we ignore these general arguments here and instead compute the leading-order contribution to the potential between two point particles explicitly using PN-EFT for the action $S = S_{\textrm{bulk}} + S_\mysc{pp} + S_\mysc{gf}$, where
$$ S_\textrm{bulk} = 2 Λ² \int \sqrt{-g} \left( R + λ_{(2)} G_2 \right)\,,$$
and $S_\mysc{pp}$, and $S_\mysc{gf}$ are the point particle action and the gauge fixing terms defined in \cref{eq:Spp,eq:GF}.
We find that each term in $G_2$ vanishes individually—a result reflected in the cancellation of the corresponding Feynman diagrams in \cref{fig:GB-contributions}.

Leading order corrections to the potential appear at order $l²$. The relevant diagrams include contributions from both the Ricci scalar (GR) and $G_2$. In terms of the Kol-Smolkin variables introduced in \eqref{eq:KS}, $G_2$ generates three vertices: a quadratic interaction in time derivatives of $ϕ$, a cubic $ϕ$ interaction, and a $ϕσ$ interaction. These combine with a cubic $ϕϕσ$ interaction from GR, leading to an exact cancellation. Below, we summarize the calculations step by step.

First, the $R^2$ contribution: expanding $2Λ² λ_{(2)} \sqrt{-g} R²$ to leading order yields
\begin{align*}
	2Λ² λ_{(2)} \sqrt{-g} R² & \supset \frac{16}{Λ} ∇²ϕ \left( ϕ∇²ϕ - (∂ϕ)² \right)λ_{(2)} \\
	   & \quad + 8 ∇²ϕ \left( ∂_i∂_jσ^{ij} - ∇²σ \right)λ_{(2)} \,,
\end{align*}
where $\sqrt{-g} = e^{-2ϕ/Λ} \sqrt{γ} = 1 - 2ϕ/Λ + σ/(2Λ) + \mathcal{O}(1/Λ²)$.
The quadratic vertex in \cref{fig:GB-2point} $\sim \mathcal{O}(v²)$, and vanishes in the static limit—a universal feature for the \cref{fig:GB-2point} diagram for all terms in $G_2$.
The GR vertex for $ϕϕσ$ is
$$ 2Λ² \sqrt{-g} \left( R - \frac{1}{4}Γ^μ Γ_μ \right) \supset \frac{2}{Λ} \left( 2∂_iϕ ∂_jϕ σ^{ij} - (∂ϕ)²σ \right)\,. $$
Combining these, \cref{fig:GB-peace} evaluates to $-4 I$, where
$$ I \coloneqq - \frac{16}{512}\frac{m_1 m_2²}{Λ^4} λ_{(2)}  \int_{p,q} \frac{(p.q)²}{p²q²(p+q)²} e^{ipr}e^{iqr}\,.$$
Meanwhile, \cref{fig:GB-peace+} contributes $+4I$, resulting in
\begin{equation}\label{eq:Rsq}
	R²: \textrm{\cref{fig:GB-2point}} + \textrm{\cref{fig:GB-peace}} + \textrm{\cref{fig:GB-peace+}} = 0 -4I +4I = 0\,.
\end{equation}
Next, doing the same for the $R_{μν}²$ term gives
\begin{align*}
	2Λ² λ_{(2)}  \sqrt{-g} R_{μν}² & \supset \frac{16}{Λ} ∇²ϕ \left( ϕ∇²ϕ - \frac{1}{2} (∂ϕ)² \right)λ_{(2)} \\
	  & \quad + 4 ∇²ϕ \left( ∂_i∂_jσ^{ij} - ∇²σ \right) λ_{(2)} \,.
\end{align*}
Again diagrams \cref{fig:GB-peace} and \cref{fig:GB-peace+} cancel exactly
\begin{equation}\label{eq:Ricsq}
	R_{μν}²: \textrm{\cref{fig:GB-2point}} + \textrm{\cref{fig:GB-peace}} + \textrm{\cref{fig:GB-peace+}} = 0 -2 I + 2 I = 0 \,.
\end{equation}
Finally, for $R_{μνρσ}²$:
\begin{align*}
	2Λ² \sqrt{-g} λ_{(2)}  R_{μν}² & \supset \frac{16}{Λ}\left[ ϕ(∇²ϕ)² -2(∂ϕ)²∇²ϕ   \right.                       \\ \nn
	   & \left. +2ϕ ∂_{ij}ϕ ∂^{ij}ϕ + 4 ∂_i ϕ ∂_j ϕ ∂^{ij}ϕ\right]λ_{(2)}              \\
	   & +\! 8 ∂^i ∂^jϕ \!\left( -∂_i ∂_j σ + 2∂_i ∂_ kσ^k_j \!-\! ∇²σ_{ij} \right)\! λ_{(2)}\,,
\end{align*}
which gives the same Feynman rules for $ϕϕσ$ as for $R²$. Computing the resulting diagrams, we get
\begin{equation}\label{eq:Riemsq}
	R_{μνρσ}²: \textrm{\cref{fig:GB-2point}} + \textrm{\cref{fig:GB-peace}} + \textrm{\cref{fig:GB-peace+}} = 0 -4 I + 4 I = 0\,.
\end{equation}

\section{Relevant mass for the scalings}\label{app:whichmass}

\subsection{Potential contribution}

Let us relax the $m_1 \sim m_2$ assumption made to arrive at expression \eqref{eq:Rp}, and see which mass regulates physical effects for non-equal mass binaries. For simplicity
we can think of the EMRI case.
For a $p$--scalar vertex, there are multiple diagrams which have different number of legs on the two worldlines: in general $n$ legs on $m_1$, and $p-n$ legs on $m_2$. The second term in the potential in \eqref{eq:Rp} is actually $\left(\frac{G\overline{m}}{rc^2}\right)^{p-2}$, where
$$\overline{m}^{p-2} \coloneqq \sum_{n=1}^{p-1} c_n(m_1^{n-1}m_2^{p-n-1} + m_2^{n-1}m_1^{p-n-1})\,,$$ where $c_n$ are numerical factors corresponding to each diagram. This gives
\begin{align}\nn
	\delta V_{(\partial^2h)^p} \sim & \ V_{\textrm {\scshape n}}\cdot\left(\frac{G \overline{m}}{r c^2}\right)^{p-2}\cdot\left(\frac{G {M_{\Lambda}}}{r c^2}\right)^{2p-2}  \overline{\alpha}_{(p)}\,                                        \\\nn
	\sim & \ V_{\textrm {\scshape n}}\cdot\left(\frac{G m}{r c^2}\right)^{p-2} \cdot \left(\frac{G {M_{\Lambda}}}{r c^2}\right)^{2p-2}  \overline{\alpha}_{(p)} \cdot \left(\frac{\overline{m}}{m}\right)^{p-2}\, \\\nn
	\sim & \ V_{\textrm {\scshape n}}\cdot\left(\frac{v^2}{c^2}\right)^{3p-4} \left(\frac{{M_{\Lambda}}}{m}\right)^{2p-2}\cdot \left(\frac{\overline{m}}{m}\right)^{p-2} \overline{\alpha}_{(p)}\,.
\end{align}
If $m_1 = m_2$, then $\overline{m} \sim m$, and we recover \eqref{eq:Rp}.
Individual masses of the binary components appear in two combinations above: $(\overline{m}/m)$, and $(M_Λ/m)$.
In the large mass ratio limit, $q \coloneqq m_1/m_2 \ll 1$, the first of these becomes
\begin{equation}\label{eq:mbar-m-scaling}
	\lim_{m_1 \ll m_2} \left(\frac{\overline{m}}{m}\right)^{p-2} = \sum_{n=1}^{p-1} c_n \left[ q^{n-1} + q^{p-n-1} \right]\,.
\end{equation}
Since $n \in [1,p-1]$, the leading contribution is $\mathcal{O}(1)$, and comes from the boundaries $n=1$, and $n=p-1$; the subleading contributions go as $q^m$, where $m \in ℤ^+$.
The $n=1$ and $n=p-1$ contributions correspond to Feynman diagrams with exactly one leg on either of the two point masses.
The other factor $M_Λ/m$ is independent of the mass ratio, and depends only on the total mass of the system.
Therefore if $m_1 \ll m_2$,
\begin{equation}
	\lim_{m_1 \ll m_2} \delta V_{(\partial^2h)^p} \sim
	V_{\textrm {\scshape n}}\cdot\left(\frac{v^2}{c^2}\right)^{3p-4} \left(\frac{{M_{\Lambda}}}{m}\right)^{2p-2} \overline{\alpha}_{(p)}\,.
\end{equation}
Thus, \emph{the correction to the potential is governed by the
total mass of the system roughly as in the comparable mass case.} Next, we will see that this is not the case for the tidal contribution.

The discussion in \cref{sec:rules-Rp-scalar,sec:rules-der} can be generalized for unequal masses in an analogous way. By defining the corresponding $\overline{m}$ in those cases, we once again conclude that at leading order in mass ratio, $\overline{m}/m \sim \mathcal{O}(1)$, analogous to \cref{eq:mbar-m-scaling}.

\subsection{Tidal contribution}
A BH in a higher derivative theory has a TLN of the form $λ_i \sim k_i m_i^5$, where $k_i \sim \left(M_Λ/m_i\right)^{2p-2}$. The effective TLN of a BH binary, $Λ$, appears in the phase of the GW as
\begin{equation}
	ψ_{\textrm{tidal}}(f) \sim ψ_{\textrm{\scshape n}} \frac{Λ}{m^5} \left(\frac{v}{c}\right)^{10}\,,
\end{equation}
where the effective TLN $Λ$ is
\begin{equation}
	Λ \sim \left( 1 + \frac{12}{q} \right) k_1 m_1^5 + \left( 1 + 12q \right)k_2 m_2^5\,.
\end{equation}
To assess what dominates the contribution to the effective TLN, let's consider the EMRI limit ($m_1 \ll m_2 \Rightarrow q \ll 1$),
\begin{equation}
	\lim_{q \ll 1} \frac{Λ}{m^5} = \left( \frac{M_Λ}{m_2} \right)^{2p-2} \frac{1}{q^{2p-6}}\,.
\end{equation}
Since $p\geq 3$, the exponents are non-negative. This suggests
\begin{align}
	\lim_{m_1 \ll m_2} ψ_{\textrm{tidal}} & \sim ψ_{\textrm{\scshape n}} \left( \frac{M_Λ}{m_2} \right)^{2p-2} \left(\frac{m_2}{m_1}\right)^{2p-6} \left(\frac{v}{c}\right)^{10} \nn \\
    & \sim ψ_{\textrm{\scshape n}} \left(\frac{M_Λ}{m_1} \right)^{2p-2} \left(\frac{m_1}{m_2}\right)^4 \left(\frac{v}{c}\right)^{10} \nn       \\
    & \sim ψ_{\textrm{\scshape n}} \left(\frac{M_Λ}{m} \right)^{2p-2} \frac{(1+q)^{2p-2}}{q^{2p-6}} \left(\frac{v}{c}\right)^{10}
\end{align}
Thus, \emph{tidal contributions, in the unequal mass case, will be dominated by the lightest mass in the binary.} This is intuitively expected
since such mass will have the largest curvature at the horizon.

\bibliography{refs}

\begin{thebibliography}{81}%
\makeatletter
\providecommand \@ifxundefined [1]{%
 \@ifx{#1\undefined}
}%
\providecommand \@ifnum [1]{%
 \ifnum #1\expandafter \@firstoftwo
 \else \expandafter \@secondoftwo
 \fi
}%
\providecommand \@ifx [1]{%
 \ifx #1\expandafter \@firstoftwo
 \else \expandafter \@secondoftwo
 \fi
}%
\providecommand \natexlab [1]{#1}%
\providecommand \enquote  [1]{``#1''}%
\providecommand \bibnamefont  [1]{#1}%
\providecommand \bibfnamefont [1]{#1}%
\providecommand \citenamefont [1]{#1}%
\providecommand \href@noop [0]{\@secondoftwo}%
\providecommand \href [0]{\begingroup \@sanitize@url \@href}%
\providecommand \@href[1]{\@@startlink{#1}\@@href}%
\providecommand \@@href[1]{\endgroup#1\@@endlink}%
\providecommand \@sanitize@url [0]{\catcode `\\12\catcode `\$12\catcode
  `\&12\catcode `\#12\catcode `\^12\catcode `\_12\catcode `\%12\relax}%
\providecommand \@@startlink[1]{}%
\providecommand \@@endlink[0]{}%
\providecommand \url  [0]{\begingroup\@sanitize@url \@url }%
\providecommand \@url [1]{\endgroup\@href {#1}{\urlprefix }}%
\providecommand \urlprefix  [0]{URL }%
\providecommand \Eprint [0]{\href }%
\providecommand \doibase [0]{https://doi.org/}%
\providecommand \selectlanguage [0]{\@gobble}%
\providecommand \bibinfo  [0]{\@secondoftwo}%
\providecommand \bibfield  [0]{\@secondoftwo}%
\providecommand \translation [1]{[#1]}%
\providecommand \BibitemOpen [0]{}%
\providecommand \bibitemStop [0]{}%
\providecommand \bibitemNoStop [0]{.\EOS\space}%
\providecommand \EOS [0]{\spacefactor3000\relax}%
\providecommand \BibitemShut  [1]{\csname bibitem#1\endcsname}%
\let\auto@bib@innerbib\@empty
\bibitem [{\citenamefont {Berti}\ \emph {et~al.}(2015)\citenamefont {Berti}
  \emph {et~al.}}]{Berti:2015itd}%
  \BibitemOpen
  \bibfield  {author} {\bibinfo {author} {\bibfnamefont {E.}~\bibnamefont
  {Berti}} \emph {et~al.},\ }\bibfield  {title} {\bibinfo {title} {{Testing
  General Relativity with Present and Future Astrophysical Observations}},\
  }\href {https://doi.org/10.1088/0264-9381/32/24/243001} {\bibfield  {journal}
  {\bibinfo  {journal} {Class. Quant. Grav.}\ }\textbf {\bibinfo {volume}
  {32}},\ \bibinfo {pages} {243001} (\bibinfo {year} {2015})},\ \Eprint
  {https://arxiv.org/abs/1501.07274} {arXiv:1501.07274 [gr-qc]} \BibitemShut
  {NoStop}%
\bibitem [{\citenamefont {Yunes}\ and\ \citenamefont
  {Siemens}(2013)}]{Yunes:2013dva}%
  \BibitemOpen
  \bibfield  {author} {\bibinfo {author} {\bibfnamefont {N.}~\bibnamefont
  {Yunes}}\ and\ \bibinfo {author} {\bibfnamefont {X.}~\bibnamefont
  {Siemens}},\ }\bibfield  {title} {\bibinfo {title} {{Gravitational-Wave Tests
  of General Relativity with Ground-Based Detectors and Pulsar
  Timing-Arrays}},\ }\href {https://doi.org/10.12942/lrr-2013-9} {\bibfield
  {journal} {\bibinfo  {journal} {Living Rev. Rel.}\ }\textbf {\bibinfo
  {volume} {16}},\ \bibinfo {pages} {9} (\bibinfo {year} {2013})},\ \Eprint
  {https://arxiv.org/abs/1304.3473} {arXiv:1304.3473 [gr-qc]} \BibitemShut
  {NoStop}%
\bibitem [{\citenamefont {Shankaranarayanan}\ and\ \citenamefont
  {Johnson}(2022)}]{Shankaranarayanan:2022wbx}%
  \BibitemOpen
  \bibfield  {author} {\bibinfo {author} {\bibfnamefont {S.}~\bibnamefont
  {Shankaranarayanan}}\ and\ \bibinfo {author} {\bibfnamefont {J.~P.}\
  \bibnamefont {Johnson}},\ }\bibfield  {title} {\bibinfo {title} {{Modified
  theories of gravity: Why, how and what?}},\ }\href
  {https://doi.org/10.1007/s10714-022-02927-2} {\bibfield  {journal} {\bibinfo
  {journal} {Gen. Rel. Grav.}\ }\textbf {\bibinfo {volume} {54}},\ \bibinfo
  {pages} {44} (\bibinfo {year} {2022})},\ \Eprint
  {https://arxiv.org/abs/2204.06533} {arXiv:2204.06533 [gr-qc]} \BibitemShut
  {NoStop}%
\bibitem [{\citenamefont {Abbott}\ \emph {et~al.}(2016)\citenamefont {Abbott}
  \emph {et~al.}}]{LIGOScientific:2016lio}%
  \BibitemOpen
  \bibfield  {author} {\bibinfo {author} {\bibfnamefont {B.~P.}\ \bibnamefont
  {Abbott}} \emph {et~al.} (\bibinfo {collaboration} {LIGO Scientific,
  Virgo}),\ }\bibfield  {title} {\bibinfo {title} {{Tests of general relativity
  with GW150914}},\ }\href {https://doi.org/10.1103/PhysRevLett.116.221101}
  {\bibfield  {journal} {\bibinfo  {journal} {Phys. Rev. Lett.}\ }\textbf
  {\bibinfo {volume} {116}},\ \bibinfo {pages} {221101} (\bibinfo {year}
  {2016})},\ \bibinfo {note} {[Erratum: Phys.Rev.Lett. 121, 129902 (2018)]},\
  \Eprint {https://arxiv.org/abs/1602.03841} {arXiv:1602.03841 [gr-qc]}
  \BibitemShut {NoStop}%
\bibitem [{\citenamefont {Abbott}\ \emph
  {et~al.}(2019{\natexlab{a}})\citenamefont {Abbott} \emph
  {et~al.}}]{LIGOScientific:2018dkp}%
  \BibitemOpen
  \bibfield  {author} {\bibinfo {author} {\bibfnamefont {B.~P.}\ \bibnamefont
  {Abbott}} \emph {et~al.} (\bibinfo {collaboration} {LIGO Scientific,
  Virgo}),\ }\bibfield  {title} {\bibinfo {title} {{Tests of General Relativity
  with GW170817}},\ }\href {https://doi.org/10.1103/PhysRevLett.123.011102}
  {\bibfield  {journal} {\bibinfo  {journal} {Phys. Rev. Lett.}\ }\textbf
  {\bibinfo {volume} {123}},\ \bibinfo {pages} {011102} (\bibinfo {year}
  {2019}{\natexlab{a}})},\ \Eprint {https://arxiv.org/abs/1811.00364}
  {arXiv:1811.00364 [gr-qc]} \BibitemShut {NoStop}%
\bibitem [{\citenamefont {Abbott}\ \emph
  {et~al.}(2019{\natexlab{b}})\citenamefont {Abbott} \emph
  {et~al.}}]{LIGOScientific:2019fpa}%
  \BibitemOpen
  \bibfield  {author} {\bibinfo {author} {\bibfnamefont {B.~P.}\ \bibnamefont
  {Abbott}} \emph {et~al.} (\bibinfo {collaboration} {LIGO Scientific,
  Virgo}),\ }\bibfield  {title} {\bibinfo {title} {{Tests of General Relativity
  with the Binary Black Hole Signals from the LIGO-Virgo Catalog GWTC-1}},\
  }\href {https://doi.org/10.1103/PhysRevD.100.104036} {\bibfield  {journal}
  {\bibinfo  {journal} {Phys. Rev. D}\ }\textbf {\bibinfo {volume} {100}},\
  \bibinfo {pages} {104036} (\bibinfo {year} {2019}{\natexlab{b}})},\ \Eprint
  {https://arxiv.org/abs/1903.04467} {arXiv:1903.04467 [gr-qc]} \BibitemShut
  {NoStop}%
\bibitem [{\citenamefont {Abbott}\ \emph
  {et~al.}(2021{\natexlab{a}})\citenamefont {Abbott} \emph
  {et~al.}}]{LIGOScientific:2020tif}%
  \BibitemOpen
  \bibfield  {author} {\bibinfo {author} {\bibfnamefont {R.}~\bibnamefont
  {Abbott}} \emph {et~al.} (\bibinfo {collaboration} {LIGO Scientific,
  Virgo}),\ }\bibfield  {title} {\bibinfo {title} {{Tests of general relativity
  with binary black holes from the second LIGO-Virgo gravitational-wave
  transient catalog}},\ }\href {https://doi.org/10.1103/PhysRevD.103.122002}
  {\bibfield  {journal} {\bibinfo  {journal} {Phys. Rev. D}\ }\textbf {\bibinfo
  {volume} {103}},\ \bibinfo {pages} {122002} (\bibinfo {year}
  {2021}{\natexlab{a}})},\ \Eprint {https://arxiv.org/abs/2010.14529}
  {arXiv:2010.14529 [gr-qc]} \BibitemShut {NoStop}%
\bibitem [{\citenamefont {Abbott}\ \emph
  {et~al.}(2021{\natexlab{b}})\citenamefont {Abbott} \emph
  {et~al.}}]{LIGOScientific:2021sio}%
  \BibitemOpen
  \bibfield  {author} {\bibinfo {author} {\bibfnamefont {R.}~\bibnamefont
  {Abbott}} \emph {et~al.} (\bibinfo {collaboration} {LIGO Scientific, VIRGO,
  KAGRA}),\ }\bibfield  {title} {\bibinfo {title} {{Tests of General Relativity
  with GWTC-3}},\ }\href@noop {} {\  (\bibinfo {year} {2021}{\natexlab{b}})},\
  \Eprint {https://arxiv.org/abs/2112.06861} {arXiv:2112.06861 [gr-qc]}
  \BibitemShut {NoStop}%
\bibitem [{\citenamefont {Perkins}\ and\ \citenamefont
  {Yunes}(2022)}]{Perkins:2022fhr}%
  \BibitemOpen
  \bibfield  {author} {\bibinfo {author} {\bibfnamefont {S.}~\bibnamefont
  {Perkins}}\ and\ \bibinfo {author} {\bibfnamefont {N.}~\bibnamefont
  {Yunes}},\ }\bibfield  {title} {\bibinfo {title} {{Are parametrized tests of
  general relativity with gravitational waves robust to unknown higher
  post-Newtonian order effects?}},\ }\href
  {https://doi.org/10.1103/PhysRevD.105.124047} {\bibfield  {journal} {\bibinfo
   {journal} {Phys. Rev. D}\ }\textbf {\bibinfo {volume} {105}},\ \bibinfo
  {pages} {124047} (\bibinfo {year} {2022})},\ \Eprint
  {https://arxiv.org/abs/2201.02542} {arXiv:2201.02542 [gr-qc]} \BibitemShut
  {NoStop}%
\bibitem [{\citenamefont {Mehta}\ \emph {et~al.}(2023)\citenamefont {Mehta},
  \citenamefont {Buonanno}, \citenamefont {Cotesta}, \citenamefont {Ghosh},
  \citenamefont {Sennett},\ and\ \citenamefont {Steinhoff}}]{Mehta:2022pcn}%
  \BibitemOpen
  \bibfield  {author} {\bibinfo {author} {\bibfnamefont {A.~K.}\ \bibnamefont
  {Mehta}}, \bibinfo {author} {\bibfnamefont {A.}~\bibnamefont {Buonanno}},
  \bibinfo {author} {\bibfnamefont {R.}~\bibnamefont {Cotesta}}, \bibinfo
  {author} {\bibfnamefont {A.}~\bibnamefont {Ghosh}}, \bibinfo {author}
  {\bibfnamefont {N.}~\bibnamefont {Sennett}},\ and\ \bibinfo {author}
  {\bibfnamefont {J.}~\bibnamefont {Steinhoff}},\ }\bibfield  {title} {\bibinfo
  {title} {{Tests of general relativity with gravitational-wave observations
  using a flexible theory-independent method}},\ }\href
  {https://doi.org/10.1103/PhysRevD.107.044020} {\bibfield  {journal} {\bibinfo
   {journal} {Phys. Rev. D}\ }\textbf {\bibinfo {volume} {107}},\ \bibinfo
  {pages} {044020} (\bibinfo {year} {2023})},\ \Eprint
  {https://arxiv.org/abs/2203.13937} {arXiv:2203.13937 [gr-qc]} \BibitemShut
  {NoStop}%
\bibitem [{\citenamefont {Kov\'acs}\ and\ \citenamefont
  {Reall}(2020)}]{Kovacs:2020ywu}%
  \BibitemOpen
  \bibfield  {author} {\bibinfo {author} {\bibfnamefont {A.~D.}\ \bibnamefont
  {Kov\'acs}}\ and\ \bibinfo {author} {\bibfnamefont {H.~S.}\ \bibnamefont
  {Reall}},\ }\bibfield  {title} {\bibinfo {title} {{Well-posed formulation of
  Lovelock and Horndeski theories}},\ }\href
  {https://doi.org/10.1103/PhysRevD.101.124003} {\bibfield  {journal} {\bibinfo
   {journal} {Phys. Rev. D}\ }\textbf {\bibinfo {volume} {101}},\ \bibinfo
  {pages} {124003} (\bibinfo {year} {2020})},\ \Eprint
  {https://arxiv.org/abs/2003.08398} {arXiv:2003.08398 [gr-qc]} \BibitemShut
  {NoStop}%
\bibitem [{\citenamefont {East}\ and\ \citenamefont
  {Ripley}(2021)}]{East:2020hgw}%
  \BibitemOpen
  \bibfield  {author} {\bibinfo {author} {\bibfnamefont {W.~E.}\ \bibnamefont
  {East}}\ and\ \bibinfo {author} {\bibfnamefont {J.~L.}\ \bibnamefont
  {Ripley}},\ }\bibfield  {title} {\bibinfo {title} {{Evolution of
  Einstein-scalar-Gauss-Bonnet gravity using a modified harmonic
  formulation}},\ }\href {https://doi.org/10.1103/PhysRevD.103.044040}
  {\bibfield  {journal} {\bibinfo  {journal} {Phys. Rev. D}\ }\textbf {\bibinfo
  {volume} {103}},\ \bibinfo {pages} {044040} (\bibinfo {year} {2021})},\
  \Eprint {https://arxiv.org/abs/2011.03547} {arXiv:2011.03547 [gr-qc]}
  \BibitemShut {NoStop}%
\bibitem [{\citenamefont {Cayuso}\ \emph {et~al.}(2023)\citenamefont {Cayuso},
  \citenamefont {Figueras}, \citenamefont {Fran{\c{c}}a},\ and\ \citenamefont
  {Lehner}}]{Cayuso:2023xbc}%
  \BibitemOpen
  \bibfield  {author} {\bibinfo {author} {\bibfnamefont {R.}~\bibnamefont
  {Cayuso}}, \bibinfo {author} {\bibfnamefont {P.}~\bibnamefont {Figueras}},
  \bibinfo {author} {\bibfnamefont {T.}~\bibnamefont {Fran{\c{c}}a}},\ and\
  \bibinfo {author} {\bibfnamefont {L.}~\bibnamefont {Lehner}},\ }\bibfield
  {title} {\bibinfo {title} {{Self-Consistent Modeling of Gravitational
  Theories beyond General Relativity}},\ }\href
  {https://doi.org/10.1103/PhysRevLett.131.111403} {\bibfield  {journal}
  {\bibinfo  {journal} {Phys. Rev. Lett.}\ }\textbf {\bibinfo {volume} {131}},\
  \bibinfo {pages} {111403} (\bibinfo {year} {2023})},\ \Eprint
  {https://arxiv.org/abs/2303.07246} {arXiv:2303.07246 [gr-qc]} \BibitemShut
  {NoStop}%
\bibitem [{\citenamefont {Corman}\ \emph {et~al.}(2024)\citenamefont {Corman},
  \citenamefont {Lehner}, \citenamefont {East},\ and\ \citenamefont
  {Dideron}}]{Corman:2024cdr}%
  \BibitemOpen
  \bibfield  {author} {\bibinfo {author} {\bibfnamefont {M.}~\bibnamefont
  {Corman}}, \bibinfo {author} {\bibfnamefont {L.}~\bibnamefont {Lehner}},
  \bibinfo {author} {\bibfnamefont {W.~E.}\ \bibnamefont {East}},\ and\
  \bibinfo {author} {\bibfnamefont {G.}~\bibnamefont {Dideron}},\ }\bibfield
  {title} {\bibinfo {title} {{Nonlinear studies of modifications to general
  relativity: Comparing different approaches}},\ }\href
  {https://doi.org/10.1103/PhysRevD.110.084048} {\bibfield  {journal} {\bibinfo
   {journal} {Phys. Rev. D}\ }\textbf {\bibinfo {volume} {110}},\ \bibinfo
  {pages} {084048} (\bibinfo {year} {2024})},\ \Eprint
  {https://arxiv.org/abs/2405.15581} {arXiv:2405.15581 [gr-qc]} \BibitemShut
  {NoStop}%
\bibitem [{\citenamefont {Figueras}\ \emph {et~al.}(2024)\citenamefont
  {Figueras}, \citenamefont {Held},\ and\ \citenamefont
  {Kov\'acs}}]{Figueras:2024bba}%
  \BibitemOpen
  \bibfield  {author} {\bibinfo {author} {\bibfnamefont {P.}~\bibnamefont
  {Figueras}}, \bibinfo {author} {\bibfnamefont {A.}~\bibnamefont {Held}},\
  and\ \bibinfo {author} {\bibfnamefont {A.~D.}\ \bibnamefont {Kov\'acs}},\
  }\bibfield  {title} {\bibinfo {title} {{Well-posed initial value formulation
  of general effective field theories of gravity}},\ }\href@noop {} {\
  (\bibinfo {year} {2024})},\ \Eprint {https://arxiv.org/abs/2407.08775}
  {arXiv:2407.08775 [gr-qc]} \BibitemShut {NoStop}%
\bibitem [{\citenamefont {Donoghue}(1995)}]{Donoghue:1995cz}%
  \BibitemOpen
  \bibfield  {author} {\bibinfo {author} {\bibfnamefont {J.~F.}\ \bibnamefont
  {Donoghue}},\ }\bibfield  {title} {\bibinfo {title} {{Introduction to the
  effective field theory description of gravity}},\ }in\ \href@noop {} {\emph
  {\bibinfo {booktitle} {{Advanced School on Effective Theories}}}}\ (\bibinfo
  {year} {1995})\ \Eprint {https://arxiv.org/abs/gr-qc/9512024}
  {arXiv:gr-qc/9512024} \BibitemShut {NoStop}%
\bibitem [{\citenamefont {Carrillo~Gonzalez}\ \emph {et~al.}(2022)\citenamefont
  {Carrillo~Gonzalez}, \citenamefont {de~Rham}, \citenamefont {Pozsgay},\ and\
  \citenamefont {Tolley}}]{CarrilloGonzalez:2022fwg}%
  \BibitemOpen
  \bibfield  {author} {\bibinfo {author} {\bibfnamefont {M.}~\bibnamefont
  {Carrillo~Gonzalez}}, \bibinfo {author} {\bibfnamefont {C.}~\bibnamefont
  {de~Rham}}, \bibinfo {author} {\bibfnamefont {V.}~\bibnamefont {Pozsgay}},\
  and\ \bibinfo {author} {\bibfnamefont {A.~J.}\ \bibnamefont {Tolley}},\
  }\bibfield  {title} {\bibinfo {title} {{Causal effective field theories}},\
  }\href {https://doi.org/10.1103/PhysRevD.106.105018} {\bibfield  {journal}
  {\bibinfo  {journal} {Phys. Rev. D}\ }\textbf {\bibinfo {volume} {106}},\
  \bibinfo {pages} {105018} (\bibinfo {year} {2022})},\ \Eprint
  {https://arxiv.org/abs/2207.03491} {arXiv:2207.03491 [hep-th]} \BibitemShut
  {NoStop}%
\bibitem [{\citenamefont {Stein}\ and\ \citenamefont
  {Yagi}(2014)}]{Stein:2013wza}%
  \BibitemOpen
  \bibfield  {author} {\bibinfo {author} {\bibfnamefont {L.~C.}\ \bibnamefont
  {Stein}}\ and\ \bibinfo {author} {\bibfnamefont {K.}~\bibnamefont {Yagi}},\
  }\bibfield  {title} {\bibinfo {title} {{Parametrizing and constraining scalar
  corrections to general relativity}},\ }\href
  {https://doi.org/10.1103/PhysRevD.89.044026} {\bibfield  {journal} {\bibinfo
  {journal} {Phys. Rev. D}\ }\textbf {\bibinfo {volume} {89}},\ \bibinfo
  {pages} {044026} (\bibinfo {year} {2014})},\ \Eprint
  {https://arxiv.org/abs/1310.6743} {arXiv:1310.6743 [gr-qc]} \BibitemShut
  {NoStop}%
\bibitem [{\citenamefont {Carullo}(2021)}]{Carullo:2021dui}%
  \BibitemOpen
  \bibfield  {author} {\bibinfo {author} {\bibfnamefont {G.}~\bibnamefont
  {Carullo}},\ }\bibfield  {title} {\bibinfo {title} {{Enhancing modified
  gravity detection from gravitational-wave observations using the parametrized
  ringdown spin expansion coeffcients formalism}},\ }\href
  {https://doi.org/10.1103/PhysRevD.103.124043} {\bibfield  {journal} {\bibinfo
   {journal} {Phys. Rev. D}\ }\textbf {\bibinfo {volume} {103}},\ \bibinfo
  {pages} {124043} (\bibinfo {year} {2021})},\ \Eprint
  {https://arxiv.org/abs/2102.05939} {arXiv:2102.05939 [gr-qc]} \BibitemShut
  {NoStop}%
\bibitem [{\citenamefont {Maselli}\ \emph {et~al.}(2024)\citenamefont
  {Maselli}, \citenamefont {Yi}, \citenamefont {Pierini}, \citenamefont
  {Vellucci}, \citenamefont {Reali}, \citenamefont {Gualtieri},\ and\
  \citenamefont {Berti}}]{Maselli:2023khq}%
  \BibitemOpen
  \bibfield  {author} {\bibinfo {author} {\bibfnamefont {A.}~\bibnamefont
  {Maselli}}, \bibinfo {author} {\bibfnamefont {S.}~\bibnamefont {Yi}},
  \bibinfo {author} {\bibfnamefont {L.}~\bibnamefont {Pierini}}, \bibinfo
  {author} {\bibfnamefont {V.}~\bibnamefont {Vellucci}}, \bibinfo {author}
  {\bibfnamefont {L.}~\bibnamefont {Reali}}, \bibinfo {author} {\bibfnamefont
  {L.}~\bibnamefont {Gualtieri}},\ and\ \bibinfo {author} {\bibfnamefont
  {E.}~\bibnamefont {Berti}},\ }\bibfield  {title} {\bibinfo {title} {{Black
  hole spectroscopy beyond Kerr: Agnostic and theory-based tests with
  next-generation interferometers}},\ }\href
  {https://doi.org/10.1103/PhysRevD.109.064060} {\bibfield  {journal} {\bibinfo
   {journal} {Phys. Rev. D}\ }\textbf {\bibinfo {volume} {109}},\ \bibinfo
  {pages} {064060} (\bibinfo {year} {2024})},\ \Eprint
  {https://arxiv.org/abs/2311.14803} {arXiv:2311.14803 [gr-qc]} \BibitemShut
  {NoStop}%
\bibitem [{\citenamefont {Dideron}\ \emph {et~al.}(2023)\citenamefont
  {Dideron}, \citenamefont {Mukherjee},\ and\ \citenamefont
  {Lehner}}]{Dideron:2022tap}%
  \BibitemOpen
  \bibfield  {author} {\bibinfo {author} {\bibfnamefont {G.}~\bibnamefont
  {Dideron}}, \bibinfo {author} {\bibfnamefont {S.}~\bibnamefont {Mukherjee}},\
  and\ \bibinfo {author} {\bibfnamefont {L.}~\bibnamefont {Lehner}},\
  }\bibfield  {title} {\bibinfo {title} {{New framework to study unmodeled
  physics from gravitational wave data}},\ }\href
  {https://doi.org/10.1103/PhysRevD.107.104023} {\bibfield  {journal} {\bibinfo
   {journal} {Phys. Rev. D}\ }\textbf {\bibinfo {volume} {107}},\ \bibinfo
  {pages} {104023} (\bibinfo {year} {2023})},\ \Eprint
  {https://arxiv.org/abs/2209.14321} {arXiv:2209.14321 [gr-qc]} \BibitemShut
  {NoStop}%
\bibitem [{\citenamefont {Payne}\ \emph {et~al.}(2024)\citenamefont {Payne},
  \citenamefont {Isi}, \citenamefont {Chatziioannou}, \citenamefont {Lehner},
  \citenamefont {Chen},\ and\ \citenamefont {Farr}}]{Payne:2024yhk}%
  \BibitemOpen
  \bibfield  {author} {\bibinfo {author} {\bibfnamefont {E.}~\bibnamefont
  {Payne}}, \bibinfo {author} {\bibfnamefont {M.}~\bibnamefont {Isi}}, \bibinfo
  {author} {\bibfnamefont {K.}~\bibnamefont {Chatziioannou}}, \bibinfo {author}
  {\bibfnamefont {L.}~\bibnamefont {Lehner}}, \bibinfo {author} {\bibfnamefont
  {Y.}~\bibnamefont {Chen}},\ and\ \bibinfo {author} {\bibfnamefont {W.~M.}\
  \bibnamefont {Farr}},\ }\bibfield  {title} {\bibinfo {title} {{Curvature
  Dependence of Gravitational-Wave Tests of General Relativity}},\ }\href
  {https://doi.org/10.1103/PhysRevLett.133.251401} {\bibfield  {journal}
  {\bibinfo  {journal} {Phys. Rev. Lett.}\ }\textbf {\bibinfo {volume} {133}},\
  \bibinfo {pages} {251401} (\bibinfo {year} {2024})},\ \Eprint
  {https://arxiv.org/abs/2407.07043} {arXiv:2407.07043 [gr-qc]} \BibitemShut
  {NoStop}%
\bibitem [{\citenamefont {Goldberger}\ and\ \citenamefont
  {Rothstein}(2006)}]{Goldberger:2004jt}%
  \BibitemOpen
  \bibfield  {author} {\bibinfo {author} {\bibfnamefont {W.~D.}\ \bibnamefont
  {Goldberger}}\ and\ \bibinfo {author} {\bibfnamefont {I.~Z.}\ \bibnamefont
  {Rothstein}},\ }\bibfield  {title} {\bibinfo {title} {{An Effective field
  theory of gravity for extended objects}},\ }\href
  {https://doi.org/10.1103/PhysRevD.73.104029} {\bibfield  {journal} {\bibinfo
  {journal} {Phys. Rev. D}\ }\textbf {\bibinfo {volume} {73}},\ \bibinfo
  {pages} {104029} (\bibinfo {year} {2006})},\ \Eprint
  {https://arxiv.org/abs/hep-th/0409156} {arXiv:hep-th/0409156} \BibitemShut
  {NoStop}%
\bibitem [{\citenamefont {Porto}(2016)}]{Porto:2016pyg}%
  \BibitemOpen
  \bibfield  {author} {\bibinfo {author} {\bibfnamefont {R.~A.}\ \bibnamefont
  {Porto}},\ }\bibfield  {title} {\bibinfo {title} {{The effective field
  theorist{\textquoteright}s approach to gravitational dynamics}},\ }\href
  {https://doi.org/10.1016/j.physrep.2016.04.003} {\bibfield  {journal}
  {\bibinfo  {journal} {Phys. Rept.}\ }\textbf {\bibinfo {volume} {633}},\
  \bibinfo {pages} {1} (\bibinfo {year} {2016})},\ \Eprint
  {https://arxiv.org/abs/1601.04914} {arXiv:1601.04914 [hep-th]} \BibitemShut
  {NoStop}%
\bibitem [{\citenamefont {Gupta}\ \emph {et~al.}(2024)\citenamefont {Gupta}
  \emph {et~al.}}]{Gupta:2024gun}%
  \BibitemOpen
  \bibfield  {author} {\bibinfo {author} {\bibfnamefont {A.}~\bibnamefont
  {Gupta}} \emph {et~al.},\ }\bibfield  {title} {\bibinfo {title} {{Possible
  Causes of False General Relativity Violations in Gravitational Wave
  Observations}},\ }\bibfield  {journal} {\bibinfo  {journal} {SciPost
  Phys.Comm.Rep}\ }\href {https://doi.org/10.21468/SciPostPhysCommRep.5}
  {10.21468/SciPostPhysCommRep.5} (\bibinfo {year} {2024}),\ \Eprint
  {https://arxiv.org/abs/2405.02197} {arXiv:2405.02197 [gr-qc]} \BibitemShut
  {NoStop}%
\bibitem [{\citenamefont {Essick}\ \emph {et~al.}(2017)\citenamefont {Essick},
  \citenamefont {Vitale},\ and\ \citenamefont {Evans}}]{Essick:2017wyl}%
  \BibitemOpen
  \bibfield  {author} {\bibinfo {author} {\bibfnamefont {R.}~\bibnamefont
  {Essick}}, \bibinfo {author} {\bibfnamefont {S.}~\bibnamefont {Vitale}},\
  and\ \bibinfo {author} {\bibfnamefont {M.}~\bibnamefont {Evans}},\ }\bibfield
   {title} {\bibinfo {title} {{Frequency-dependent responses in third
  generation gravitational-wave detectors}},\ }\href
  {https://doi.org/10.1103/PhysRevD.96.084004} {\bibfield  {journal} {\bibinfo
  {journal} {Phys. Rev. D}\ }\textbf {\bibinfo {volume} {96}},\ \bibinfo
  {pages} {084004} (\bibinfo {year} {2017})},\ \Eprint
  {https://arxiv.org/abs/1708.06843} {arXiv:1708.06843 [gr-qc]} \BibitemShut
  {NoStop}%
\bibitem [{\citenamefont {Kanti}\ \emph {et~al.}(1996)\citenamefont {Kanti},
  \citenamefont {Mavromatos}, \citenamefont {Rizos}, \citenamefont {Tamvakis},\
  and\ \citenamefont {Winstanley}}]{Kanti:1995vq}%
  \BibitemOpen
  \bibfield  {author} {\bibinfo {author} {\bibfnamefont {P.}~\bibnamefont
  {Kanti}}, \bibinfo {author} {\bibfnamefont {N.~E.}\ \bibnamefont
  {Mavromatos}}, \bibinfo {author} {\bibfnamefont {J.}~\bibnamefont {Rizos}},
  \bibinfo {author} {\bibfnamefont {K.}~\bibnamefont {Tamvakis}},\ and\
  \bibinfo {author} {\bibfnamefont {E.}~\bibnamefont {Winstanley}},\ }\bibfield
   {title} {\bibinfo {title} {{Dilatonic black holes in higher curvature string
  gravity}},\ }\href {https://doi.org/10.1103/PhysRevD.54.5049} {\bibfield
  {journal} {\bibinfo  {journal} {Phys. Rev. D}\ }\textbf {\bibinfo {volume}
  {54}},\ \bibinfo {pages} {5049} (\bibinfo {year} {1996})},\ \Eprint
  {https://arxiv.org/abs/hep-th/9511071} {arXiv:hep-th/9511071} \BibitemShut
  {NoStop}%
\bibitem [{\citenamefont {Alexander}\ and\ \citenamefont
  {Yunes}(2009)}]{Alexander:2009tp}%
  \BibitemOpen
  \bibfield  {author} {\bibinfo {author} {\bibfnamefont {S.}~\bibnamefont
  {Alexander}}\ and\ \bibinfo {author} {\bibfnamefont {N.}~\bibnamefont
  {Yunes}},\ }\bibfield  {title} {\bibinfo {title} {{Chern-Simons Modified
  General Relativity}},\ }\href {https://doi.org/10.1016/j.physrep.2009.07.002}
  {\bibfield  {journal} {\bibinfo  {journal} {Phys. Rept.}\ }\textbf {\bibinfo
  {volume} {480}},\ \bibinfo {pages} {1} (\bibinfo {year} {2009})},\ \Eprint
  {https://arxiv.org/abs/0907.2562} {arXiv:0907.2562 [hep-th]} \BibitemShut
  {NoStop}%
\bibitem [{\citenamefont {Bueno}\ and\ \citenamefont
  {Cano}(2016)}]{Bueno:2016xff}%
  \BibitemOpen
  \bibfield  {author} {\bibinfo {author} {\bibfnamefont {P.}~\bibnamefont
  {Bueno}}\ and\ \bibinfo {author} {\bibfnamefont {P.~A.}\ \bibnamefont
  {Cano}},\ }\bibfield  {title} {\bibinfo {title} {{Einsteinian cubic
  gravity}},\ }\href {https://doi.org/10.1103/PhysRevD.94.104005} {\bibfield
  {journal} {\bibinfo  {journal} {Phys. Rev. D}\ }\textbf {\bibinfo {volume}
  {94}},\ \bibinfo {pages} {104005} (\bibinfo {year} {2016})},\ \Eprint
  {https://arxiv.org/abs/1607.06463} {arXiv:1607.06463 [hep-th]} \BibitemShut
  {NoStop}%
\bibitem [{\citenamefont {Silva}\ \emph {et~al.}(2023)\citenamefont {Silva},
  \citenamefont {Ghosh},\ and\ \citenamefont {Buonanno}}]{Silva:2022srr}%
  \BibitemOpen
  \bibfield  {author} {\bibinfo {author} {\bibfnamefont {H.~O.}\ \bibnamefont
  {Silva}}, \bibinfo {author} {\bibfnamefont {A.}~\bibnamefont {Ghosh}},\ and\
  \bibinfo {author} {\bibfnamefont {A.}~\bibnamefont {Buonanno}},\ }\bibfield
  {title} {\bibinfo {title} {{Black-hole ringdown as a probe of
  higher-curvature gravity theories}},\ }\href
  {https://doi.org/10.1103/PhysRevD.107.044030} {\bibfield  {journal} {\bibinfo
   {journal} {Phys. Rev. D}\ }\textbf {\bibinfo {volume} {107}},\ \bibinfo
  {pages} {044030} (\bibinfo {year} {2023})},\ \Eprint
  {https://arxiv.org/abs/2205.05132} {arXiv:2205.05132 [gr-qc]} \BibitemShut
  {NoStop}%
\bibitem [{\citenamefont {Maenaut}\ \emph {et~al.}(2024)\citenamefont
  {Maenaut}, \citenamefont {Carullo}, \citenamefont {Cano}, \citenamefont
  {Liu}, \citenamefont {Cardoso}, \citenamefont {Hertog},\ and\ \citenamefont
  {Li}}]{Maenaut:2024oci}%
  \BibitemOpen
  \bibfield  {author} {\bibinfo {author} {\bibfnamefont {S.}~\bibnamefont
  {Maenaut}}, \bibinfo {author} {\bibfnamefont {G.}~\bibnamefont {Carullo}},
  \bibinfo {author} {\bibfnamefont {P.~A.}\ \bibnamefont {Cano}}, \bibinfo
  {author} {\bibfnamefont {A.}~\bibnamefont {Liu}}, \bibinfo {author}
  {\bibfnamefont {V.}~\bibnamefont {Cardoso}}, \bibinfo {author} {\bibfnamefont
  {T.}~\bibnamefont {Hertog}},\ and\ \bibinfo {author} {\bibfnamefont
  {T.~G.~F.}\ \bibnamefont {Li}},\ }\bibfield  {title} {\bibinfo {title}
  {{Ringdown Analysis of Rotating Black Holes in Effective Field Theory
  Extensions of General Relativity}},\ }\href@noop {} {\  (\bibinfo {year}
  {2024})},\ \Eprint {https://arxiv.org/abs/2411.17893} {arXiv:2411.17893
  [gr-qc]} \BibitemShut {NoStop}%
\bibitem [{\citenamefont {Foffa}\ and\ \citenamefont
  {Sturani}(2014)}]{Foffa:2013qca}%
  \BibitemOpen
  \bibfield  {author} {\bibinfo {author} {\bibfnamefont {S.}~\bibnamefont
  {Foffa}}\ and\ \bibinfo {author} {\bibfnamefont {R.}~\bibnamefont
  {Sturani}},\ }\bibfield  {title} {\bibinfo {title} {{Effective field theory
  methods to model compact binaries}},\ }\href
  {https://doi.org/10.1088/0264-9381/31/4/043001} {\bibfield  {journal}
  {\bibinfo  {journal} {Class. Quant. Grav.}\ }\textbf {\bibinfo {volume}
  {31}},\ \bibinfo {pages} {043001} (\bibinfo {year} {2014})},\ \Eprint
  {https://arxiv.org/abs/1309.3474} {arXiv:1309.3474 [gr-qc]} \BibitemShut
  {NoStop}%
\bibitem [{\citenamefont {Levi}(2020)}]{Levi:2018nxp}%
  \BibitemOpen
  \bibfield  {author} {\bibinfo {author} {\bibfnamefont {M.}~\bibnamefont
  {Levi}},\ }\bibfield  {title} {\bibinfo {title} {{Effective Field Theories of
  Post-Newtonian Gravity: A comprehensive review}},\ }\href
  {https://doi.org/10.1088/1361-6633/ab12bc} {\bibfield  {journal} {\bibinfo
  {journal} {Rept. Prog. Phys.}\ }\textbf {\bibinfo {volume} {83}},\ \bibinfo
  {pages} {075901} (\bibinfo {year} {2020})},\ \Eprint
  {https://arxiv.org/abs/1807.01699} {arXiv:1807.01699 [hep-th]} \BibitemShut
  {NoStop}%
\bibitem [{\citenamefont {Kol}\ and\ \citenamefont
  {Smolkin}(2008)}]{Kol:2007bc}%
  \BibitemOpen
  \bibfield  {author} {\bibinfo {author} {\bibfnamefont {B.}~\bibnamefont
  {Kol}}\ and\ \bibinfo {author} {\bibfnamefont {M.}~\bibnamefont {Smolkin}},\
  }\bibfield  {title} {\bibinfo {title} {{Non-Relativistic Gravitation: From
  Newton to Einstein and Back}},\ }\href
  {https://doi.org/10.1088/0264-9381/25/14/145011} {\bibfield  {journal}
  {\bibinfo  {journal} {Class. Quant. Grav.}\ }\textbf {\bibinfo {volume}
  {25}},\ \bibinfo {pages} {145011} (\bibinfo {year} {2008})},\ \Eprint
  {https://arxiv.org/abs/0712.4116} {arXiv:0712.4116 [hep-th]} \BibitemShut
  {NoStop}%
\bibitem [{\citenamefont {Damour}\ and\ \citenamefont
  {Nagar}(2009)}]{Damour:2009vw}%
  \BibitemOpen
  \bibfield  {author} {\bibinfo {author} {\bibfnamefont {T.}~\bibnamefont
  {Damour}}\ and\ \bibinfo {author} {\bibfnamefont {A.}~\bibnamefont {Nagar}},\
  }\bibfield  {title} {\bibinfo {title} {{Relativistic tidal properties of
  neutron stars}},\ }\href {https://doi.org/10.1103/PhysRevD.80.084035}
  {\bibfield  {journal} {\bibinfo  {journal} {Phys. Rev. D}\ }\textbf {\bibinfo
  {volume} {80}},\ \bibinfo {pages} {084035} (\bibinfo {year} {2009})},\
  \Eprint {https://arxiv.org/abs/0906.0096} {arXiv:0906.0096 [gr-qc]}
  \BibitemShut {NoStop}%
\bibitem [{\citenamefont {Binnington}\ and\ \citenamefont
  {Poisson}(2009)}]{Binnington:2009bb}%
  \BibitemOpen
  \bibfield  {author} {\bibinfo {author} {\bibfnamefont {T.}~\bibnamefont
  {Binnington}}\ and\ \bibinfo {author} {\bibfnamefont {E.}~\bibnamefont
  {Poisson}},\ }\bibfield  {title} {\bibinfo {title} {{Relativistic theory of
  tidal Love numbers}},\ }\href {https://doi.org/10.1103/PhysRevD.80.084018}
  {\bibfield  {journal} {\bibinfo  {journal} {Phys. Rev. D}\ }\textbf {\bibinfo
  {volume} {80}},\ \bibinfo {pages} {084018} (\bibinfo {year} {2009})},\
  \Eprint {https://arxiv.org/abs/0906.1366} {arXiv:0906.1366 [gr-qc]}
  \BibitemShut {NoStop}%
\bibitem [{\citenamefont {Kol}\ and\ \citenamefont
  {Smolkin}(2012)}]{Kol:2011vg}%
  \BibitemOpen
  \bibfield  {author} {\bibinfo {author} {\bibfnamefont {B.}~\bibnamefont
  {Kol}}\ and\ \bibinfo {author} {\bibfnamefont {M.}~\bibnamefont {Smolkin}},\
  }\bibfield  {title} {\bibinfo {title} {{Black hole stereotyping: Induced
  gravito-static polarization}},\ }\href
  {https://doi.org/10.1007/JHEP02(2012)010} {\bibfield  {journal} {\bibinfo
  {journal} {JHEP}\ }\textbf {\bibinfo {volume} {02}},\ \bibinfo {pages}
  {010}},\ \Eprint {https://arxiv.org/abs/1110.3764} {arXiv:1110.3764 [hep-th]}
  \BibitemShut {NoStop}%
\bibitem [{\citenamefont {Thorne}(1980)}]{Thorne:1980ru}%
  \BibitemOpen
  \bibfield  {author} {\bibinfo {author} {\bibfnamefont {K.~S.}\ \bibnamefont
  {Thorne}},\ }\bibfield  {title} {\bibinfo {title} {{Multipole Expansions of
  Gravitational Radiation}},\ }\href
  {https://doi.org/10.1103/RevModPhys.52.299} {\bibfield  {journal} {\bibinfo
  {journal} {Rev. Mod. Phys.}\ }\textbf {\bibinfo {volume} {52}},\ \bibinfo
  {pages} {299} (\bibinfo {year} {1980})}\BibitemShut {NoStop}%
\bibitem [{\citenamefont {Endlich}\ \emph {et~al.}(2017)\citenamefont
  {Endlich}, \citenamefont {Gorbenko}, \citenamefont {Huang},\ and\
  \citenamefont {Senatore}}]{Endlich:2017tqa}%
  \BibitemOpen
  \bibfield  {author} {\bibinfo {author} {\bibfnamefont {S.}~\bibnamefont
  {Endlich}}, \bibinfo {author} {\bibfnamefont {V.}~\bibnamefont {Gorbenko}},
  \bibinfo {author} {\bibfnamefont {J.}~\bibnamefont {Huang}},\ and\ \bibinfo
  {author} {\bibfnamefont {L.}~\bibnamefont {Senatore}},\ }\bibfield  {title}
  {\bibinfo {title} {{An effective formalism for testing extensions to General
  Relativity with gravitational waves}},\ }\href
  {https://doi.org/10.1007/JHEP09(2017)122} {\bibfield  {journal} {\bibinfo
  {journal} {JHEP}\ }\textbf {\bibinfo {volume} {2017}}\bibfield  {number}
  {\bibinfo  {number} { (09)},\ \bibinfo {pages} {122}},\ }\Eprint
  {https://arxiv.org/abs/1704.01590} {arXiv:1704.01590 [gr-qc]} \BibitemShut
  {NoStop}%
\bibitem [{\citenamefont {Diedrichs}\ \emph {et~al.}(2024)\citenamefont
  {Diedrichs}, \citenamefont {Schmitt},\ and\ \citenamefont
  {Sagunski}}]{Diedrichs:2023foj}%
  \BibitemOpen
  \bibfield  {author} {\bibinfo {author} {\bibfnamefont {R.~F.}\ \bibnamefont
  {Diedrichs}}, \bibinfo {author} {\bibfnamefont {D.}~\bibnamefont {Schmitt}},\
  and\ \bibinfo {author} {\bibfnamefont {L.}~\bibnamefont {Sagunski}},\
  }\bibfield  {title} {\bibinfo {title} {{Binary systems in massive
  scalar-tensor theories: Next-to-leading order gravitational wave phase from
  effective field theory}},\ }\href
  {https://doi.org/10.1103/PhysRevD.110.104073} {\bibfield  {journal} {\bibinfo
   {journal} {Phys. Rev. D}\ }\textbf {\bibinfo {volume} {110}},\ \bibinfo
  {pages} {104073} (\bibinfo {year} {2024})},\ \Eprint
  {https://arxiv.org/abs/2311.04274} {arXiv:2311.04274 [gr-qc]} \BibitemShut
  {NoStop}%
\bibitem [{\citenamefont {Almeida}\ and\ \citenamefont
  {Zhou}(2024)}]{Almeida:2024cqz}%
  \BibitemOpen
  \bibfield  {author} {\bibinfo {author} {\bibfnamefont {G.~L.}\ \bibnamefont
  {Almeida}}\ and\ \bibinfo {author} {\bibfnamefont {S.-Y.}\ \bibnamefont
  {Zhou}},\ }\bibfield  {title} {\bibinfo {title} {{Post-Newtonian dynamics of
  spinning black hole binaries in Einstein-scalar-Gauss-Bonnet gravity}},\
  }\href {https://doi.org/10.1103/PhysRevD.110.124016} {\bibfield  {journal}
  {\bibinfo  {journal} {Phys. Rev. D}\ }\textbf {\bibinfo {volume} {110}},\
  \bibinfo {pages} {124016} (\bibinfo {year} {2024})},\ \Eprint
  {https://arxiv.org/abs/2408.14196} {arXiv:2408.14196 [gr-qc]} \BibitemShut
  {NoStop}%
\bibitem [{\citenamefont {Lins}\ and\ \citenamefont
  {Sturani}(2021)}]{Lins:2020omt}%
  \BibitemOpen
  \bibfield  {author} {\bibinfo {author} {\bibfnamefont {A.~N.}\ \bibnamefont
  {Lins}}\ and\ \bibinfo {author} {\bibfnamefont {R.}~\bibnamefont {Sturani}},\
  }\bibfield  {title} {\bibinfo {title} {{Effects of Short-Distance
  Modifications to General Relativity in Spinning Binary Systems}},\ }\href
  {https://doi.org/10.1103/PhysRevD.103.084030} {\bibfield  {journal} {\bibinfo
   {journal} {Phys. Rev. D}\ }\textbf {\bibinfo {volume} {103}},\ \bibinfo
  {pages} {084030} (\bibinfo {year} {2021})},\ \Eprint
  {https://arxiv.org/abs/2011.02124} {arXiv:2011.02124 [gr-qc]} \BibitemShut
  {NoStop}%
\bibitem [{\citenamefont {Cano}\ and\ \citenamefont
  {Ruip\'erez}(2019)}]{Cano:2019ore}%
  \BibitemOpen
  \bibfield  {author} {\bibinfo {author} {\bibfnamefont {P.~A.}\ \bibnamefont
  {Cano}}\ and\ \bibinfo {author} {\bibfnamefont {A.}~\bibnamefont
  {Ruip\'erez}},\ }\bibfield  {title} {\bibinfo {title} {{Leading
  higher-derivative corrections to Kerr geometry}},\ }\href
  {https://doi.org/10.1007/JHEP05(2019)189} {\bibfield  {journal} {\bibinfo
  {journal} {JHEP}\ }\textbf {\bibinfo {volume} {05}},\ \bibinfo {pages}
  {189}},\ \bibinfo {note} {[Erratum: JHEP 03, 187 (2020)]},\ \Eprint
  {https://arxiv.org/abs/1901.01315} {arXiv:1901.01315 [gr-qc]} \BibitemShut
  {NoStop}%
\bibitem [{\citenamefont {Emond}\ and\ \citenamefont
  {Moynihan}(2019)}]{Emond:2019crr}%
  \BibitemOpen
  \bibfield  {author} {\bibinfo {author} {\bibfnamefont {W.~T.}\ \bibnamefont
  {Emond}}\ and\ \bibinfo {author} {\bibfnamefont {N.}~\bibnamefont
  {Moynihan}},\ }\bibfield  {title} {\bibinfo {title} {{Scattering Amplitudes,
  Black Holes and Leading Singularities in Cubic Theories of Gravity}},\ }\href
  {https://doi.org/10.1007/JHEP12(2019)019} {\bibfield  {journal} {\bibinfo
  {journal} {JHEP}\ }\textbf {\bibinfo {volume} {12}},\ \bibinfo {pages}
  {019}},\ \Eprint {https://arxiv.org/abs/1905.08213} {arXiv:1905.08213
  [hep-th]} \BibitemShut {NoStop}%
\bibitem [{\citenamefont {Brandhuber}\ and\ \citenamefont
  {Travaglini}(2020)}]{Brandhuber:2019qpg}%
  \BibitemOpen
  \bibfield  {author} {\bibinfo {author} {\bibfnamefont {A.}~\bibnamefont
  {Brandhuber}}\ and\ \bibinfo {author} {\bibfnamefont {G.}~\bibnamefont
  {Travaglini}},\ }\bibfield  {title} {\bibinfo {title} {{On higher-derivative
  effects on the gravitational potential and particle bending}},\ }\href
  {https://doi.org/10.1007/JHEP01(2020)010} {\bibfield  {journal} {\bibinfo
  {journal} {JHEP}\ }\textbf {\bibinfo {volume} {01}},\ \bibinfo {pages}
  {010}},\ \Eprint {https://arxiv.org/abs/1905.05657} {arXiv:1905.05657
  [hep-th]} \BibitemShut {NoStop}%
\bibitem [{\citenamefont {Accettulli~Huber}\ \emph {et~al.}(2021)\citenamefont
  {Accettulli~Huber}, \citenamefont {Brandhuber}, \citenamefont {De~Angelis},\
  and\ \citenamefont {Travaglini}}]{AccettulliHuber:2020dal}%
  \BibitemOpen
  \bibfield  {author} {\bibinfo {author} {\bibfnamefont {M.}~\bibnamefont
  {Accettulli~Huber}}, \bibinfo {author} {\bibfnamefont {A.}~\bibnamefont
  {Brandhuber}}, \bibinfo {author} {\bibfnamefont {S.}~\bibnamefont
  {De~Angelis}},\ and\ \bibinfo {author} {\bibfnamefont {G.}~\bibnamefont
  {Travaglini}},\ }\bibfield  {title} {\bibinfo {title} {{From amplitudes to
  gravitational radiation with cubic interactions and tidal effects}},\ }\href
  {https://doi.org/10.1103/PhysRevD.103.045015} {\bibfield  {journal} {\bibinfo
   {journal} {Phys. Rev. D}\ }\textbf {\bibinfo {volume} {103}},\ \bibinfo
  {pages} {045015} (\bibinfo {year} {2021})},\ \Eprint
  {https://arxiv.org/abs/2012.06548} {arXiv:2012.06548 [hep-th]} \BibitemShut
  {NoStop}%
\bibitem [{\citenamefont {Wilson-Gerow}(2025)}]{Wilson-Gerow:2025xhr}%
  \BibitemOpen
  \bibfield  {author} {\bibinfo {author} {\bibfnamefont {J.}~\bibnamefont
  {Wilson-Gerow}},\ }\bibfield  {title} {\bibinfo {title} {{Conservative
  Dynamics of Relativistic Binaries Beyond Einstein Gravity}},\ }\href@noop {}
  {\  (\bibinfo {year} {2025})},\ \Eprint {https://arxiv.org/abs/2503.02867}
  {arXiv:2503.02867 [hep-th]} \BibitemShut {NoStop}%
\bibitem [{\citenamefont {Cannella}\ \emph {et~al.}(2009)\citenamefont
  {Cannella}, \citenamefont {Foffa}, \citenamefont {Maggiore}, \citenamefont
  {Sanctuary},\ and\ \citenamefont {Sturani}}]{Cannella:2009he}%
  \BibitemOpen
  \bibfield  {author} {\bibinfo {author} {\bibfnamefont {U.}~\bibnamefont
  {Cannella}}, \bibinfo {author} {\bibfnamefont {S.}~\bibnamefont {Foffa}},
  \bibinfo {author} {\bibfnamefont {M.}~\bibnamefont {Maggiore}}, \bibinfo
  {author} {\bibfnamefont {H.}~\bibnamefont {Sanctuary}},\ and\ \bibinfo
  {author} {\bibfnamefont {R.}~\bibnamefont {Sturani}},\ }\bibfield  {title}
  {\bibinfo {title} {{Extracting the three and four-graviton vertices from
  binary pulsars and coalescing binaries}},\ }\href
  {https://doi.org/10.1103/PhysRevD.80.124035} {\bibfield  {journal} {\bibinfo
  {journal} {Phys. Rev. D}\ }\textbf {\bibinfo {volume} {80}},\ \bibinfo
  {pages} {124035} (\bibinfo {year} {2009})},\ \Eprint
  {https://arxiv.org/abs/0907.2186} {arXiv:0907.2186 [gr-qc]} \BibitemShut
  {NoStop}%
\bibitem [{\citenamefont {Sotiriou}\ and\ \citenamefont
  {Zhou}(2014)}]{Sotiriou:2013qea}%
  \BibitemOpen
  \bibfield  {author} {\bibinfo {author} {\bibfnamefont {T.~P.}\ \bibnamefont
  {Sotiriou}}\ and\ \bibinfo {author} {\bibfnamefont {S.-Y.}\ \bibnamefont
  {Zhou}},\ }\bibfield  {title} {\bibinfo {title} {{Black hole hair in
  generalized scalar-tensor gravity}},\ }\href
  {https://doi.org/10.1103/PhysRevLett.112.251102} {\bibfield  {journal}
  {\bibinfo  {journal} {Phys. Rev. Lett.}\ }\textbf {\bibinfo {volume} {112}},\
  \bibinfo {pages} {251102} (\bibinfo {year} {2014})},\ \Eprint
  {https://arxiv.org/abs/1312.3622} {arXiv:1312.3622 [gr-qc]} \BibitemShut
  {NoStop}%
\bibitem [{\citenamefont {Juli{\'e}}\ and\ \citenamefont
  {Berti}(2019)}]{Julie:2019sab}%
  \BibitemOpen
  \bibfield  {author} {\bibinfo {author} {\bibfnamefont {F.-L.}\ \bibnamefont
  {Juli{\'e}}}\ and\ \bibinfo {author} {\bibfnamefont {E.}~\bibnamefont
  {Berti}},\ }\bibfield  {title} {\bibinfo {title} {{Post-Newtonian dynamics
  and black hole thermodynamics in Einstein-scalar-Gauss-Bonnet gravity}},\
  }\href {https://doi.org/10.1103/PhysRevD.100.104061} {\bibfield  {journal}
  {\bibinfo  {journal} {Phys. Rev. D}\ }\textbf {\bibinfo {volume} {100}},\
  \bibinfo {pages} {104061} (\bibinfo {year} {2019})},\ \Eprint
  {https://arxiv.org/abs/1909.05258} {arXiv:1909.05258 [gr-qc]} \BibitemShut
  {NoStop}%
\bibitem [{\citenamefont {Juli\'e}\ \emph {et~al.}(2022)\citenamefont
  {Juli\'e}, \citenamefont {Silva}, \citenamefont {Berti},\ and\ \citenamefont
  {Yunes}}]{Julie:2022huo}%
  \BibitemOpen
  \bibfield  {author} {\bibinfo {author} {\bibfnamefont {F.-L.}\ \bibnamefont
  {Juli\'e}}, \bibinfo {author} {\bibfnamefont {H.~O.}\ \bibnamefont {Silva}},
  \bibinfo {author} {\bibfnamefont {E.}~\bibnamefont {Berti}},\ and\ \bibinfo
  {author} {\bibfnamefont {N.}~\bibnamefont {Yunes}},\ }\bibfield  {title}
  {\bibinfo {title} {{Black hole sensitivities in Einstein-scalar-Gauss-Bonnet
  gravity}},\ }\href {https://doi.org/10.1103/PhysRevD.105.124031} {\bibfield
  {journal} {\bibinfo  {journal} {Phys. Rev. D}\ }\textbf {\bibinfo {volume}
  {105}},\ \bibinfo {pages} {124031} (\bibinfo {year} {2022})},\ \Eprint
  {https://arxiv.org/abs/2202.01329} {arXiv:2202.01329 [gr-qc]} \BibitemShut
  {NoStop}%
\bibitem [{\citenamefont {Juli{\'e}}(2023)}]{Julie:2023ncq}%
  \BibitemOpen
  \bibfield  {author} {\bibinfo {author} {\bibfnamefont {F.-L.}\ \bibnamefont
  {Juli{\'e}}},\ }\bibfield  {title} {\bibinfo {title} {{Dynamical
  scalarization in Schwarzschild binary inspirals}},\ }\href@noop {} {\
  (\bibinfo {year} {2023})},\ \Eprint {https://arxiv.org/abs/2312.16764}
  {arXiv:2312.16764 [gr-qc]} \BibitemShut {NoStop}%
\bibitem [{\citenamefont {Bernard}(2020)}]{Bernard:2019yfz}%
  \BibitemOpen
  \bibfield  {author} {\bibinfo {author} {\bibfnamefont {L.}~\bibnamefont
  {Bernard}},\ }\bibfield  {title} {\bibinfo {title} {{Dipolar tidal effects in
  scalar-tensor theories}},\ }\href
  {https://doi.org/10.1103/PhysRevD.101.021501} {\bibfield  {journal} {\bibinfo
   {journal} {Phys. Rev. D}\ }\textbf {\bibinfo {volume} {101}},\ \bibinfo
  {pages} {021501} (\bibinfo {year} {2020})},\ \bibinfo {note} {[Erratum:
  Phys.Rev.D 107, 069901 (2023)]},\ \Eprint {https://arxiv.org/abs/1906.10735}
  {arXiv:1906.10735 [gr-qc]} \BibitemShut {NoStop}%
\bibitem [{\citenamefont {Kuntz}\ \emph {et~al.}(2019)\citenamefont {Kuntz},
  \citenamefont {Piazza},\ and\ \citenamefont {Vernizzi}}]{Kuntz:2019zef}%
  \BibitemOpen
  \bibfield  {author} {\bibinfo {author} {\bibfnamefont {A.}~\bibnamefont
  {Kuntz}}, \bibinfo {author} {\bibfnamefont {F.}~\bibnamefont {Piazza}},\ and\
  \bibinfo {author} {\bibfnamefont {F.}~\bibnamefont {Vernizzi}},\ }\bibfield
  {title} {\bibinfo {title} {{Effective field theory for gravitational
  radiation in scalar-tensor gravity}},\ }\href
  {https://doi.org/10.1088/1475-7516/2019/05/052} {\bibfield  {journal}
  {\bibinfo  {journal} {JCAP}\ }\textbf {\bibinfo {volume} {05}},\ \bibinfo
  {pages} {052}},\ \Eprint {https://arxiv.org/abs/1902.04941} {arXiv:1902.04941
  [gr-qc]} \BibitemShut {NoStop}%
\bibitem [{\citenamefont {Ruhdorfer}\ \emph {et~al.}(2020)\citenamefont
  {Ruhdorfer}, \citenamefont {Serra},\ and\ \citenamefont
  {Weiler}}]{Ruhdorfer:2019qmk}%
  \BibitemOpen
  \bibfield  {author} {\bibinfo {author} {\bibfnamefont {M.}~\bibnamefont
  {Ruhdorfer}}, \bibinfo {author} {\bibfnamefont {J.}~\bibnamefont {Serra}},\
  and\ \bibinfo {author} {\bibfnamefont {A.}~\bibnamefont {Weiler}},\
  }\bibfield  {title} {\bibinfo {title} {{Effective Field Theory of Gravity to
  All Orders}},\ }\href {https://doi.org/10.1007/JHEP05(2020)083} {\bibfield
  {journal} {\bibinfo  {journal} {JHEP}\ }\textbf {\bibinfo {volume} {05}},\
  \bibinfo {pages} {083}},\ \Eprint {https://arxiv.org/abs/1908.08050}
  {arXiv:1908.08050 [hep-ph]} \BibitemShut {NoStop}%
\bibitem [{\citenamefont {Aguilar-Gutierrez}\ \emph {et~al.}(2023)\citenamefont
  {Aguilar-Gutierrez}, \citenamefont {Bueno}, \citenamefont {Cano},
  \citenamefont {Hennigar},\ and\ \citenamefont
  {Llorens}}]{Aguilar-Gutierrez:2023kfn}%
  \BibitemOpen
  \bibfield  {author} {\bibinfo {author} {\bibfnamefont {S.~E.}\ \bibnamefont
  {Aguilar-Gutierrez}}, \bibinfo {author} {\bibfnamefont {P.}~\bibnamefont
  {Bueno}}, \bibinfo {author} {\bibfnamefont {P.~A.}\ \bibnamefont {Cano}},
  \bibinfo {author} {\bibfnamefont {R.~A.}\ \bibnamefont {Hennigar}},\ and\
  \bibinfo {author} {\bibfnamefont {Q.}~\bibnamefont {Llorens}},\ }\bibfield
  {title} {\bibinfo {title} {{Aspects of higher-curvature gravities with
  covariant derivatives}},\ }\href
  {https://doi.org/10.1103/PhysRevD.108.124075} {\bibfield  {journal} {\bibinfo
   {journal} {Phys. Rev. D}\ }\textbf {\bibinfo {volume} {108}},\ \bibinfo
  {pages} {124075} (\bibinfo {year} {2023})},\ \Eprint
  {https://arxiv.org/abs/2310.09333} {arXiv:2310.09333 [hep-th]} \BibitemShut
  {NoStop}%
\bibitem [{\citenamefont {{Horndeski}}(1974)}]{1974IJTP...10..363H}%
  \BibitemOpen
  \bibfield  {author} {\bibinfo {author} {\bibfnamefont {G.~W.}\ \bibnamefont
  {{Horndeski}}},\ }\bibfield  {title} {\bibinfo {title} {{Second-order
  scalar-tensor field equations in a four-dimensional space}},\ }\href
  {https://doi.org/10.1007/BF01807638} {\bibfield  {journal} {\bibinfo
  {journal} {International Journal of Theoretical Physics}\ }\textbf {\bibinfo
  {volume} {10}},\ \bibinfo {pages} {363} (\bibinfo {year} {1974})}\BibitemShut
  {NoStop}%
\bibitem [{\citenamefont {de~Rham}\ and\ \citenamefont
  {Tolley}(2020{\natexlab{a}})}]{deRham:2020zyh}%
  \BibitemOpen
  \bibfield  {author} {\bibinfo {author} {\bibfnamefont {C.}~\bibnamefont
  {de~Rham}}\ and\ \bibinfo {author} {\bibfnamefont {A.~J.}\ \bibnamefont
  {Tolley}},\ }\bibfield  {title} {\bibinfo {title} {{Causality in curved
  spacetimes: The speed of light and gravity}},\ }\href
  {https://doi.org/10.1103/PhysRevD.102.084048} {\bibfield  {journal} {\bibinfo
   {journal} {Phys. Rev. D}\ }\textbf {\bibinfo {volume} {102}},\ \bibinfo
  {pages} {084048} (\bibinfo {year} {2020}{\natexlab{a}})},\ \Eprint
  {https://arxiv.org/abs/2007.01847} {arXiv:2007.01847 [hep-th]} \BibitemShut
  {NoStop}%
\bibitem [{\citenamefont {Solomon}\ and\ \citenamefont
  {Trodden}(2018)}]{Solomon:2017nlh}%
  \BibitemOpen
  \bibfield  {author} {\bibinfo {author} {\bibfnamefont {A.~R.}\ \bibnamefont
  {Solomon}}\ and\ \bibinfo {author} {\bibfnamefont {M.}~\bibnamefont
  {Trodden}},\ }\bibfield  {title} {\bibinfo {title} {{Higher-derivative
  operators and effective field theory for general scalar-tensor theories}},\
  }\href {https://doi.org/10.1088/1475-7516/2018/02/031} {\bibfield  {journal}
  {\bibinfo  {journal} {JCAP}\ }\textbf {\bibinfo {volume} {02}},\ \bibinfo
  {pages} {031}},\ \Eprint {https://arxiv.org/abs/1709.09695} {arXiv:1709.09695
  [hep-th]} \BibitemShut {NoStop}%
\bibitem [{\citenamefont {Held}\ and\ \citenamefont
  {Lim}(2021)}]{Held:2021pht}%
  \BibitemOpen
  \bibfield  {author} {\bibinfo {author} {\bibfnamefont {A.}~\bibnamefont
  {Held}}\ and\ \bibinfo {author} {\bibfnamefont {H.}~\bibnamefont {Lim}},\
  }\bibfield  {title} {\bibinfo {title} {{Nonlinear dynamics of quadratic
  gravity in spherical symmetry}},\ }\href
  {https://doi.org/10.1103/PhysRevD.104.084075} {\bibfield  {journal} {\bibinfo
   {journal} {Phys. Rev. D}\ }\textbf {\bibinfo {volume} {104}},\ \bibinfo
  {pages} {084075} (\bibinfo {year} {2021})},\ \Eprint
  {https://arxiv.org/abs/2104.04010} {arXiv:2104.04010 [gr-qc]} \BibitemShut
  {NoStop}%
\bibitem [{\citenamefont {Doneva}\ \emph {et~al.}(2023)\citenamefont {Doneva},
  \citenamefont {Arest{\'e}~Sal{\'o}}, \citenamefont {Clough}, \citenamefont
  {Figueras},\ and\ \citenamefont {Yazadjiev}}]{Doneva:2023oww}%
  \BibitemOpen
  \bibfield  {author} {\bibinfo {author} {\bibfnamefont {D.~D.}\ \bibnamefont
  {Doneva}}, \bibinfo {author} {\bibfnamefont {L.}~\bibnamefont
  {Arest{\'e}~Sal{\'o}}}, \bibinfo {author} {\bibfnamefont {K.}~\bibnamefont
  {Clough}}, \bibinfo {author} {\bibfnamefont {P.}~\bibnamefont {Figueras}},\
  and\ \bibinfo {author} {\bibfnamefont {S.~S.}\ \bibnamefont {Yazadjiev}},\
  }\bibfield  {title} {\bibinfo {title} {{Testing the limits of
  scalar-Gauss-Bonnet gravity through nonlinear evolutions of spin-induced
  scalarization}},\ }\href {https://doi.org/10.1103/PhysRevD.108.084017}
  {\bibfield  {journal} {\bibinfo  {journal} {Phys. Rev. D}\ }\textbf {\bibinfo
  {volume} {108}},\ \bibinfo {pages} {084017} (\bibinfo {year} {2023})},\
  \Eprint {https://arxiv.org/abs/2307.06474} {arXiv:2307.06474 [gr-qc]}
  \BibitemShut {NoStop}%
\bibitem [{\citenamefont {Martin-Garcia}\ \emph {et~al.}(2008)\citenamefont
  {Martin-Garcia}, \citenamefont {Yllanes},\ and\ \citenamefont
  {Portugal}}]{Martin-Garcia:2008yei}%
  \BibitemOpen
  \bibfield  {author} {\bibinfo {author} {\bibfnamefont {J.~M.}\ \bibnamefont
  {Martin-Garcia}}, \bibinfo {author} {\bibfnamefont {D.}~\bibnamefont
  {Yllanes}},\ and\ \bibinfo {author} {\bibfnamefont {R.}~\bibnamefont
  {Portugal}},\ }\bibfield  {title} {\bibinfo {title} {{The Invar tensor
  package: Differential invariants of Riemann}},\ }\href
  {https://doi.org/10.1016/j.cpc.2008.04.018} {\bibfield  {journal} {\bibinfo
  {journal} {Comput. Phys. Commun.}\ }\textbf {\bibinfo {volume} {179}},\
  \bibinfo {pages} {586} (\bibinfo {year} {2008})},\ \Eprint
  {https://arxiv.org/abs/0802.1274} {arXiv:0802.1274 [cs.SC]} \BibitemShut
  {NoStop}%
\bibitem [{\citenamefont {de~Rham}\ and\ \citenamefont
  {Tolley}(2020{\natexlab{b}})}]{deRham:2019ctd}%
  \BibitemOpen
  \bibfield  {author} {\bibinfo {author} {\bibfnamefont {C.}~\bibnamefont
  {de~Rham}}\ and\ \bibinfo {author} {\bibfnamefont {A.~J.}\ \bibnamefont
  {Tolley}},\ }\bibfield  {title} {\bibinfo {title} {{Speed of gravity}},\
  }\href {https://doi.org/10.1103/PhysRevD.101.063518} {\bibfield  {journal}
  {\bibinfo  {journal} {Phys. Rev. D}\ }\textbf {\bibinfo {volume} {101}},\
  \bibinfo {pages} {063518} (\bibinfo {year} {2020}{\natexlab{b}})},\ \Eprint
  {https://arxiv.org/abs/1909.00881} {arXiv:1909.00881 [hep-th]} \BibitemShut
  {NoStop}%
\bibitem [{\citenamefont {Yagi}\ \emph
  {et~al.}(2012{\natexlab{a}})\citenamefont {Yagi}, \citenamefont {Stein},
  \citenamefont {Yunes},\ and\ \citenamefont {Tanaka}}]{Yagi:2011xp}%
  \BibitemOpen
  \bibfield  {author} {\bibinfo {author} {\bibfnamefont {K.}~\bibnamefont
  {Yagi}}, \bibinfo {author} {\bibfnamefont {L.~C.}\ \bibnamefont {Stein}},
  \bibinfo {author} {\bibfnamefont {N.}~\bibnamefont {Yunes}},\ and\ \bibinfo
  {author} {\bibfnamefont {T.}~\bibnamefont {Tanaka}},\ }\bibfield  {title}
  {\bibinfo {title} {{Post-Newtonian, Quasi-Circular Binary Inspirals in
  Quadratic Modified Gravity}},\ }\href
  {https://doi.org/10.1103/PhysRevD.85.064022} {\bibfield  {journal} {\bibinfo
  {journal} {Phys. Rev. D}\ }\textbf {\bibinfo {volume} {85}},\ \bibinfo
  {pages} {064022} (\bibinfo {year} {2012}{\natexlab{a}})},\ \bibinfo {note}
  {[Erratum: Phys.Rev.D 93, 029902 (2016)]},\ \Eprint
  {https://arxiv.org/abs/1110.5950} {arXiv:1110.5950 [gr-qc]} \BibitemShut
  {NoStop}%
\bibitem [{\citenamefont {Shiralilou}\ \emph {et~al.}(2022)\citenamefont
  {Shiralilou}, \citenamefont {Hinderer}, \citenamefont {Nissanke},
  \citenamefont {Ortiz},\ and\ \citenamefont {Witek}}]{Shiralilou:2021mfl}%
  \BibitemOpen
  \bibfield  {author} {\bibinfo {author} {\bibfnamefont {B.}~\bibnamefont
  {Shiralilou}}, \bibinfo {author} {\bibfnamefont {T.}~\bibnamefont
  {Hinderer}}, \bibinfo {author} {\bibfnamefont {S.~M.}\ \bibnamefont
  {Nissanke}}, \bibinfo {author} {\bibfnamefont {N.}~\bibnamefont {Ortiz}},\
  and\ \bibinfo {author} {\bibfnamefont {H.}~\bibnamefont {Witek}},\ }\bibfield
   {title} {\bibinfo {title} {{Post-Newtonian gravitational and scalar waves in
  scalar-Gauss{\textendash}Bonnet gravity}},\ }\href
  {https://doi.org/10.1088/1361-6382/ac4196} {\bibfield  {journal} {\bibinfo
  {journal} {Class. Quant. Grav.}\ }\textbf {\bibinfo {volume} {39}},\ \bibinfo
  {pages} {035002} (\bibinfo {year} {2022})},\ \Eprint
  {https://arxiv.org/abs/2105.13972} {arXiv:2105.13972 [gr-qc]} \BibitemShut
  {NoStop}%
\bibitem [{\citenamefont {Yagi}\ \emph
  {et~al.}(2012{\natexlab{b}})\citenamefont {Yagi}, \citenamefont {Yunes},\
  and\ \citenamefont {Tanaka}}]{Yagi:2012vf}%
  \BibitemOpen
  \bibfield  {author} {\bibinfo {author} {\bibfnamefont {K.}~\bibnamefont
  {Yagi}}, \bibinfo {author} {\bibfnamefont {N.}~\bibnamefont {Yunes}},\ and\
  \bibinfo {author} {\bibfnamefont {T.}~\bibnamefont {Tanaka}},\ }\bibfield
  {title} {\bibinfo {title} {{Gravitational Waves from Quasi-Circular Black
  Hole Binaries in Dynamical Chern-Simons Gravity}},\ }\href
  {https://doi.org/10.1103/PhysRevLett.116.169902} {\bibfield  {journal}
  {\bibinfo  {journal} {Phys. Rev. Lett.}\ }\textbf {\bibinfo {volume} {109}},\
  \bibinfo {pages} {251105} (\bibinfo {year} {2012}{\natexlab{b}})},\ \bibinfo
  {note} {[Erratum: Phys.Rev.Lett. 116, 169902 (2016), Erratum: Phys.Rev.Lett.
  124, 029901 (2020)]},\ \Eprint {https://arxiv.org/abs/1208.5102}
  {arXiv:1208.5102 [gr-qc]} \BibitemShut {NoStop}%
\bibitem [{\citenamefont {Cardoso}\ \emph {et~al.}(2017)\citenamefont
  {Cardoso}, \citenamefont {Franzin}, \citenamefont {Maselli}, \citenamefont
  {Pani},\ and\ \citenamefont {Raposo}}]{Cardoso:2017cfl}%
  \BibitemOpen
  \bibfield  {author} {\bibinfo {author} {\bibfnamefont {V.}~\bibnamefont
  {Cardoso}}, \bibinfo {author} {\bibfnamefont {E.}~\bibnamefont {Franzin}},
  \bibinfo {author} {\bibfnamefont {A.}~\bibnamefont {Maselli}}, \bibinfo
  {author} {\bibfnamefont {P.}~\bibnamefont {Pani}},\ and\ \bibinfo {author}
  {\bibfnamefont {G.}~\bibnamefont {Raposo}},\ }\bibfield  {title} {\bibinfo
  {title} {{Testing strong-field gravity with tidal Love numbers}},\ }\href
  {https://doi.org/10.1103/PhysRevD.95.084014} {\bibfield  {journal} {\bibinfo
  {journal} {Phys. Rev. D}\ }\textbf {\bibinfo {volume} {95}},\ \bibinfo
  {pages} {084014} (\bibinfo {year} {2017})},\ \bibinfo {note} {[Addendum:
  Phys.Rev.D 95, 089901 (2017)]},\ \Eprint {https://arxiv.org/abs/1701.01116}
  {arXiv:1701.01116 [gr-qc]} \BibitemShut {NoStop}%
\bibitem [{\citenamefont {Tahura}\ and\ \citenamefont
  {Yagi}(2018)}]{Tahura:2018zuq}%
  \BibitemOpen
  \bibfield  {author} {\bibinfo {author} {\bibfnamefont {S.}~\bibnamefont
  {Tahura}}\ and\ \bibinfo {author} {\bibfnamefont {K.}~\bibnamefont {Yagi}},\
  }\bibfield  {title} {\bibinfo {title} {{Parameterized Post-Einsteinian
  Gravitational Waveforms in Various Modified Theories of Gravity}},\ }\href
  {https://doi.org/10.1103/PhysRevD.98.084042} {\bibfield  {journal} {\bibinfo
  {journal} {Phys. Rev. D}\ }\textbf {\bibinfo {volume} {98}},\ \bibinfo
  {pages} {084042} (\bibinfo {year} {2018})},\ \bibinfo {note} {[Erratum:
  Phys.Rev.D 101, 109902 (2020)]},\ \Eprint {https://arxiv.org/abs/1809.00259}
  {arXiv:1809.00259 [gr-qc]} \BibitemShut {NoStop}%
\bibitem [{\citenamefont {van Gemeren}\ \emph {et~al.}(2023)\citenamefont {van
  Gemeren}, \citenamefont {Shiralilou},\ and\ \citenamefont
  {Hinderer}}]{vanGemeren:2023rhh}%
  \BibitemOpen
  \bibfield  {author} {\bibinfo {author} {\bibfnamefont {I.}~\bibnamefont {van
  Gemeren}}, \bibinfo {author} {\bibfnamefont {B.}~\bibnamefont {Shiralilou}},\
  and\ \bibinfo {author} {\bibfnamefont {T.}~\bibnamefont {Hinderer}},\
  }\bibfield  {title} {\bibinfo {title} {{Dipolar tidal effects in
  gravitational waves from scalarized black hole binary inspirals in quadratic
  gravity}},\ }\href {https://doi.org/10.1103/PhysRevD.108.024026} {\bibfield
  {journal} {\bibinfo  {journal} {Phys. Rev. D}\ }\textbf {\bibinfo {volume}
  {108}},\ \bibinfo {pages} {024026} (\bibinfo {year} {2023})},\ \bibinfo
  {note} {[Erratum: Phys.Rev.D 109, 089901 (2024)]},\ \Eprint
  {https://arxiv.org/abs/2302.08480} {arXiv:2302.08480 [gr-qc]} \BibitemShut
  {NoStop}%
\bibitem [{\citenamefont {Creci}\ \emph {et~al.}(2025)\citenamefont {Creci},
  \citenamefont {van Gemeren}, \citenamefont {Hinderer},\ and\ \citenamefont
  {Steinhoff}}]{Creci:2024wfu}%
  \BibitemOpen
  \bibfield  {author} {\bibinfo {author} {\bibfnamefont {G.}~\bibnamefont
  {Creci}}, \bibinfo {author} {\bibfnamefont {I.}~\bibnamefont {van Gemeren}},
  \bibinfo {author} {\bibfnamefont {T.}~\bibnamefont {Hinderer}},\ and\
  \bibinfo {author} {\bibfnamefont {J.}~\bibnamefont {Steinhoff}},\ }\bibfield
  {title} {\bibinfo {title} {{Tidal effects in gravitational waves from neutron
  stars in scalar-tensor theories of gravity}},\ }\href
  {https://doi.org/10.21468/SciPostPhysCore.8.2.042} {\bibfield  {journal}
  {\bibinfo  {journal} {SciPost Phys. Core}\ }\textbf {\bibinfo {volume} {8}},\
  \bibinfo {pages} {042} (\bibinfo {year} {2025})},\ \Eprint
  {https://arxiv.org/abs/2412.06620} {arXiv:2412.06620 [gr-qc]} \BibitemShut
  {NoStop}%
\bibitem [{\citenamefont {Jain}\ and\ \citenamefont
  {Rettegno}(2025)}]{Jain:2024lie}%
  \BibitemOpen
  \bibfield  {author} {\bibinfo {author} {\bibfnamefont {T.}~\bibnamefont
  {Jain}}\ and\ \bibinfo {author} {\bibfnamefont {P.}~\bibnamefont
  {Rettegno}},\ }\bibfield  {title} {\bibinfo {title} {{Angular momentum flux
  in scalar-tensor theories up to 1.5 post-Newtonian order}},\ }\href
  {https://doi.org/10.1103/PhysRevD.111.104038} {\bibfield  {journal} {\bibinfo
   {journal} {Phys. Rev. D}\ }\textbf {\bibinfo {volume} {111}},\ \bibinfo
  {pages} {104038} (\bibinfo {year} {2025})},\ \Eprint
  {https://arxiv.org/abs/2407.10908} {arXiv:2407.10908 [gr-qc]} \BibitemShut
  {NoStop}%
\bibitem [{\citenamefont {Barausse}\ \emph {et~al.}(2013)\citenamefont
  {Barausse}, \citenamefont {Palenzuela}, \citenamefont {Ponce},\ and\
  \citenamefont {Lehner}}]{Barausse:2012da}%
  \BibitemOpen
  \bibfield  {author} {\bibinfo {author} {\bibfnamefont {E.}~\bibnamefont
  {Barausse}}, \bibinfo {author} {\bibfnamefont {C.}~\bibnamefont
  {Palenzuela}}, \bibinfo {author} {\bibfnamefont {M.}~\bibnamefont {Ponce}},\
  and\ \bibinfo {author} {\bibfnamefont {L.}~\bibnamefont {Lehner}},\
  }\bibfield  {title} {\bibinfo {title} {{Neutron-star mergers in scalar-tensor
  theories of gravity}},\ }\href {https://doi.org/10.1103/PhysRevD.87.081506}
  {\bibfield  {journal} {\bibinfo  {journal} {Phys. Rev. D}\ }\textbf {\bibinfo
  {volume} {87}},\ \bibinfo {pages} {081506} (\bibinfo {year} {2013})},\
  \Eprint {https://arxiv.org/abs/1212.5053} {arXiv:1212.5053 [gr-qc]}
  \BibitemShut {NoStop}%
\bibitem [{\citenamefont {Wagle}\ \emph {et~al.}(2019)\citenamefont {Wagle},
  \citenamefont {Saffer},\ and\ \citenamefont {Yunes}}]{Wagle:2019mdq}%
  \BibitemOpen
  \bibfield  {author} {\bibinfo {author} {\bibfnamefont {P.}~\bibnamefont
  {Wagle}}, \bibinfo {author} {\bibfnamefont {A.}~\bibnamefont {Saffer}},\ and\
  \bibinfo {author} {\bibfnamefont {N.}~\bibnamefont {Yunes}},\ }\bibfield
  {title} {\bibinfo {title} {{Polarization modes of gravitational waves in
  Quadratic Gravity}},\ }\href {https://doi.org/10.1103/PhysRevD.100.124007}
  {\bibfield  {journal} {\bibinfo  {journal} {Phys. Rev. D}\ }\textbf {\bibinfo
  {volume} {100}},\ \bibinfo {pages} {124007} (\bibinfo {year} {2019})},\
  \Eprint {https://arxiv.org/abs/1910.04800} {arXiv:1910.04800 [gr-qc]}
  \BibitemShut {NoStop}%
\bibitem [{\citenamefont {Chia}\ \emph {et~al.}(2024)\citenamefont {Chia},
  \citenamefont {Edwards}, \citenamefont {Wadekar}, \citenamefont {Zimmerman},
  \citenamefont {Olsen}, \citenamefont {Roulet}, \citenamefont {Venumadhav},
  \citenamefont {Zackay},\ and\ \citenamefont {Zaldarriaga}}]{Chia:2023tle}%
  \BibitemOpen
  \bibfield  {author} {\bibinfo {author} {\bibfnamefont {H.~S.}\ \bibnamefont
  {Chia}}, \bibinfo {author} {\bibfnamefont {T.~D.~P.}\ \bibnamefont
  {Edwards}}, \bibinfo {author} {\bibfnamefont {D.}~\bibnamefont {Wadekar}},
  \bibinfo {author} {\bibfnamefont {A.}~\bibnamefont {Zimmerman}}, \bibinfo
  {author} {\bibfnamefont {S.}~\bibnamefont {Olsen}}, \bibinfo {author}
  {\bibfnamefont {J.}~\bibnamefont {Roulet}}, \bibinfo {author} {\bibfnamefont
  {T.}~\bibnamefont {Venumadhav}}, \bibinfo {author} {\bibfnamefont
  {B.}~\bibnamefont {Zackay}},\ and\ \bibinfo {author} {\bibfnamefont
  {M.}~\bibnamefont {Zaldarriaga}},\ }\bibfield  {title} {\bibinfo {title} {{In
  pursuit of Love numbers: First templated search for compact objects with
  large tidal deformabilities in the LIGO-Virgo data}},\ }\href
  {https://doi.org/10.1103/PhysRevD.110.063007} {\bibfield  {journal} {\bibinfo
   {journal} {Phys. Rev. D}\ }\textbf {\bibinfo {volume} {110}},\ \bibinfo
  {pages} {063007} (\bibinfo {year} {2024})},\ \Eprint
  {https://arxiv.org/abs/2306.00050} {arXiv:2306.00050 [gr-qc]} \BibitemShut
  {NoStop}%
\bibitem [{\citenamefont {Agathos}\ \emph {et~al.}(2014)\citenamefont
  {Agathos}, \citenamefont {Del~Pozzo}, \citenamefont {Li}, \citenamefont {Van
  Den~Broeck}, \citenamefont {Veitch},\ and\ \citenamefont
  {Vitale}}]{Agathos:2013upa}%
  \BibitemOpen
  \bibfield  {author} {\bibinfo {author} {\bibfnamefont {M.}~\bibnamefont
  {Agathos}}, \bibinfo {author} {\bibfnamefont {W.}~\bibnamefont {Del~Pozzo}},
  \bibinfo {author} {\bibfnamefont {T.~G.~F.}\ \bibnamefont {Li}}, \bibinfo
  {author} {\bibfnamefont {C.}~\bibnamefont {Van Den~Broeck}}, \bibinfo
  {author} {\bibfnamefont {J.}~\bibnamefont {Veitch}},\ and\ \bibinfo {author}
  {\bibfnamefont {S.}~\bibnamefont {Vitale}},\ }\bibfield  {title} {\bibinfo
  {title} {{TIGER: A data analysis pipeline for testing the strong-field
  dynamics of general relativity with gravitational wave signals from
  coalescing compact binaries}},\ }\href
  {https://doi.org/10.1103/PhysRevD.89.082001} {\bibfield  {journal} {\bibinfo
  {journal} {Phys. Rev. D}\ }\textbf {\bibinfo {volume} {89}},\ \bibinfo
  {pages} {082001} (\bibinfo {year} {2014})},\ \Eprint
  {https://arxiv.org/abs/1311.0420} {arXiv:1311.0420 [gr-qc]} \BibitemShut
  {NoStop}%
\bibitem [{\citenamefont {Saleem}\ \emph {et~al.}(2022)\citenamefont {Saleem},
  \citenamefont {Datta}, \citenamefont {Arun},\ and\ \citenamefont
  {Sathyaprakash}}]{Saleem:2021nsb}%
  \BibitemOpen
  \bibfield  {author} {\bibinfo {author} {\bibfnamefont {M.}~\bibnamefont
  {Saleem}}, \bibinfo {author} {\bibfnamefont {S.}~\bibnamefont {Datta}},
  \bibinfo {author} {\bibfnamefont {K.~G.}\ \bibnamefont {Arun}},\ and\
  \bibinfo {author} {\bibfnamefont {B.~S.}\ \bibnamefont {Sathyaprakash}},\
  }\bibfield  {title} {\bibinfo {title} {{Parametrized tests of post-Newtonian
  theory using principal component analysis}},\ }\href
  {https://doi.org/10.1103/PhysRevD.105.084062} {\bibfield  {journal} {\bibinfo
   {journal} {Phys. Rev. D}\ }\textbf {\bibinfo {volume} {105}},\ \bibinfo
  {pages} {084062} (\bibinfo {year} {2022})},\ \Eprint
  {https://arxiv.org/abs/2110.10147} {arXiv:2110.10147 [gr-qc]} \BibitemShut
  {NoStop}%
\bibitem [{\citenamefont {Saini}\ \emph {et~al.}(2022)\citenamefont {Saini},
  \citenamefont {Favata},\ and\ \citenamefont {Arun}}]{Saini:2022igm}%
  \BibitemOpen
  \bibfield  {author} {\bibinfo {author} {\bibfnamefont {P.}~\bibnamefont
  {Saini}}, \bibinfo {author} {\bibfnamefont {M.}~\bibnamefont {Favata}},\ and\
  \bibinfo {author} {\bibfnamefont {K.~G.}\ \bibnamefont {Arun}},\ }\bibfield
  {title} {\bibinfo {title} {{Systematic bias on parametrized tests of general
  relativity due to neglect of orbital eccentricity}},\ }\href
  {https://doi.org/10.1103/PhysRevD.106.084031} {\bibfield  {journal} {\bibinfo
   {journal} {Phys. Rev. D}\ }\textbf {\bibinfo {volume} {106}},\ \bibinfo
  {pages} {084031} (\bibinfo {year} {2022})},\ \Eprint
  {https://arxiv.org/abs/2203.04634} {arXiv:2203.04634 [gr-qc]} \BibitemShut
  {NoStop}%
\bibitem [{\citenamefont {Porto}(2008)}]{Porto:2007qi}%
  \BibitemOpen
  \bibfield  {author} {\bibinfo {author} {\bibfnamefont {R.~A.}\ \bibnamefont
  {Porto}},\ }\bibfield  {title} {\bibinfo {title} {{Absorption effects due to
  spin in the worldline approach to black hole dynamics}},\ }\href
  {https://doi.org/10.1103/PhysRevD.77.064026} {\bibfield  {journal} {\bibinfo
  {journal} {Phys. Rev. D}\ }\textbf {\bibinfo {volume} {77}},\ \bibinfo
  {pages} {064026} (\bibinfo {year} {2008})},\ \Eprint
  {https://arxiv.org/abs/0710.5150} {arXiv:0710.5150 [hep-th]} \BibitemShut
  {NoStop}%
\bibitem [{\citenamefont {Metsaev}\ and\ \citenamefont
  {Tseytlin}(1987)}]{Metsaev:1986yb}%
  \BibitemOpen
  \bibfield  {author} {\bibinfo {author} {\bibfnamefont {R.~R.}\ \bibnamefont
  {Metsaev}}\ and\ \bibinfo {author} {\bibfnamefont {A.~A.}\ \bibnamefont
  {Tseytlin}},\ }\bibfield  {title} {\bibinfo {title} {{Curvature Cubed Terms
  in String Theory Effective Actions}},\ }\href
  {https://doi.org/10.1016/0370-2693(87)91527-9} {\bibfield  {journal}
  {\bibinfo  {journal} {Phys. Lett. B}\ }\textbf {\bibinfo {volume} {185}},\
  \bibinfo {pages} {52} (\bibinfo {year} {1987})}\BibitemShut {NoStop}%
\bibitem [{\citenamefont {Wang}(2025)}]{louhan}%
  \BibitemOpen
  \bibfield  {author} {\bibinfo {author} {\bibfnamefont {L.}~\bibnamefont
  {Wang}},\ }\emph {\bibinfo {title} {Testing beyond GR theories and the PN-EFT
  approach}},\ \href@noop {} {Master's thesis},\ \bibinfo  {school} {Perimeter
  Scholars International Master's Program} (\bibinfo {year} {2025})\BibitemShut
  {NoStop}%
\bibitem [{\citenamefont {Liu}\ and\ \citenamefont
  {Yunes}(2025)}]{Liu:2024atc}%
  \BibitemOpen
  \bibfield  {author} {\bibinfo {author} {\bibfnamefont {H.}~\bibnamefont
  {Liu}}\ and\ \bibinfo {author} {\bibfnamefont {N.}~\bibnamefont {Yunes}},\
  }\bibfield  {title} {\bibinfo {title} {{Robust and improved constraints on
  higher-curvature gravitational effective-field-theory with the GW170608
  event}},\ }\href {https://doi.org/10.1103/PhysRevD.111.084049} {\bibfield
  {journal} {\bibinfo  {journal} {Phys. Rev. D}\ }\textbf {\bibinfo {volume}
  {111}},\ \bibinfo {pages} {084049} (\bibinfo {year} {2025})},\ \Eprint
  {https://arxiv.org/abs/2407.08929} {arXiv:2407.08929 [gr-qc]} \BibitemShut
  {NoStop}%
\end{thebibliography}%

\end{document}